\pdfoutput=1

\documentclass[12pt, oneside]{book}

\usepackage{fullpage}

\usepackage{threeparttable}
\usepackage{lscape}

\usepackage{graphicx}
\usepackage{amsmath}
\usepackage{longtable}
\usepackage[none]{hyphenat}
 \usepackage{setspace}
 \usepackage{url}
 \usepackage{caption}
 \usepackage{subcaption}
 \usepackage{chapterbib}
 \usepackage{tocloft}

\cftpagenumbersoff{table}

\usepackage{caption}

\usepackage{natbib}
\bibpunct[; ]{(}{)}{,}{a}{}{;}

\newcommand{\Dmel}{\emph{D. melanogaster}}

\newcommand{\Dyak}{\emph{D. yakuba}}
\newcommand{\FullDyak}{\emph{Drosophila yakuba}}
\newcommand{\Dros}{\emph{Drosophila}}
\newcommand{\superscript}[1]{\ensuremath{^{\textrm{#1}}}}

\newcommand{\Dsim}{\emph{D. simulans}}
\newcommand{\FullDsim}{\emph{Drosophila simulans}}

\newcommand{\Dere}{\emph{D. erecta}}

\newcommand{\jgw}{\emph{jgw}}

\newcommand{\fiveP}{5$'$}
\newcommand{\thrP}{3$'$}

\newcommand{\Rebekah}{Rebek\mbox{}ah }

\usepackage{textcomp}

\begin{document}

\author{\Rebekah L. Rogers$^1$, Julie M. Cridland$^{1,2}$, Ling Shao$^1$, Tina T. Hu$^3$, \\ Peter Andolfatto$^3$, and Kevin R. Thornton$^1$}

\title{Landscape of standing variation for tandem duplications in \FullDyak {} and \FullDsim }
\date{}

{\hfill \textbf{Accepted- Molecular Biology and Evolution}
\let\newpage\relax\maketitle}

\begin{center} \Large Research Article \end{center}
\vspace{0.25in}

\noindent 1) Ecology and Evolutionary Biology, University of California, Irvine \\
\noindent 2) Ecology and Evolutionary Biology, University of California, Davis \\
\noindent 3) Ecology and Evolutionary Biology and the Lewis Stigler Institute for Integrative Genomics, Princeton University \\

\noindent \textbf{Running head: } Tandem duplications in non-model \Dros

\vspace{0.25in}


\vspace{0.25in}

\noindent \textbf{Corresponding author:} Rebekah L. Rogers,  Dept. of Ecology and Evolutionary Biology, 5323 McGaugh Hall, University of California, Irvine, CA 92697 \\
\\
\noindent \textbf{Phone:}  949-824-0614

\noindent \textbf{Fax:}  949-824-2181

\noindent \textbf{Email:} rogersrl@uci.edu \\

\newpage
%

\newpage
\chapter*{}

\renewcommand{\thefigure}{\arabic{figure}}
\renewcommand{\thetable}{\arabic{table}}
\setcounter{figure}{0}
\section*{Abstract}

We have used whole genome paired-end Illumina sequence data to identify tandem duplications in 20 isofemale lines of \Dyak, and 20 isofemale lines of \Dsim  {} and performed genome wide validation with PacBio long molecule sequencing.  We identify 1,415 tandem duplications that are segregating in \Dyak {} as well as 975 duplications in \Dsim, indicating greater variation in \Dyak.  Additionally, we observe high rates of secondary deletions at duplicated sites, with 8\% of duplicated sites in \Dsim {} and 17\% of sites in \Dyak {} modified with deletions.  These secondary deletions are consistent with the action of the large loop mismatch repair system acting to remove polymorphic tandem duplication, resulting in rapid dynamics of gain and loss in duplicated alleles and a richer substrate of genetic novelty than has been previously reported.  Most duplications are present in only single strains, suggesting deleterious impacts are common.  \Dsim {} shows larger numbers of whole gene duplications in comparison to larger proportions of gene fragments in \Dyak.   \Dsim {} displays an excess of high frequency variants on the X chromosome, consistent with adaptive evolution through duplications on the \Dsim {} X or demographic forces driving duplicates to high frequency.  We identify 78 chimeric genes in \Dyak {} and 38 chimeric genes in \Dsim, as well as 143 cases of recruited non-coding sequence in \Dyak {} and 96 in \Dsim, in agreement with rates of chimeric gene origination in \Dmel.   Together, these results suggest that tandem duplications often result in complex variation beyond whole gene duplications that offers a rich substrate of standing variation that is likely to contribute both to detrimental phenotypes and disease, as well as to adaptive evolutionary change.

\newpage

\section*{Introduction}

Gene duplications are an essential source of genetic novelty that can be useful in adaptation and in the origins of developmental complexity across phyla \citep{WolfeReview}.  Additionally, duplicate sequences are commonly found in mammalian stem cells \citep{Liang2008}, cancer cell lines \citep{Inaki2012}, and are associated with autoimmune disease, HIV susceptibility, Crohn's disease, asthma, allergies, and autism \citep{Ionita2009}.  Distinguishing the propensity with which gene duplications serve as causative disease factors as opposed to a source of favorable variation depends heavily on accurate ascertainment of their occurance and frequencies in the population.  

In the \Dros {} there is substantial variation in the number and types of duplicate genes that are present in the sequenced reference genomes \citep{Hahn2007}, though the extent to which selection might drive rapid fixation of duplicate genes or whether mutation rates differ across species remains uncertain.   Furthermore, these surveys of single strains from each species may not be representative of the variation present in populations and offer only limited opportunities to study their role in adaptation.  The advent of Illumina sequencing has made population genomics of complex mutations in non-model \Dros {} readily tractable. Paired-end Illumina sequencing offers the opportunity to survey copy number variation using definitive sequence-based comparisons that are free from complications related to sole use of coverage or hybridization intensities.  Through the identification of paired-end reads that map in abnormal orientations, we can identify a high-confidence dataset describing tandem duplications in sample populations \citep{JCridland, Korbel2007, Tuzun2005}. 
 
\Dyak {} and \Dsim {} offer the opportunity to compare the role of tandem duplications in species that have high levels of nucleotide diversity and large effective population sizes of $N_e \approx 10^6$ \citep{Bachtrog2006, Sawyer1992, EyreWalker2002}, allowing us to compare mutational and adaptive processes in independent systems where neutral forces of genetic drift should be minimal.

If different chromosomes produce tandem duplications at different rates, we may expect them to contribute differentially to adaptive changes.  In \Dmel, the X chromosome contains greater repetitive content \citep{CridlandTEs}, displays different gene density \citep{DmelRef}, has potentially smaller population sizes \citep{Wright1931, Andolfatto2001}, lower levels of background selection \citep{Charlesworth}, and an excess of genes involved in female-specific expression \citep{Ranz2003} in comparison to the autosomes.  Furthermore, the X chromosome is hemizygous in males, exposing recessive mutations to the full effects of selection more often than comparable loci on the autosomes \citep{Charlesworth1987}.  Hence, the incidence of duplications on the X and the types of genes affected may differ from the autosomes, and thereby produce different impacts on phenotypic evolution.

Many copy number variants are thought to be non-neutral \citep{Emerson2008, Cardoso2011, Hu1992}, especially when they capture partial gene sequences or create chimeric gene structures \citep{RapidEvolution} or result in recruitment of non-coding sequences \citep{Lee2012}.  Such modifications are likely to change gene regulatory profiles \citep{RapidEvolution}, increasing the likelihood of non-neutral phenotypes.  Surveys in \Dmel {} have identified large numbers of such variants \citep{Emerson2008, Cardoso2011, Cardoso2012, Lee2012, RBH}.  Establishing profiles of partial gene duplication, whole gene duplication, chimera formation, and recruitment of non-coding sequence are essential to a complete understanding of the roles tandem duplicates play in beneficial and detrimental phenotypic changes across species.  

Here we describe the number, types, and genomic locations of tandem duplications segregating in 20 strains of \Dyak {} and in 20 strains of \Dsim {} and discuss differences across species and across chromosomes, as well as their potential to create novel gene constructs.  


\section*{Results}
 We have sequenced the complete genomes of 20 isofemale lines of \Dyak {} and 20 isofemale lines \Dsim {} each inbred in the lab for 9-12 generations to produce effectively haploid samples, as well as the reference genome stocks of each species (as a control for genome quality and false positives) \citep{TwelveGenomes,SimRef}.  Genomes are sequenced to high coverage of 50-150X for a total of 42 complete genomes (Table \ref{sfig:laneSummYak}-\ref{InsertSize}, see Methods).   We have used mapping orientation of paired-end reads to identify recently-derived, segregating duplications in these samples less than 25 kb in length that are supported by 3 or more divergently-oriented read pairs (see Methods, Text S1, Table \ref{ByStrain}-\ref{ByStrainWithRefs}).  We limit analysis to regions of the genome which can be assayed with coverage depth of 3 or more reads across all strains, corresponding to the detection limit for tandem duplicates.  We identify 1,415 segregating tandem duplications in \Dyak {} and 975  segregating tandem duplications in \Dsim {} (Figure \ref{Summary}), including large numbers of gene duplications (Table \ref{GeneDups}) with a low false positive rate (Table \ref{tab:confirmations}).  We assess the numbers and types of gene duplications, differences in duplication rates and sizes across chromosomes, and describe evidence of secondary modification through deletions, which will influence the extent to which these variants can serve as a source of genetic novelty. 

\subsection*{Genotyping and Quality Control}  

\begin{figure}[h]
\begin{center}
\clearpage
\includegraphics[scale=0.8]{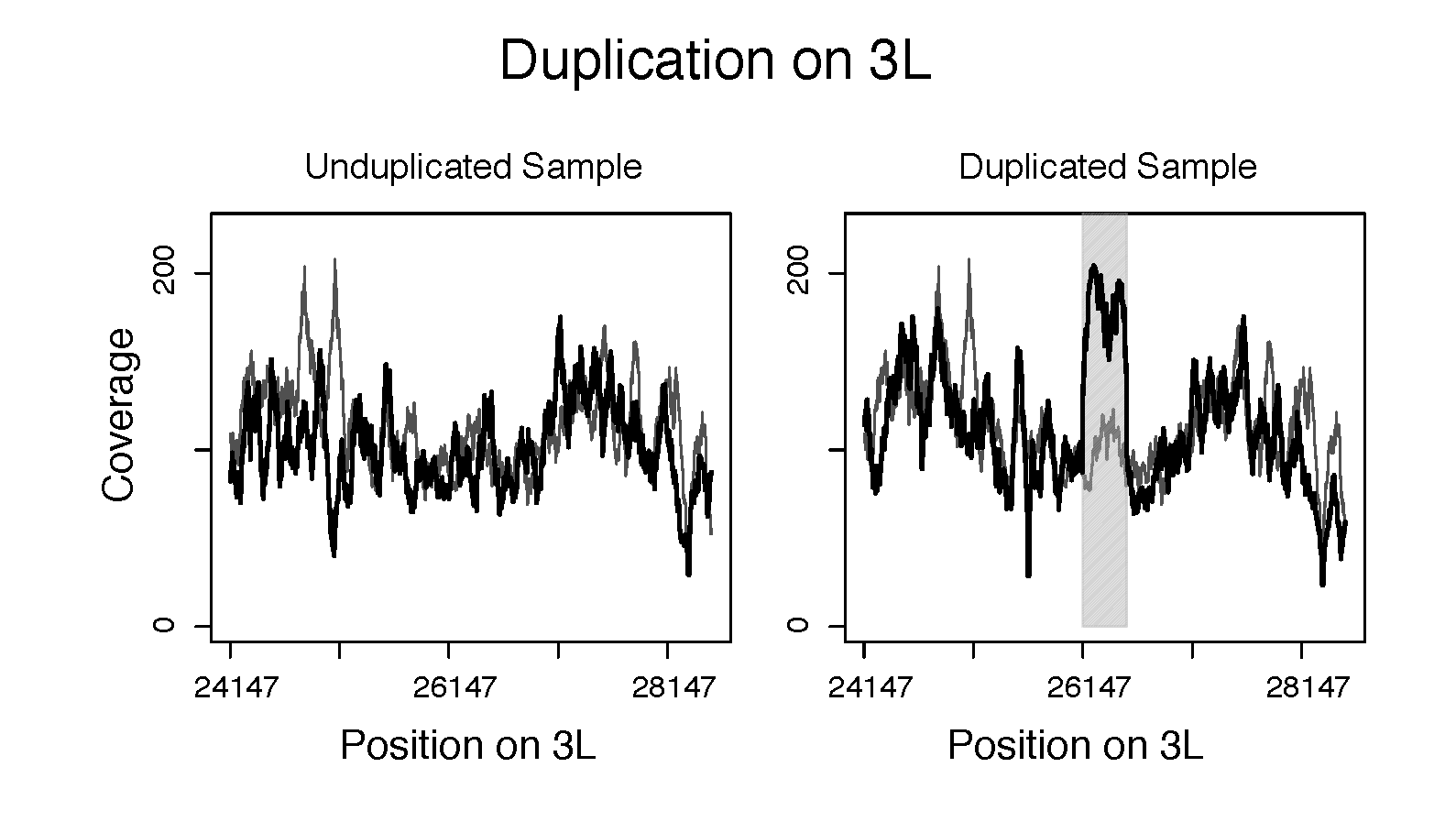}
\end{center}
\caption{Coverage change for a duplication on chromosome 3L in Line 9 of \Dyak.  Regions spanned by divergently-oriented reads are shaded.  Sample coverage is shown in black, while reference genome coverage is shown in grey.    }

\label{CoverageChange}
\end{figure}

Divergently-oriented paired-end reads are effective indicators of tandem duplications \citep{JCridland, Zichner2013, Mills2011, Tuzun2005}.  We have used paired-end read orientation (Figure \ref{DupMapping}) combined with increased coverage in genomic sequencing (Figure \ref{CoverageChange}) to identify tandem duplications in population samples of \Dyak {} and \Dsim.  Divergently-oriented reads  indicative of putative tandem duplications were clustered within a single strain, with three or more divergently-oriented read pairs within the strain required to define each tandem duplication (see Methods).  Duplications were then clustered across strains with coordinates defined as the maximum span of divergent reads across all strains.  The distribution of supporting read pairs is highly skewed, with 3-4 supporting read pairs for many calls (Figure \ref{DIVCov}).   To account for duplications which may be undetected, we additionally included variants that showed two-fold increases (Figure \ref{CoverageChange}) in quantile normalized coverage (Figure \ref{sfig:DsimNcov}-\ref{sfig:DyakNcov}) and which are supported by one or more divergently-oriented read pairs were also identified as having duplications if the duplicate was present in a second strain, thereby correcting sample frequency estimates for false negatives (see Methods, Text S1).  We retained only those tandem duplicates which are not present in outgroup reference genomes of \Dmel, \Dere, and \Dyak {} or \Dsim {} as defined in a BLAST search (see Methods) suggesting recent origins.  

Using divergently-oriented paired-end reads, we have identified 1,415 segregating tandem duplications across 20 sample strains of \Dyak, in comparison to 975 segregating tandem duplications in 20 lines of \Dsim, with significantly more duplicates identified in \Dyak {} than in \Dsim {} (one-sided $t$-test $t=-3.8126$, $df = 24.593$, $P= 0.0004089$).  More variants are identified in \Dyak {} in spite of higher coverage in \Dsim {} (Table \ref{tab:rawCovSumm}), suggesting that the difference is likely to be biological rather than technical.  In fact, the number of variants identified is only weakly correlated with coverage per strain (Figure \ref{CovCorr}) a product of sequencing to the saturation point of coverage (Table \ref{Downsample}).  Downsampling reads from \Dyak {} CY17C, which was sequenced to 151X, we find that the portion of the genome covered with 3 or more reads (the detection limit of our assay) plateaus at roughly 45X (Table \ref{Downsample}), though lower coverage data used in previous studies \citep{Sudmant2010, Zichner2013, Mills2011, Alkan2009} will be far from this plateau.  The tandem duplications identified across these sample strains cover 2.574\% of the assayable genome of the X and 4 major autosomes in \Dyak {} and 1.837\% of the assayable genome of the X and 4 major autosomes in \Dsim.  We are able to identify tandem duplications as small as 66 bp in \Dyak {} and 78 bp in \Dsim.     

 We have successfully PCR amplified $\frac{23}{46}$ randomly chosen variants in \Dyak {} and $\frac{35}{42}$ variants in \Dsim.  The rate of PCR confirmation in \Dsim {} is not significantly different from previous studies of CNVs, but we observe significant differences between \Dyak {} and all other confirmation rates (Table \ref{FET}).  In view of this disparity, combined with difficulties of PCR primer design for variants whose precise structures are unknown, we generated PacBio long molecule sequencing data for 4 strains of \Dyak {} in order to more reliably estimate the false positive rate (Table \ref{ReadLengths}). PacBio long molecule sequencing has recently been used to validate targeted duplications in human genome data \citep{EichlerPacBio}. We extend this approach to genome-wide identification and validation of tandem duplications, and have generated PacBio reads for four different sample strains of \Dyak.  Across these four strains, we observe confirmation of 661 out of 688 mutations, for a maximum false-positive rate of 3.9\% (Table \ref{tab:confirmations}), though some variants may be unconfirmed due to low clone coverage in a region.  Hence, the duplicates identified with paired-end reads in high coverage genomic sequence data are extremely accurate and comparable to or better than previous methods or attempts to identify and validate duplicates using lower coverage genomic sequences or microarrays \citep{Mills2011, Alkan2009, Cardoso2011, Zichner2013, Sudmant2010}.  Split read mapping with short Illumina reads performed poorly in comparison and failed to confirm 88.3\% of variants, and breakpoint assembly was possible for less than 60\% of variants in spite of high rates of confirmation with PacBio (See SI Text).  Thus, requiring these criteria would exclude the majority of variant calls and would likely be biased against duplicates with formation facilitated by repetitive sequences.  Where duplicate breakpoints contain repetitive or low complexity sequences, or where subsequent modification of alleles through deletion has altered surrounding sequence, PCRs are likely to fail, and we would suggest that confirmation using long molecule sequencing is far more reliable in the face of complex structures.  Further description of genomic sequences, tandem duplications, and discussion of paired-end read performance in high coverage genomic sequencing data in comparison to other methods is available in Text S1.  

\subsection*{Complex variation}
We identified deletions that have occurred in duplicated alleles using long-spanning read pairs 600 bp or longer, corresponding approximately to the 99.9th percentile of fragment lengths in the reference genomes (Table \ref{InsertSize}).  Out of 880 duplications $\geq$600 bp  in length in \Dyak {} which could be surveyed for deletions within duplications using long-spanning reads, 151 (17\%) contain long spanning read pairs covering 50\% or more of the duplicate sequence in one or more strain, indicative of subsequent deletion, multiple independent short-range dispersed duplications, or incomplete duplication (Figure \ref{Complex}).  In \Dsim, $\frac{39}{486}$ (8\%) duplications $\geq$ 600 bp contain long-spanning reads covering 50\% or more of duplicated sequence in one or more strains.   Among 69 such modified variants in \Dyak {} that are present in multiple strains, 66 have at least one strain that lacks these long-spanning reads, whereas 12 out of 14 variants in \Dsim {} lack long spanning reads in one or more strains.  Given large numbers of unaltered duplicates the most parsimonious explanation is that deletions are most often secondary modification and that the majority of these constructs form through full length duplication and subsequent deletion rather than independent dispersed duplications.  

\begin{figure}[!h]
  \begin{center}
\includegraphics[scale=1.15]{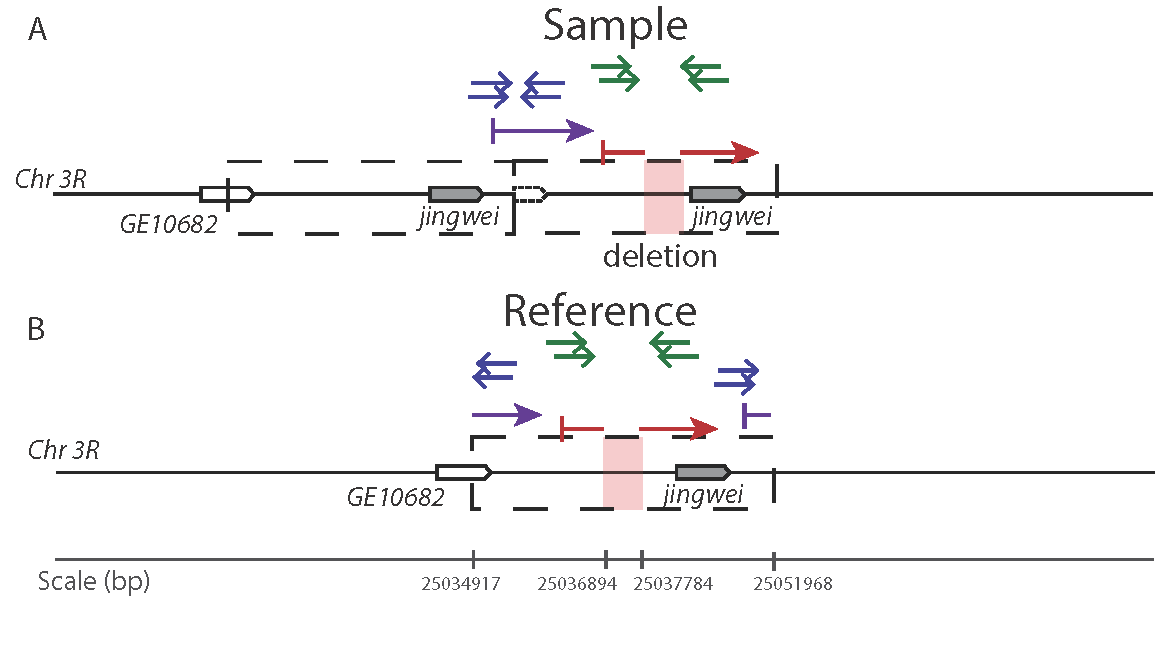}
  \caption{\label{ComplexMapping} Read mapping patterns indicative of a modified duplication surrounding \emph{jingwei} in \Dyak {} line NY66-2.  Duplications are indicated with divergently-oriented paired-end reads (blue) as well as with split read mapping of long molecule sequencing (purple).  Deletions in one copy are suggested by gapped read mapping of long molecule reads (red) as well as multiple long-spanning read pairs at the tail of mapping distances in paired-end read sequencing (green) just upstream from \emph{jgw}.  Up to 20\% of duplicates observed have long-spanning read pairs indicative of putative deletions in one or more alleles in the population.  }
\end{center}
\end{figure}

In one well-characterized example, we have identified a duplication which spans the chimeric retrogene \emph{jingwei} (\jgw) \citep{MLong1993}, which houses a deletion upstream from \jgw {} (Figure \ref{ComplexMapping}).  The duplication is defined by 10 divergent read pairs and confirmed by split read mapping in PacBio long molecule sequencing, whereas the deletion is supported by 20 long-spanning read pairs in line NY66-2 and gapped alignment in PacBio reads (Figure \ref{ComplexMapping}).  The same duplication and deletion are independently confirmed with PacBio sequences in line CY21B3.  The duplication spanning \jgw {} is found at a frequency of $\frac{5}{20}$ strains, while the deletion shown is observed only in CY21B3 and NY66-2, suggesting that the deletions is a secondary modification. A second independent duplication spans \jgw {} in $\frac{4}{20}$ strains and is confirmed in PacBio data, indicating that the region has been modified multiple times in different strains.

\begin{figure}[h]
\begin{center}

\includegraphics[scale=0.75]{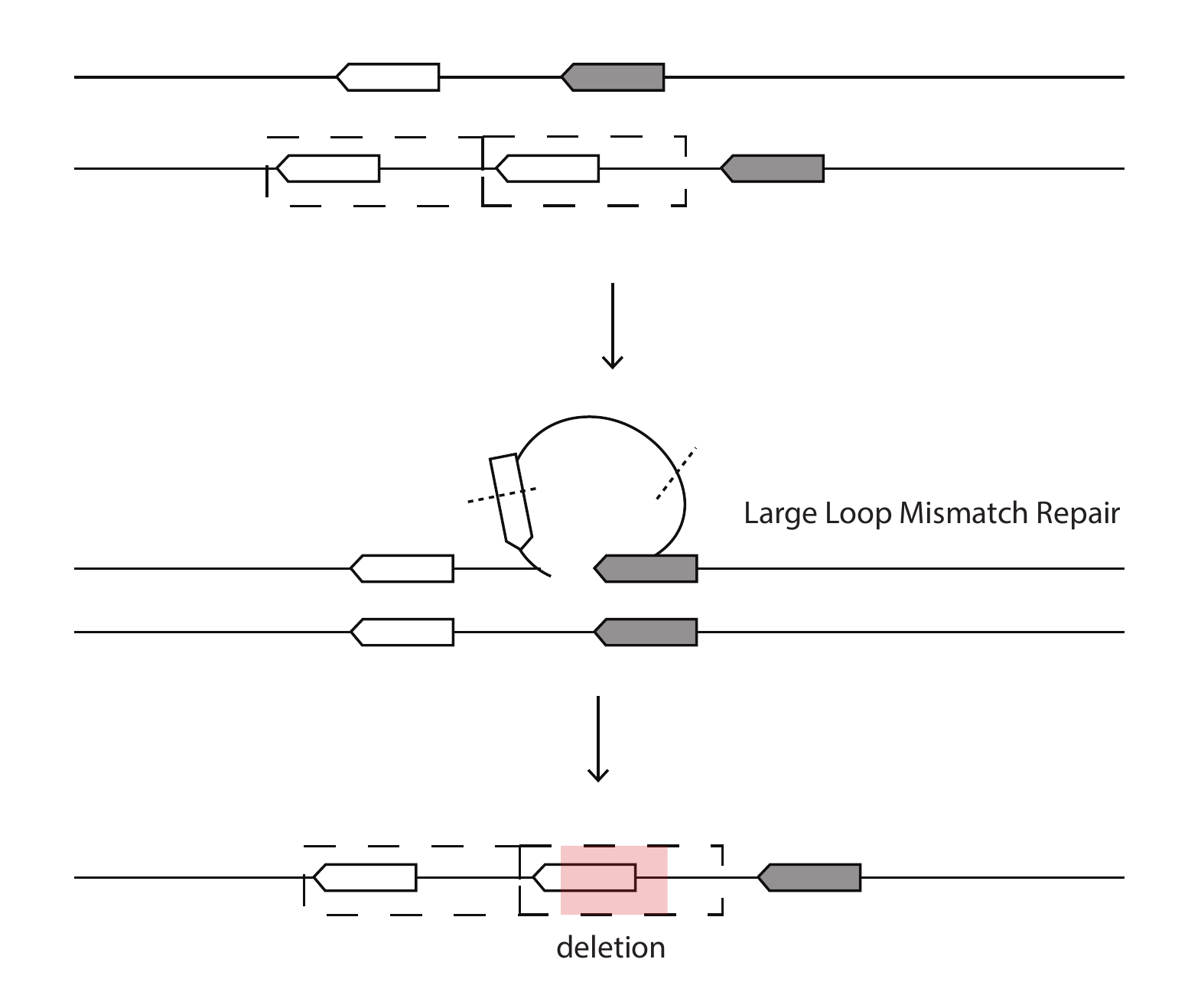}
\end{center}
\caption{Secondary Deletion via large loop mismatch repair.  A tandem duplication forms via ectopic recombination or replication slippage.  At some point prior to fixation in the population the duplication pairs with an unduplicated chromatid in meiosis or mitosis, invoking the action of the large loop mismatch repair system.  Imprecise excision results in a modified duplicate with partially deleted sequence.  Large loop mismatch repair requires that duplications are polymorphic, and would therefore produce secondary modification over short timescales, resulting in rapid modification of tandem duplicates.  }

\label{LLMRs}
\end{figure}

Deletions are exceptionally common in \Dros {} \citep{Petrov1996}, and several genetic mechanisms might offer means of excision in a short time frame after duplication.   The large loop mismatch repair system can facilitate deletions of duplicated sequence to modify duplicated sequence as long as variants are polymorphic. The presence of unpaired duplicated DNA during meiosis or mitosis would commonly invoke the action of the large loop mismatch repair system, which if resolved imprecisely, could result in the construct observed (Figure \ref{LLMRs}).  Deletions lying within a duplication have a median size of 3.6 kb in \Dyak {} and 1.8 kb in \Dsim. Such large deletions are well outside the norm for genome wide large deletions in mutation accumulation lines of \Dmel, which show an average 409 bp and maximum of 2.6 kb \citep{Schrider2013}.  Deletions of this size however are consistent with the size of excised fragments in large loop mismatch repair of several kilobases \citep{Kearney2001}.   Deletion during non-homologous end joining or homology mediated replication slippage might produce deletions as well though it is unclear whether mutation rates are naturally high enough to operate in short time frames.  Thus, we would expect modification of duplicated alleles to be extremely common, especially in deletion-biased \Dros.  

\subsection*{Differences in gene duplications across species}
Duplicated coding sequences can diverge to produce novel peptides, novel regulatory profiles, or specialized subfunctions \citep{WolfeReview}.  In order to determine the extent to which genes are likely to be duplicated and whether  particular categories of gene duplications are more likely to be favored, we identified coding sequences captured by tandem duplications.  We find large numbers of segregating gene duplications in both \Dyak {} and \Dsim {} including hundreds of whole gene duplications (Table \ref{GeneDups}).  We used the maximum span of divergently-oriented reads across all strains to identify tandem duplications that capture gene sequences in \Dyak {} or \Dsim {} and to determine their propensity to capture whole and partial gene sequences. 

We find that 47.3\% of tandem duplications in \Dyak{} and 40.8\% in \Dsim {} capture coding sequences. The average duplicated gene in \Dyak {} covers 45.9\% of the gene sequence and  60.5\% of the gene sequence in \Dsim.   There are 670 duplications that capture gene sequences, spanning 845 different genes in \Dyak, while 398  duplications span 478 genes in \Dsim.  Some 103  genes in \Dyak {}  and 65 genes in \Dsim {} are captured in multiple independent duplications, with some genes falling in as many as 6 independent putative duplications as defined by divergently-oriented reads in \Dyak {} and 32 independent putative duplications in \Dsim.  Such high rate of independent duplications in \Dsim {} is consistent with previous studies using microarrays \citep{Cardoso2012}.  In total 993 gene fragments in \Dyak {} and 758 gene fragments in \Dsim {} exist as segregating copy number variants in the population, and 56 genes are duplicated in both species.    

Assuming that unmodified duplications without deletions represent the original mutated state, in \Dyak, $\frac{274}{845}$ (27.6\%) of duplicated gene fragments span more than 80\% of gene sequence and capture the translation start site while 65 (7.7\%) capture 20\% or less and the translation start site.  In  \Dsim {} $\frac{317}{478}$ (66.3\%) duplicated gene sequences capture 80\% or more of the gene sequence and the translation start, and 34 (7.1\%) capture 20\% or less and include the translation start.  Based on a resampling of gene duplications in \Dyak, \Dsim {} houses an overabundance of whole or nearly whole gene duplications ($P<10^{-7}$) and an underrepresentation of small fragments ($P=0.00291$), suggesting differences in the occurrence of whole gene duplications across species due either to mutational pressures or selection.
  
\subsection*{Duplicate genes and rapidly evolving phenotypes} 

Biases in the rates at which duplications form in different genomic regions or a greater propensity for selection to favor duplications in specific functional classes can result in a bias in gene ontology categories among duplicated genes.  We used DAVID gene ontology analysis software to identify overrepresented functions among duplicate genes in \Dyak {} and \Dsim.  In \Dyak {} we observe 678 duplicated genes with orthologs in \Dmel.    Overrepresented functional categories include immunoglobulins, extracellular matrix, chitins and aminoglycans, immune response and wound healing, drug and hormone metabolism, chorion development, chemosensory response and development and morphogenesis (Table \ref{GOBias}).  In \Dsim {} we observe 478 duplicated genes with orthologs in \Dmel.  Overrepresented gene ontology categories include cytochromes and oxidoreductases plus toxin metabolism,  immune response to microbes, phosopholipid metabolism, chemosensory processing, carboxylesterases, glutathion transferase and drug metabolism, and sarcomeres  (Table \ref{GOBias}). In \Dsim {} 65 genes were involved in multiple independent duplications that have distinct breakpoints.  Overrepresented gene ontology categories include immune response to bacteria, chorion development and oogenesis, chemosensory perception, and organic cation membrane transport (Table \ref{IndependentDuplication}).  In \Dyak {} among 72 genes duplicated independently, chorion development and oogenesis, cell signaling, immune response, sensory processing, and development are overrepresented (Table \ref{IndependentDuplication}).

There are 25 high frequency variants found at a sample frequency of $\frac{17}{20}$ or greater in \Dsim, including lipases and endopeptidases expressed in male accessory glands and several genes involved in immune response to microbes  (Table \ref{HighFreq}).  One gene arose independently and has reached high frequency twice in \Dsim.  In \Dyak, we observe 13 high frequency variants, including endopeptidases and AMP dependent ligases (Table \ref{HighFreq}).  Both male reproductive proteins \citep{Wong2012} and immune response to pathogens \citep{Lazarro2012} are known for their  rapid evolution, and therefore these genes are strong candidates to search for evidence of ongoing selective sweeps.  Though mutational biases can produce similarities in gene ontology categories, the overabundance of toxin metabolism genes and immune response peptides in both species as well as the overrepresentation of chemoreceptors, chitin cuticle genes, and oogenesis factors suggests that duplications are likely key players in rapidly evolving systems.

\subsection*{Chimeric Genes and Altered Coding Sequences}
\begin{figure}[h]
\begin{center}

\hspace{-0.25in}
\includegraphics[scale=.5]{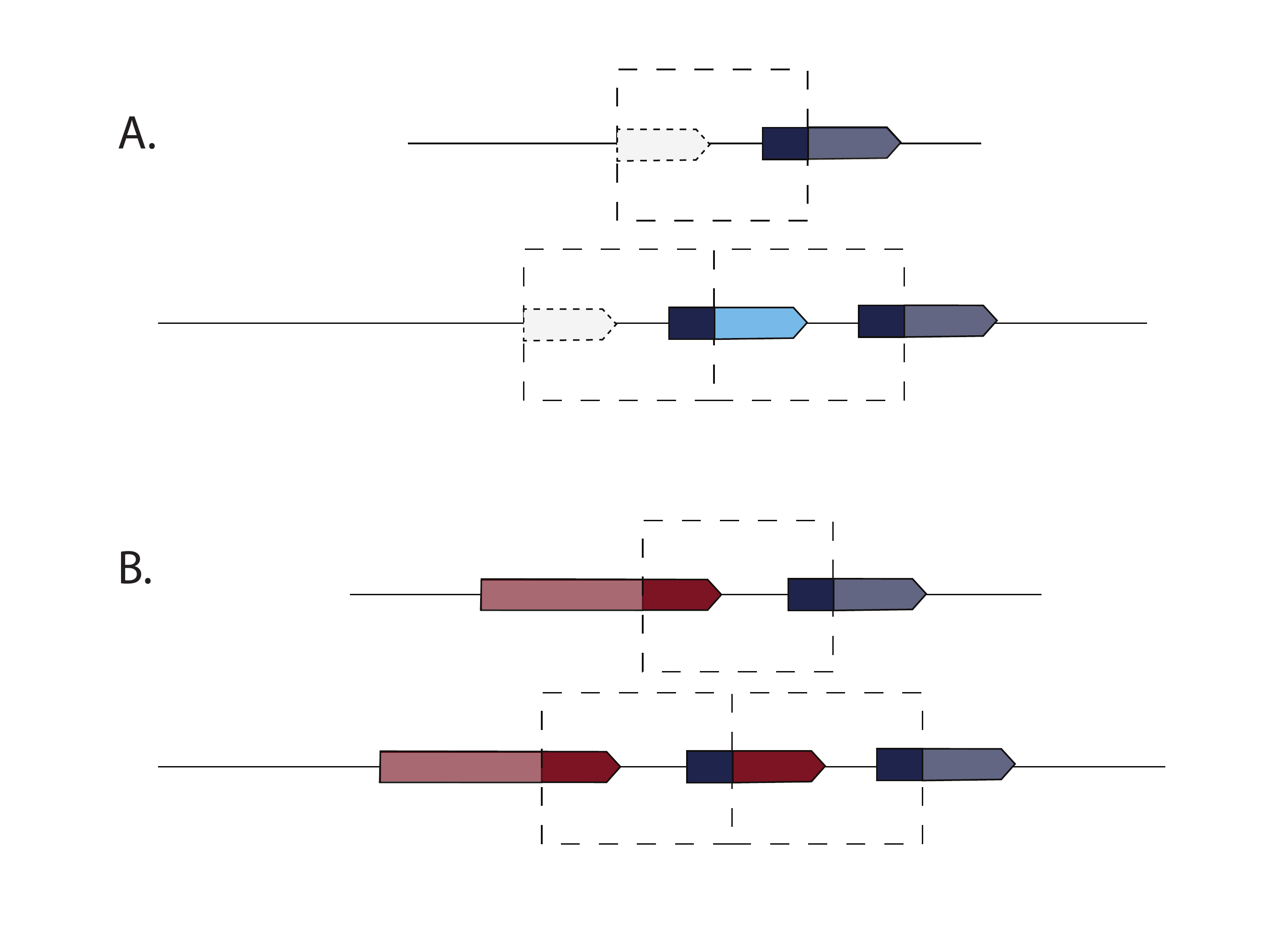}
\end{center}
\caption{\label{ChimeraFig} Abnormal Gene Structures.  Duplicated sequence is highlighted with bold colors and is framed by the dashed box.  A) The partial duplication of a coding sequence (blue) results in the recruitment of previously upstream non coding sequence (dashed lines) to create a novel open reading frame (blue and turquoise).  B)  Tandem duplication where both boundaries fall within coding sequences results in a chimeric gene.  }
\end{figure}

If only one boundary of a tandem duplication falls within a coding sequence and thereby copies the \fiveP {} end of a gene, the resulting construct will recruit formerly non-coding sequence to form the \thrP {} end of the coding sequence (Figure \ref{ChimeraFig}A).  We observe 143 cases of recruitment of non-coding sequence in \Dyak {} (Table \ref{DyakRecruitedI}-\ref{DyakRecruitedII}) and 96 cases in \Dsim {} (Table \ref{DsimRecruitedI}-\ref{DsimRecruitedII}).  Several of these are found at moderate frequencies greater than 50\%.  Overrepresented GO categories among genes in \Dsim {} include immune defense and sarcomeres, whereas genes with recruited sequence in \Dyak {} show an overrepresentation of genes involved in locomotory behavior. We observe one high frequency variant in \Dyak {} at a sample frequency of $\frac{17}{20}$, an Adenylate cyclase involved in locomotor rhythm as well as two high frequency variants in \Dsim {} LysB, an antimicrobial humoral response gene and a gene of unknown function.    These high frequency chimeras are strong candidates for selective sweeps.  
 
\begin{figure}[h]
\begin{center}
\hspace{-0.25in}
\includegraphics[scale=.5]{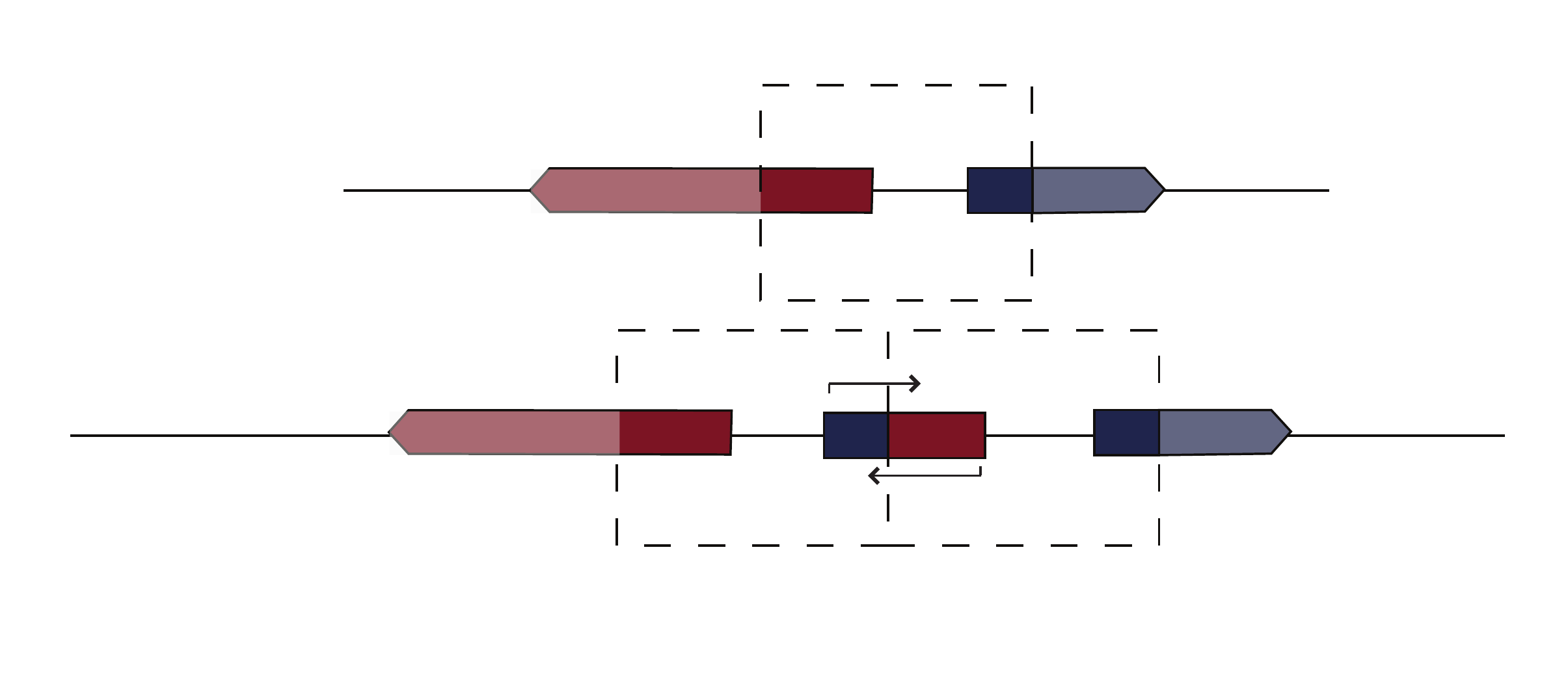}
\end{center}
\caption{\label{DualPromFig} Dual promoter genes.  Duplicated sequence is highlighted with bold colors and is framed by the dashed box.    Tandem duplication where both boundaries fall within coding sequences results in a chimeric gene which contains two promoters, one which facilitates transcription in one direction, the other facilitating transcription from the opposite strand.  The chimera is capable of making partial anti-sense transcripts.  }
\end{figure}

If both boundaries fall within different coding sequences, tandem duplications can create chimeric genes (Figure \ref{ChimeraFig}B) \citep{RBH}.  We find 130 tandem duplications in \Dyak {}, and  76 in \Dsim {}  where both breakpoints fall within non-overlapping coding sequences.  Some 11 of the 130 duplications in \Dyak {} and 30 of 76 in \Dsim {} with both breakpoints in gene sequences face one another and as such are not expected to create new open reading frames, as the constructs will lack promoters.  Another 40 of the 130 duplications in \Dyak {} and 8 of 76 in \Dsim {} with both breakpoints in gene sequences, and will have promoters that can potentially transcribe sequences from both strands of DNA (Figure \ref{DualPromFig}, Table \ref{DyakDualProm}-\ref{DsimDualProm}).  Only 78 chimeric coding sequences in \Dyak {}  (Table \ref{DyakChimI}-\ref{DyakChimII}) and 38 chimeric genes in \Dsim {}  (Table \ref{DsimChim}) have parental genes in parallel orientation.  Among the parental genes of these chimeras, cytochromes and genes involved in drug metabolism are overrepresented in \Dyak.  Other functional categories which are present but not overrepresented include endopeptidases, signaling glycopeptides, and sensory signal transduction peptides.  Among parental genes in \Dsim, Cytochromes and insecticide metabolism genes, sensory perception genes, and endopeptidase genes are overrepresented.  Additional categories present include signal peptides, endocytosis genes, and oogenesis genes.  Several such constructs are found at moderate frequencies above 10/20, suggesting that they are at least not detrimental.  However, two chimeras in \Dyak {} are found at high frequency.  One formed from a combination of \emph{GE12441} and \emph{GE12442} is at a frequency of 16/20, and one formed from \emph{GE12353} and \emph{GE12354} is at a frequency of 19/20. In \Dsim {} one chimera, formed from \emph{CG11598} and \emph{CG11608}, is at a frequency of 20/20.    All of these genes are lipases or endopeptidases.  These high frequency variants are strong candidates for  selective sweeps.   

Compared to the number of tandem duplications that capture coding sequences, the number of duplications which form chimeric genes indicates that chimeric constructs derived from parental genes in parallel orientation form as a result of 10.4\% of tandem duplications that capture genes in \Dyak {} and 9.5\% of tandem duplications that capture coding sequences in \Dsim.  These numbers are in general agreement with rates of chimeric genes formation estimated from a within-genome study of \Dmel {} of 16.0\% compared to the rate of formation of duplicate genes \citep{RBH}.  

 \subsection*{Association with transposable elements and direct repeats}  

Repetitive sequences are known to facilitate ectopic recombination events that commonly yield tandem duplications \citep{Lim1994}.   In \Dyak, 179  (12.7\%) tandem duplications fall within 1 kb of a TE in at least one sample strain that has a duplication and 52 (3.7\%) fall within 100 bp of a TE (Table \ref{RepContent}).  In \Dsim, 122  (12.5\%) lie within 1 kb of a TE and 53 (5.4\%) fall within 100 bp of a TE (Table \ref{RepContent}).  Additionally, 125 (8.8\%) of duplications in \Dyak {} have 100 bp or more of direct repeated sequence in the 500 bp up and downstream of duplication boundaries and 237 (16.7\%) have 30 bp or more in the reference sequence as identified in a BLASTn comparison of regions flanking divergently-oriented read spans at an E-value $\leq 10^{-5}$ (Table \ref{RepContent}).  In \Dsim, 56 (5.7\%) have 100 bp or more of direct repeated sequence in the 500 bp up and downstream of duplication boundaries in the reference and 150 (14.4\%) have 30 bp or more of repeated sequence (Table \ref{RepContent}).  In total 371 duplications in \Dyak {} and 243 duplications in \Dsim {} either lie within 1 kb of a transposable element in at least one strain or are flanked by 30 bp or more of direct repeated sequence.  Hence, a maximum of 26.2\% of duplications identified in \Dyak {} and 24.9\% of duplications identified in \Dsim {} may have been facilitated by ectopic recombination between large repeats, consistent with previous estimates from single genome studies of 30\% in \Dmel {} but somewhat higher than those in \Dyak {} of 12\% \citep{QZhou2008}.

In \Dyak, 14.4\% of duplications with 100 bp or more of repetitive sequence and 21.1\% of duplications with 30 bp or more are located on the X.   In contrast, 46.4\% of duplications in \Dsim {} with 100 bp or more of direct repeated sequence in the reference and 44.7\% with more than 30 bp of repeated sequence in the \Dsim {} reference lie on the X chromosome.  Based on a resampling of randomly chosen duplications, duplications on the X chromosome are overrepresented among duplications with direct repeats ($P<10^{-7}$) but the same is not true of duplicates with direct repeats in \Dyak {} ($P=0.248$).  A genome wide blastn comparison shows that direct repeats are not overrepresented on the \Dsim {} X chromosome and cannot explain the observed association (Table \ref{SmallRep})  Hence, duplication via ectopic recombination may be exceptionally common on the X chromosome in \Dsim.

\subsection*{Excess of duplications on the \Dsim {} X}

\begin{figure}[h]
\begin{center}

\hspace{-0.25in}
\includegraphics[scale=0.65]{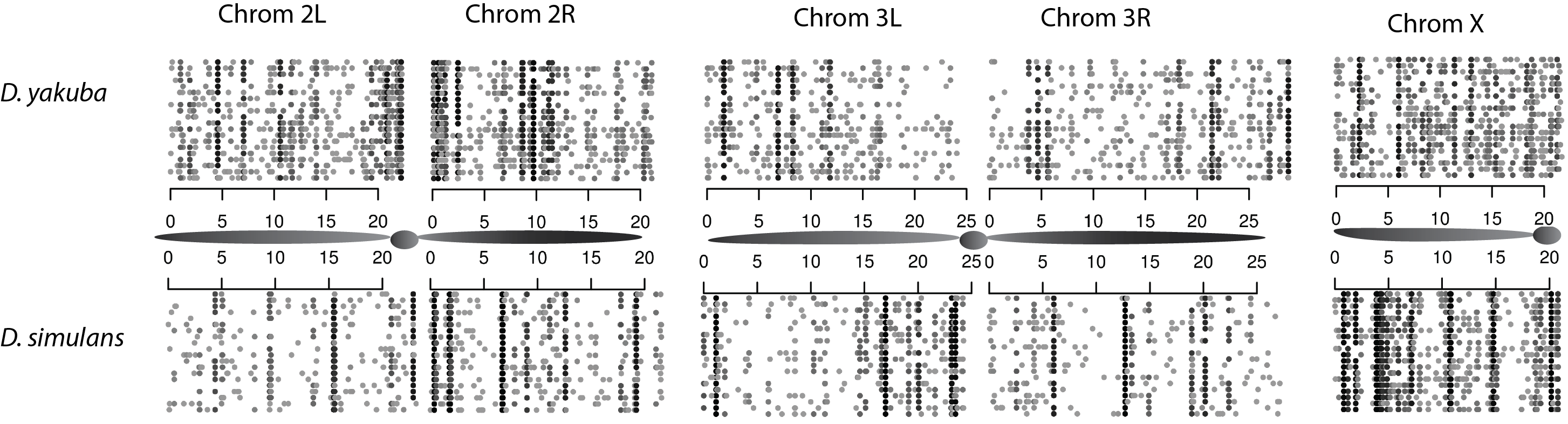}
\end{center}
\caption{Tandem duplications in 20 sample strains of \Dyak.  Regions spanned by divergently-oriented reads are shown with sample strains plotted on different rows, whereas axes list genomic location in Mbp.  Duplications are more common around the centromeres, especially on chromosome 2. Frequencies are shaded in greyscale according to frequency, with high frequency variants shown in solid black.  The \Dsim {} X chromosome appears to have an excess of high frequency variants in comparison to the \Dsim {} autosomes and the \Dyak {} X chromosome.  }

\label{Summary}
\end{figure}

\begin{figure}[h]
\begin{center}
\includegraphics[scale=0.5]{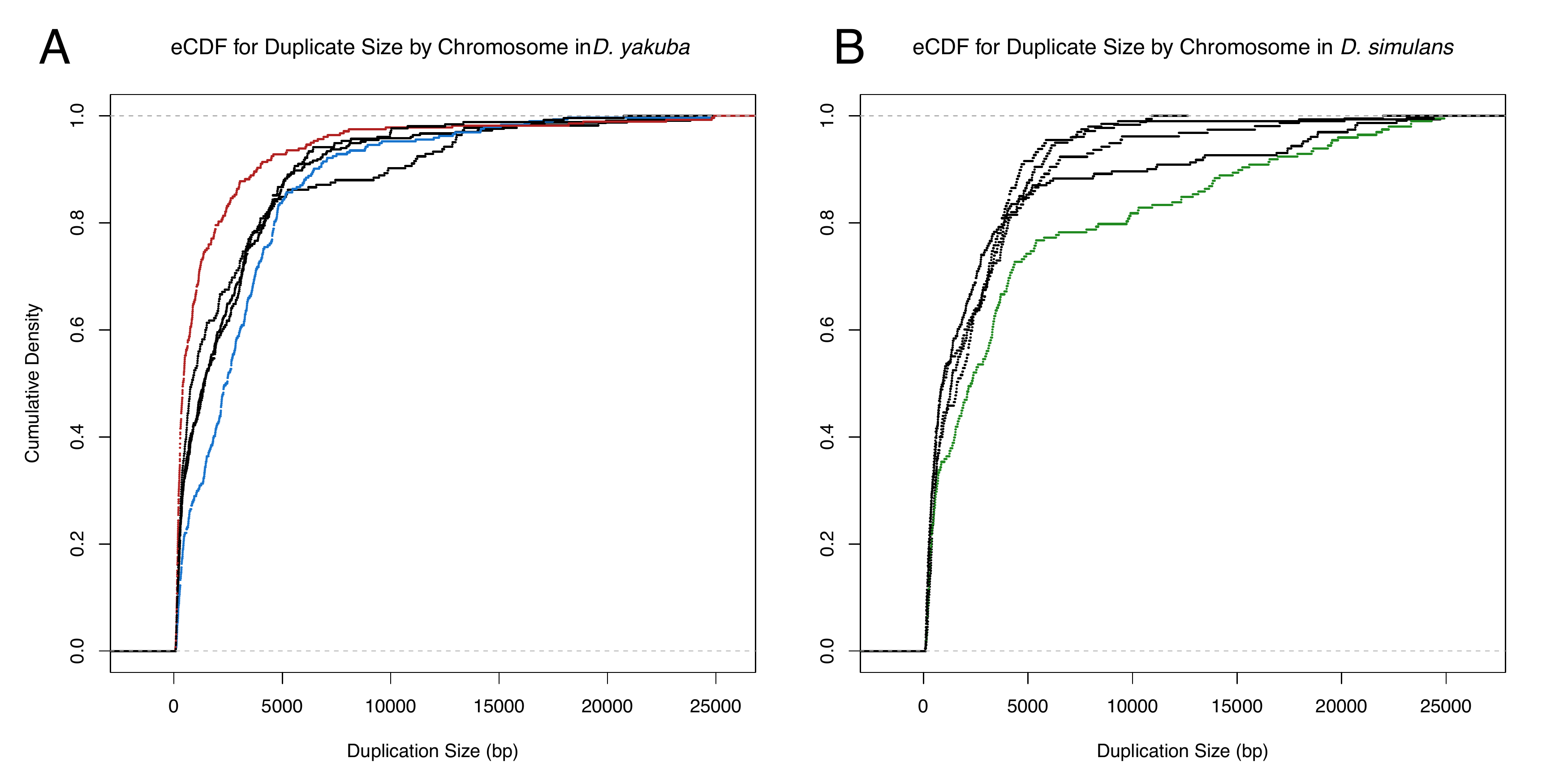}
\end{center}
\caption{Cumulative distribution function for duplication sizes for the X and 4 major autosomal arms in A) \Dyak {} and B) \Dsim.  The X chromosome in \Dyak {} (red) is significantly different from all autosomes ($P< 10^{-3}$) due to a large number of small duplications 500 bp or less.  Chromosome 2R (blue) is also different from Chr2L and 3R ($P< 0.05$).  In \Dsim {} chromosome 3L (green) is significantly different 2L, 3R, and the X.}

\label{SizeDist}
\end{figure}

The distribution of duplication sizes was calculated for each major chromosomal arm in each species.  Average duplicate size is 2,518 bp, in close agreement with that observed in mutation accumulation lines in \Dmel {} \citep{Schrider2013} but somewhat larger than that observed using microarrays in \Dsim {} \citep{Cardoso2011}.  The X chromosome in \Dyak {} displays an overabundance of small duplications in comparison to each of the autosomes in a Tukey's HSD test after correction for multiple testing with ~27\% of duplicates 500 bp or less ($P \leq 6.8\times 10^{-5}$, Figure \ref{Summary}, Figure \ref{SizeDist}A, Table \ref{SizeDistStats}).  Chromosome 2R is also significantly different from the other three major autosomal arms ($P \leq 2.95\times 10^{-3}$, Figure \ref{SizeDist}A, Table \ref{SizeDistStats}). However, in \Dsim {} there is no significant difference between the X and 2R, 2L and 3R, even though the X houses a greater density of duplications (Figure \ref{SizeDist}B). The \Dsim{} chromosome 3L is different from 2L ($P=0.029$, Figure \ref{SizeDist}B, Table \ref{SizeDistStats}). 

\begin{figure}[h]
\begin{center}

\includegraphics{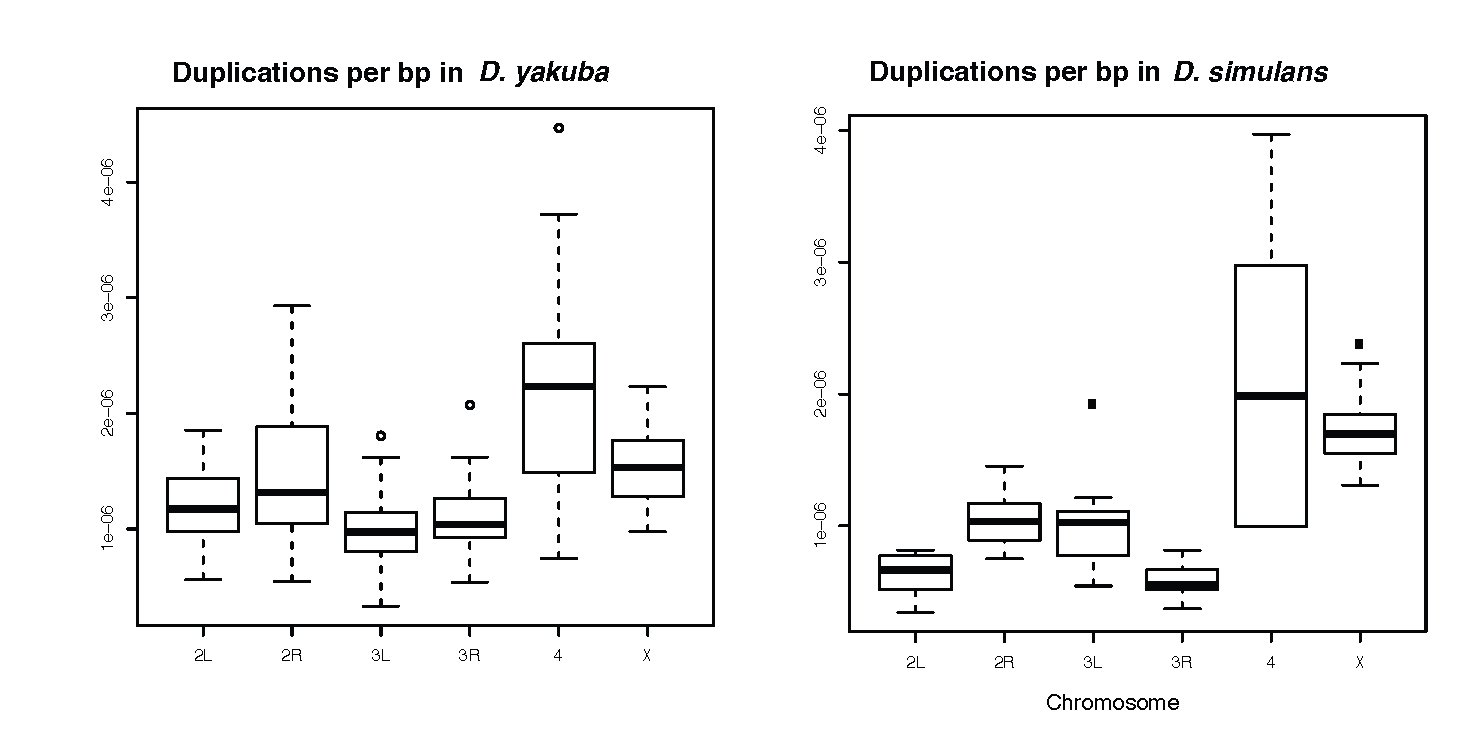}
\end{center}   
\caption{Number of Duplications per bp on the X and major autosomes in \Dyak {} and \Dsim.  The X chromosome in \Dsim {} contains an excess of duplications in comparison with the autosomes.  Chromosome 2R also contains more duplications per bp than chromosome 3R but no other autosomes are significantly different. Chromosome 4, the dot chromosome, has more duplications per mapped bp in both \Dyak {} and \Dsim {} than any other chromosome. }
\label{BoxPlot}
\end{figure}

We observe a significant effect in the number of duplications per mapped base pair by chromosome in both \Dyak {} ($F(5,109)=8.321$, $P=1.09\times10^{6}$) and \Dsim {} ($F(5,113)=36.74$,  $P<2 \times 10^{-16}$) .  In a post-hoc Tukey's HSD test  with correction for multiple testing, the \Dsim {} X chromosome contains more duplications per mapped base pair than any of the autosomes, with 316 duplications ($P \leq 1.0537 \times 10^{-4}$, Table \ref{TukeyHSDPerChrom}, Figure \ref{Summary}, Figure \ref{BoxPlot}).    Chromosome 2R contains an excess of duplicates in comparison to chromosome 3R ($P=0.032$), but all other pairwise comparisons of the four major autosomes are not significant.  Chromosome 4 contains an excess of duplications per mapped base pair in comparison to all other chromosomes in both \Dsim {} and \Dyak.  In \Dyak {} the X is different from 3L ($P=0.039$) but not from any other autosome. Some 11 of the 25  duplicates in \Dsim {} are at a frequency of $\frac{17}{20}$ strains or greater (44\%) on the X (Figure \ref{Summary}).  In comparison, only 2 of the 13 high frequency duplications in \Dyak {} (15.4\%) are located on the X, nor do we see a comparable overabundance of duplications on the \Dyak {} X.  These results point to a clear excess of duplications on the X chromosome in \Dsim {} in comparison to the autosomes, as well as an overabundance of duplications on the fourth  chromosome in both species.

Given the excess of duplications associated with repetitive content on the \Dsim {} X, repetitive elements may be an important factor in forming the observed overabundance of duplications on the \Dsim {} X.  While mutational and selective processes can lead to a bias in the number of duplications that form on different chromosomes, the excess of high frequency variants on the \Dsim {} X at a frequency of 20 out of 20 would suggest that at least some of the overabundance on the \Dsim {} X is due to selective forces or demography resulting in duplicates spreading through the population.    

\section*{Discussion}
We have used paired-end reads to describe tandem duplications in sample strains of \Dyak {} and \Dsim, their sample frequencies, and the genes that they affect.  We use high coverage Illumina genomic sequencing data of 50X or greater to successfully identify tandem duplications among individually sequenced isofemale lines derived from natural populations.  We have filtered tandem duplications to include recently-derived segregating tandem duplicates that are not present in the reference genome of each species for genomic regions that have coverage of three or more read pairs across all sequenced strains.  We show high rates of confirmation using long molecule PacBio sequences with 96.1\% of variants showing evidence of confirmation.  

We identify 1415 tandem duplicates in \Dyak {} and 975 in \Dsim, indicating that there is substantial standing variation segregating in populations that may contribute adaptive evolution and the instance of detrimental phenotypes.  We identify hundreds of chimeric genes and cases where genes recruit formerly non-coding sequence.  We have shown an excess of duplications on the \Dsim {} X chromosome as well as an overabundance of whole gene duplications in \Dsim, suggestive of selection acting on duplications. 

\subsection*{Rapid modification of duplicated alleles}

\begin{figure}[!h]
\begin{center}
\includegraphics[scale=0.25]{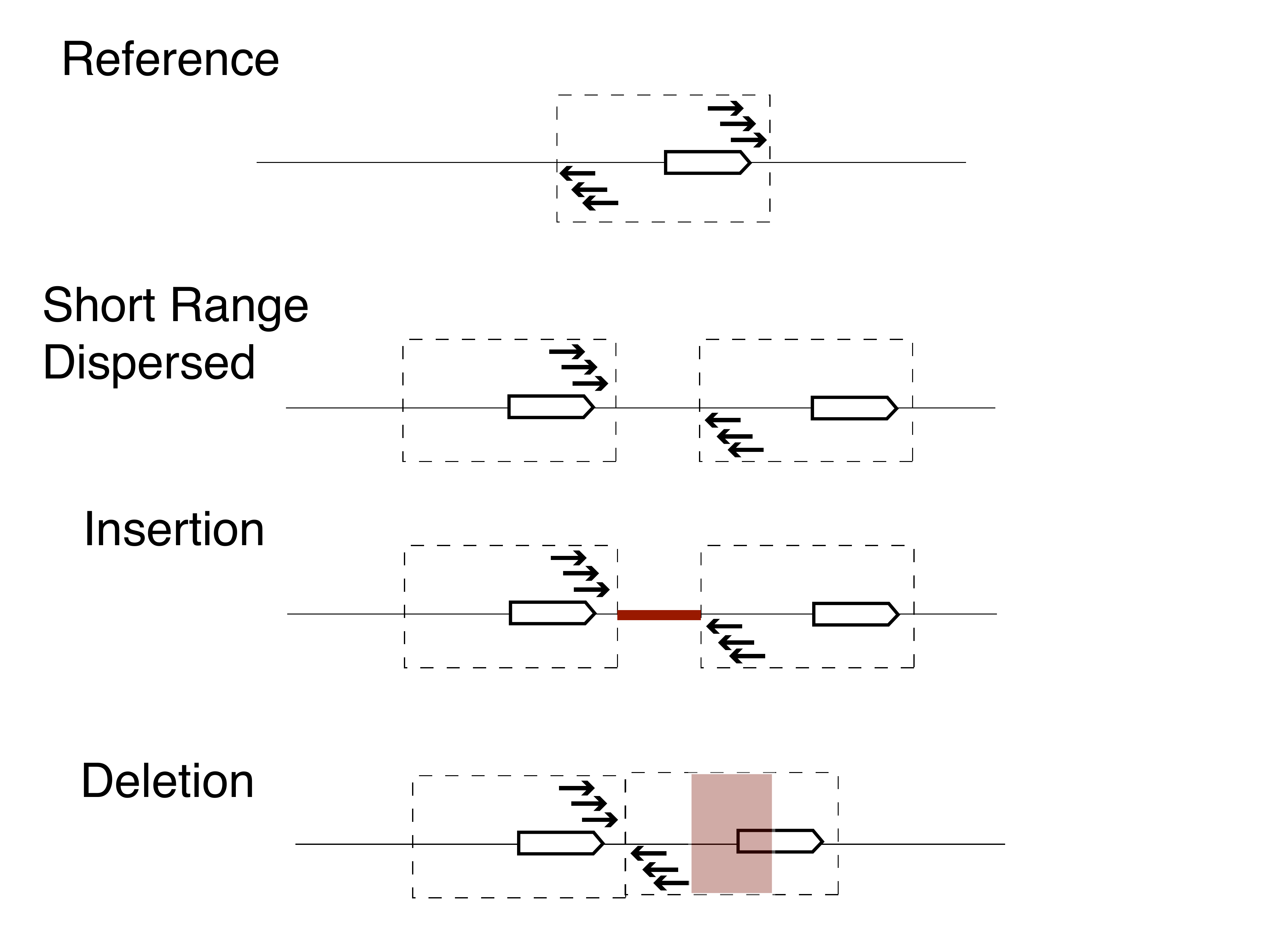}
  \caption{\label{Complex} Complex breakpoints and subsequent modification of tandem duplications.  Short range dispersed duplications, duplication with insertion of novel sequence, and duplication with subsequent deletion will all display the same signals of divergently-oriented reads.  While all of these indicate duplication has occurred, signals solely from short sequence read pairs are unlikely to capture the full complexity of duplication events. }
\end{center}
\end{figure}

Standing variation is expected to play a major role in adaptation and evolutionary change \citep{Barrett2008}.  If the span of standing variation in populations is limited, the dynamics, genomic content, and variability of standing variation in populations is likely to play a defining role in evolutionary outcomes.  The observed span of tandem duplications across strains is limited, with 2.574\% of the assayable genome of the X and 4 major autosomes in \Dyak {} and 1.837\% of the assayable genome of the X and 4 major autosomes in \Dsim.  Yet, the variation that is observed portrays a dynamic picture of gains and losses with evidence that duplications can induce subsequent deletions through large loop mismatch repair, suggesting that regions that are duplicated create genomic instability.  The resulting expansion and contraction of genomic sequences will contribute to greater variability in these limited regions than has been suggested to date, offering wider variation upon which selection can act.  Up to 17\% of duplications in \Dyak {} and 8\% of duplications in \Dsim {} show long spanning reads in one or more strains, indicative of complex changes such as subsequent deletion, insertion of foreign sequence, or incomplete or short-range dispersed duplication (Figure \ref{Complex}).  These results are consistent with complex breakpoints previously observed in \Dmel {} \citep{JCridland}.  Moreover, coverage changes for certain variants are consistent with duplication followed by subsequent deletion in one or both copies (Figure \ref{CoverageFails}).  Hence, the current pool of genetic diversity will in fact be far greater than simple interpretation of divergently-oriented reads or split read mapping might indicate. The majority of such changes have one or more strains with no signs of modification, suggesting that these variants are primarily duplications followed by deletions. 

Secondary deletions of recently duplicated alleles may be exceptionally common, especially in deletion-biased genomes like \Dros.  Deletion of excess unpaired DNA for polymorphic duplicates during large loop mismatch repair, excision of transposable elements, replication slippage, and deletion during non-homologous end joining all offer common mechanisms that are likely to remove portions of duplicated alleles.   Among these mechanisms, the large loop mismatch repair system specifically targets newly added DNA and is likely to be a driving force in the rapid modification of duplicated alleles. In the ideal case, precise excision would simply return the construct to singleton state resulting in a rapid cycle of mutations and reversions. However, when such removal is imprecise, these subsequent deletions are likely to modify duplicated sequences leaving incompletely duplicated segments.  The average distance spanned by these putative deletions is over 2 kb, well above the mean for deletions observed in mutation accumulation lines of \Dmel {} \citep{Schrider2013}, but in agreement with the amount of DNA that can be efficiently removed by large loop mismatch repair \citep{Kearney2001}.  

Duplications have the potential to induce secondary deletions quickly, while variants remain polymorphic, thereby offering mechanisms for rapid and potentially drastic genomic change that can potentially alter gene content, dosage, and regulation.  Variation in populations, while limited in its genomic scope, may offer multiple variant forms at individual duplication sites.  Thus, the substrate that is present for selection and adaptation will be far richer than a simple duplication or single copy, but rather can take on these complex forms of modified variants that remain largely unexplored in terms of their molecular and evolutionary impacts.   Thus, although the observed amount of variation is limited to only a fraction of the genome, the level of variation at these duplicated sites portrays an exceptionally dynamic flux of duplications and deletions at these sites that will result in changes in the content and organization of the genome and therefore is expected to have a strong influence on evolutionary outcomes.  

\Dyak {} displays 1.5 times as many duplications in comparison to \Dsim, as well as a two-fold enrichment in the percentage of variants with signals of deletion in one or more copy and higher population level mutation rates.   The rapid flux of duplication and deletion observed in \Dyak {} has produced a wider array of standing variation, which is expected to have a significant effect on evolutionary trajectories.  \Dyak {} will likely display a greater tendency toward pathogenic phenotypes associated with tandem duplicates \citep{Emerson2008, Hu1992} but also a greater source of standing variation that can be useful in adaptation and the development of novel traits \citep{WolfeReview}.   Estimates of $N_e$ in \Dyak {} are higher than in \Dsim.  We would thus expect greater instances of detrimental duplicates to be higher in \Dsim {} than in \Dyak, but neutral mutations will collect more quickly across populations of \Dyak {} due to high $N_e$.  Hence, we suggest that the overabundance of duplicates in \Dyak {} is not due to drift.  Neither do we observe an excess of high frequency variants in \Dyak {} that might be suggestive of selection, especially with respect to polymorphic variants.   

Based on birth-death models of gene families, \Dsim {} is suggested to have high rates of duplication, whereas \Dyak {} showed only moderate rates of gene family evolution \citep{Hahn2007}.   This may in fact be influenced by the overabundance of whole gene duplicates in \Dsim {} and not a reflection of genome wide mutation rates.  The dichotomy between reference genomes and genome wide polymorphic variants might putatively be driven by selection for whole gene duplicates in \Dsim {} or mutational biases toward whole gene duplications. 

\subsection*{Chimeric genes}
Chimeric genes are a known source of genetic novelty that are more likely to produce regulatory changes, alterations in cellular targeting and membrane bound domains, as well as selective sweeps in comparison to whole gene duplications \citep{RapidEvolution}.  Chimeric genes have been known to produce peptides with novel functions in \Dros {} \citep{MLong1993,Zhang2004, Ranz2003} and in humans \citep{Zhang2009, Oshima2010} and many are associated with adaptive bursts of amino acid substitutions \citep{Jones2005a, Jones2005b}.   We observe large numbers of recently-derived chimeric constructs within populations, with 222 chimeric genes or genes that recruit non-coding sequence in \Dyak {} and 134 in \Dsim, even in a limited sample of 20 strains per species.  

In spite of their known role in adaptation, the majority of copy number variants are thought to be detrimental \citep{Emerson2008, Cardoso2011,Hu1992}. Chimeric genes are associated with human cancers, and the molecular changes associated with chimera formation may contribute to their role as causative factors in human disease.  The molecular changes that are facilitated by chimera formation \citep{RapidEvolution} likely contribute to their detrimental impacts on organisms and their role in disease as well as their potential for adaptation.  We observe large numbers of chimeric genes that are identified as single variants in the population.  Thus, chimeras may play the dual role of key players in adaptation to novel environments and as agents of detrimental phenotypic changes.  The large amounts of standing variation observed may therefore contribute to disease alleles in populations, and proper identification is likely to be important for studies in human health.  

\subsection*{Breakpoint determination}

Many variants have breakpoints that cannot be assembled \emph{de novo} from Illumina sequences (Table \ref{Reconstruct}).  Yet, we observe a 96.1\% confirmation rate using PacBio reads.  These results imply that breakpoints are often repetitive, low-complexity sequence or contain novel insertions and secondary events that are difficult to determine from paired-end read mapping alone or current naive \emph{de novo} assembly methods.  Hence, although paired-end Illumina read mapping is highly accurate, it cannot ascertain breakpoints to single base pair resolution in the majority of cases.  Moreover, requiring breakpoint assembly to identify duplications will produce a strong ascertainment bias against up to 50\% of all variants. This bias is more severe for small variants, even in \Dros, which have compact genomes and few repeats in comparison to plants or vertebrates.  Thus, short high-throughput Illumina reads orientation mapping offers an accurate but incomplete picture of variation present in the population, which can now be clarified with low coverage long read sequencing data. 

Microarrays and coverage are subject to affects of mis-probing, mis-mapping, and large amounts of noise relative to signal (Figure \ref{CoverageFails}).  However, where accurate, arrays may reflect the span of duplicated segments more accurately than divergent reads alone, as they would accurately reflect deletion after duplication.  However, the presence of complex events such as subsequent deletions may, if not properly identified and accounted for, overestimate of the mutation rate of duplications and underestimate their frequency in the population by claiming a modified variant as an independent duplication.  Here, the directional nature and spatial relationships of read pair mapping shows advantages: divergently-oriented reads distinguish duplication whereas long spanning properly oriented reads can indicate a deletion with greater clarity and properly identify subsequent modification.   Identifying putative deletions in duplicated sequences requires a tight distribution of insert sizes during library preparation (Table \ref{InsertSize}) but offers a far more complete picture of variation that is segregating in populations and more accurate estimation of variant frequencies that is well worth the effort.  
\subsection*{The X chromosome}

The \Dsim {} X chromosome appears to be unusual in that it contains an excess of duplications per mapped base pair in comparison to the autosomes, and an overabundance of duplications associated with long repeats.  Within-genome surveys of non-synonymous mutations in the \Dsim {} reference  \citep{Andolfatto2011} and large numbers of high frequency derived variants among non-synonymous sites and UTRs in \Dsim {} \citep{Haddrill2008} indicate widespread selective sweeps acting on the X chromosome.  Similarly, we identify an excess of variants identified at high frequency on the \Dsim {} X, consistent with previous work using microarrays \citep{Cardoso2011}.  In \Dmel, the X chromosome contains greater repetitive content \citep{CridlandTEs}, displays different gene density \citep{DmelRef}, has potentially smaller population sizes \citep{Wright1931, Andolfatto2001}, lower levels of background selection \citep{Charlesworth}, and an excess of genes involved in female-specific expression \citep{Ranz2003} in comparison to the autosomes. Moreover, X chromosomes are subject to selfish genetic elements and often play a role in speciation \citep{Presgraves2008}.  Thus, the X chromosome may be exceptionally subject to widespread selection, and the role of tandem duplicates as responders to selective pressures deserves future exploration.  

Similar patterns of high frequency variation have not been observed in \Dyak, suggesting that evolution proceeds differently across the different species. The \Dyak {} X chromosome has an excess of small duplications, which might potentially indicate selection acting against large duplications on the \Dyak {} X.  Tandem duplications are known to be detrimental, and given the hemizygous state of the X chromosome in males, we would expect purifying selection to act quickly on the X.  The extent to which these patterns observed in \Dyak {} may be driven by selection or demography remains to be seen, and is an important open question deserving of future study. 

\subsection*{Methodological approach}

Originally, high throughput detection of copy number variation relied on microarrays or SNP chips available at the time \citep{Emerson2008, Dopman, Ionita2009, Conrad2010},  and therefore suffers from problems of mis-probing, variable hybridization intensities, and dye effects, producing large amounts of noise relative to signal \citep{Ionita2009}.  More recent studies have focused solely on changes in Illumina or 454 coverage analogous to changes in hybridization intensity \citep{Sudmant2010}.  We have used coverage changes in combination with divergently oriented reads to identify variants using comparisons to a resequenced reference and quantile normalized coverage data to correct for stochastic coverage changes, repetitive content, GC bias and low complexity sequence, resulting in robust variant calls.  Furthermore, the common practice of retaining only variants that are present in multiple samples or at particular genotype frequencies \citep{Conrad2010}, detected by multiple independent methods \citep{Mills2011, Zichner2013, Alkan2009}, or that are larger than several kilobases \citep{Alkan2009, Xie2009, Sudmant2010} can lead to severe ascertainment bias both with respect to the types of variants that are present, the estimation of their prevalence within populations and contribution to disease phenotypes, and the evolutionary impacts of duplicated sequences. 
 
 With more recent improvements in Illumina sequencing technology, we have been able to sequence strains individually to 50X or greater coverage.  The use of paired-end reads in extremely high coverage data provides clear advantages over previous work using roughly  3.6X-25X with subsets sequenced to 30-42X \citep{Sudmant2010, Zichner2013, Mills2011}, or 16X \citep{Alkan2009}.  The high coverage data presented here covers the majority of the assayable genome and we exclude only a few percent of variant calls due to low coverage across strains (Table \ref{Downsample}).   Thus, we are able to identify hundreds of events that have 3 or more supporting divergent read pairs that might be missed at lower coverage (Figure \ref{DIVCov}).  We have additionally filtered sequences to exclude ancestral duplications, similarly to previous work in humans \citep{Mills2011} allowing for identification of derived mutations.   The result is a high confidence dataset for recently-derived tandem duplications in data that effectively surveys the majority of the assayable genome.  This high coverage is essential in ensuring valid results from the sole use of paired-end read orientation.  However, the types of duplications that can be detected is highly dependent on sequencing library insert size.  For breakpoints with large amounts of repetitive sequence, reads separated by 300 bp may not be sufficient to overcome difficulties of read mapping.  Additionally, the minimum duplication size is also limited by insert size.  Capturing these types of constructs using paired-end reads, especially in organisms with large amounts of repetitive content, including nested transposable elements, will require a more diverse range of library insert sizes, a factor that is likely to be important for surveys of larger, more repetitive genomes.  

We are able to identify copy number variants as small as 66 bp using divergently-oriented paired-end reads, even in cases where nucleotide divergence between paralogs or partially repetitive sequence might otherwise complicate their discovery through coverage changes or  split read mapping using short reads.  Divergently-oriented reads are additionally reliable to detect duplications in regions where distributions of coverage are too irregular to allow for automated detection of duplicates.  We observe a high 96.1\% confirmation rate of variants among four sample strains of \Dyak {} using PacBio reads up to 24 kb in length, suggesting that paired-end reads in high coverage genomic sequencing will grossly outperform previous methods which show high false positive and false negative rates.  Moreover, the use of the newly annotated \Dsim {} genome \citep{SimRef} based on a single isofemale line will result in improved accuracy in comparison to previous studies in \Dsim {} \citep{Cardoso2011}.  The general principles of the paired-end approach should be broadly applicable and similar methods already have been used to identify chromosomal inversions in natural populations \citep{Corbett2012}.  PacBio reads can confirm structural variants with high rates of success, even given low coverage and nucleotide high error rates \citep{EichlerPacBio} and we have extended this approach to our genome wide survey.  However, the ambiguity of split read mapping in the face of repetitive elements can still complicate \emph{de novo} duplicate discovery using split read mapping of long reads.  Furthermore, for the present, generating high coverage genomic sequencing equivalent to that of our paired-end Illumina data is not cost-effective.  However, this technology and similar long read approaches are likely to offer advantages in confirming or discovering structural variants such as tandem duplications, as well as \emph{de novo} assembly, that is worth future exploration.   

We are able to identify and confirm a large number of complex gene structures such as chimeric genes, recruitment of adjacent non-coding sequence, potential coding sequence disruption, and potential selective silencing of expression.  These complex mutations are often associated with cancer and other diseases  \citep{Ionita2009, Inaki2012} and are most likely to cause pathogenic outcomes.  Hence the methods described here will be broadly applicable in GWAS and clinical studies as well as in evolutionary genetics of non-model systems where next generation sequencing has so recently made population genomics readily tractable. 

\subsection*{Conclusions}
  Here, we have described the landscape of standing variation for tandem duplications in isofemale lines derived from natural populations of \Dyak {} and \Dsim {} with high accuracy.  The resulting portrait of hundreds to thousands of variants, including large numbers of complex breakpoints, modifying deletions, cases of recruited non-coding sequence, and dozens of chimeric genes per species reveals a rich substrate of segregating variation across populations.   We show that although the span of duplications across the genome is quite limited, duplicates can induce secondary mutations and result in dynamic changes, resulting in greater variation across mutated sites that offers more abundant variation for use in adaptation than has been previously portrayed.  The ways in which this variation influences adaptive evolution and produces molecular changes  will clarify the extent to which mutational profiles define evolutionary outcomes and the ways in which molecular changes associated with tandem duplications serve as causative factors in disease.

\section*{Materials and Methods}
\subsection*{Population Samples}
We surveyed variation in 10 lines of \Dyak {} from Nairobi, Kenya and 10 from Nguti, Cameroon (collected by P. Andolfatto 2002) as well as 10 lines of \Dsim {} from Madagascar (collected by B. Ballard in 2002) and 10 strains from Nairobi, Kenya (collected by P. Andolfatto in 2006).  Flies from these isofemale lines (i.e. descendants from a single wild-caught female) were inbred in the lab for 9-12 generations of sibling mating.  These should provide effectively haploid samples of allelic variation representative of natural populations. 

In addition to the 20 inbred lines derived from wild-caught flies, we also sequenced the reference strains for each species.  For \Dsim, the reference strain is the $w^{501}$ stock (UCSD stock center 14021-0251.011), whose sequence is described in \citep{SimRef}.  For \Dyak, the reference strain is UCSD stock center 14021-0261.01, and the genome sequence is previously described in \citep{TwelveGenomes}.  The majority of the wild-caught strains and the \Dyak {} reference stock were sequenced with three lanes of paired-end sequencing at the UC Irvine Genomics High Throughput Facility (\url{http://dmaf.biochem.uci.edu}).  The sequencing of the \Dsim {} reference strain was described in \citep{SimRef}. The number of lanes and read lengths per lane are summarized in Tables \ref{sfig:laneSummSim} and \ref{sfig:laneSummYak}.

\subsection*{Alignment to reference}

The sequencing reads were aligned to the appropriate reference genome \citep{TwelveGenomes,SimRef} using \texttt{bwa} version 0.5.9 \citep{BWA} with the following parameters, \texttt{bwa aln -l 13  -m 50000000  -R 5000}.  The resulting paired-end mappings were resolved via the ``sampe'' module of \texttt{bwa} (\texttt{bwa sampe -a 5000 -N 5000 -n 500}), and the output was sorted and converted into a ``bam'' file using samtools version 0.1.18 \citep{samtools}.  In the alignment and resolution commands, \texttt{-l} is the hash sized used for seeding alignments, and the \texttt{-R,-a,-N} and \texttt{-n} refer to how many alignments are recorded for reads mapping to multiple locations in the reference.  After the paired-end mapping resolution, the bam files from each lane of sequencing were merged into a single bam file sorted according to position along the reference genome.  A second bam file, sorted by read name, was then created for use as input into our clustering software.  

\subsection*{Clustering abnormal mapping events}

\begin{figure}[h]
\begin{center}

\hspace{-0.25in}
\includegraphics[scale=0.4]{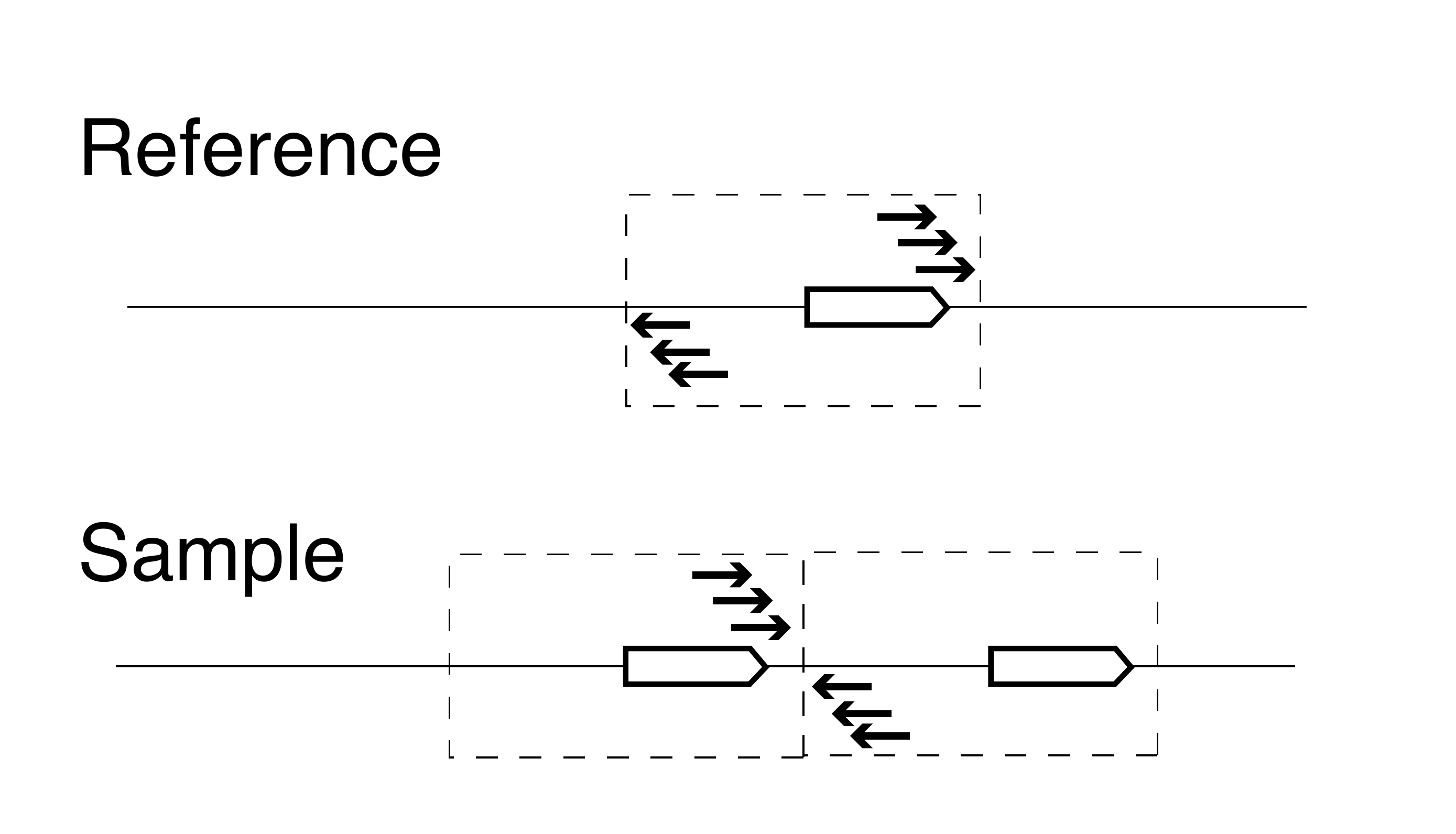}
\end{center}
\caption{A tandem duplication in a sample that was then used to generate paired-end Illumina libraries.  Duplications should be apparent through divergently-oriented read pairs when mapping onto the reference genome.  Tandem duplications require a minimum of three divergently-oriented read pairs.  Duplication span is recorded as the minimum and maximum coordinates spanned by divergent reads.}

\label{DupMapping}
\end{figure}

Tandem duplications should be readily apparent among mapped reads as sequenced read pairs that map in divergent orientations \citep{JCridland, Zichner2013, Mills2011, Tuzun2005}, provided that tandem duplications with respect to a reference genome result in a single novel junction.  Figure \ref{DupMapping} shows a putative genomic sample that contains a tandem duplication of a gene that was used to generate paired-end sequencing reads.  We allowed for up to two mismatches within mapped reads in order to capture divergent read calls in sample strains that have moderate numbers of nucleotide differences. Reads were required to map uniquely, and so if duplication breakpoints contain entirely repetitive sequences with no divergence across copies in the genome, they will not be found.  These limitations will, however, have minimal effects (see Text S1).

 Sets of read pairs in the same strain that are located within the 99.9th quantile of the mapping distance between properly mapped pairs from one another were clustered together into a single duplication.  In practice, this threshold distance is roughly 1 kb.  Tandem duplications were identified as regions where 3 or more divergently-oriented read pairs cluster to the same location in a single strain.  Allowing fewer divergent reads leads to a large number of false positive duplication calls due to cloning and sequencing errors. Further detail is offered in Text S1.   

\subsection*{PacBio alignment and analysis}
FASTQ files of PacBio reads were aligned to the \emph{D. yakuba} reference \citep{TwelveGenomes} using \texttt{blasr} \citep{Chaisson2012}, available from \url{https://github.com/PacificBiosciences/blasr}, with default options and storing the resulting alignments in a bam file.  Alignments from regions within 1 kb of putative tandem duplications called using short read data (divergent read orientation plus an increase in coverage) were extracted from the bam files using \texttt{samtools} 0.1.18 \citep{samtools}.  Reads falling within these regions were then pulled and realigned to the reference using a BLASTn \citep{BLAST} with low-complexity filters turned off (-F F) at an $E$-value cutoff of 0.1 to allow for short alignments given the high error rate of PacBio sequencing.  Alignment using \texttt{blastn} proved important because it revealed cases where the bam files resulting from \texttt{blasr} alignments failed to record secondary hits for a read, especially in cases where alignments are on the order of hundreds of base pairs.  Here, confirmation benefits from long sequences which can anchor reads uniquely to a region, producing greater confidence in split-read alignments.  Variants were considered confirmed if two segments of a single PacBio subread which do not overlap by more than 20\% align to overlapping sections of the reference (Figure \ref{fig:SMclasses}B) or if a single read aligns in split formation with the downstream end of the read aligning to an upstream region of the reference (Figure \ref{fig:SMclasses}A,C-E).  An event was considered not confirmed if there were no long reads showing any of the alignment patterns in Figure \ref{fig:SMclasses} within 1 kb of the variant.  Variants were considered definite false positives if clearly contradicted by at least one read spanning the entire putatively duplicated region as defined by Illumina read mapping and adjacent reference.  Some of these unconfirmed variants may be due to low clone coverage in the region or lack of sufficient read lengths rather than false positives.

\subsection*{HMM and coverage changes}
To detect increases and decreases relative to reference resequencing, we quantile normalized coverage for each strain in R so that coverage displayed an equal median and variance across all strains \citep{Bolstad2003}.  Such normalization renders tests of differing coverage robust in the face of differing sequence depth across samples or across sites and is essential for reliable confirmation of tandem duplicate calls \citep{Bolstad2003}.
 
We developed a Hidden Markov Model to identify statistically significant increases in coverage at duplicated sites.  In the HMM, hidden states are defined by copy number and they act to effect differential emission probabilities for the observed outcomes of coverage depth at duplicated or non duplicated sites.  We modeled differences in coverage for sample strains relative to the reference as the difference in two normal distributed random variables each with a mean and variance corresponding to the observed mean and variance in the reference in the given window.  Detailed methods of the HMM and decoding are provided in Text S1.  

\subsection*{Deletions and complex duplication events}
In order to determine the extent to which secondary deletions, incomplete duplications, or short range dispersed duplicates result in complex structures that are not representative of classic duplications, we identified long-spanning read pairs that lie within tandem duplications .  We identified read-pairs with an estimated template length of $\geq$ 600 bp, corresponding to roughly the 99.9th percentile of fragment lengths in the reference genome (Table \ref{InsertSize}).  Long spanning reads whose end points lie within 200 bp of one another and which are fully contained within the duplication endpoints were clustered together.  Clusters supported by five or more read pairs (corresponding to a putative $P$-value of $10^{-15}$) were recorded as signs of deletion or other complex rearrangement.  

\subsection*{Sample frequency of variants}
Complex mutations that are present in two different strains and have divergent reads spanning regions on both the \fiveP {} and \thrP {} ends that fall within 100 bp of one another in the two strains are considered to be equivalent duplications.  While divergently-oriented reads reliably identify duplications, the true boundaries of the duplication may differ from the divergent read spans by several bases, especially in cases where repetitive or low complexity sequence lies at breakpoints, a complication which is not addressed in recent work \citep{Zichner2013, Mills2011}.  For each duplication the minimum of divergent read starts and the maximum of divergent read stops across all strains was recorded to indicate the span of the duplication.  These breakpoints were assembled and confirmed \emph{in silico} using phrap and lastz (Text S1).  

In order to correct frequencies for the effects of false negatives, we used a combination of coverage and divergently-oriented reads to identify tandem duplicates in additional strains.  For each duplication present based on three or more pairs divergent reads in at least one strain, a different strain which had at least one divergently-oriented read-pair and displayed two-fold or greater coverage increases as determined by the HMM across at least 75\% of the duplication's span was also recorded as having that particular duplication as well.  

We retained tandem duplicates with divergent read pairs where the minimum start and the maximum stop after clustering were less than 25 kb apart.  While there may be some tandem duplications larger than 25 kb, divergent read pairs at a greater distance are substantially less likely to display two-fold coverage changes across the entire span and are therefore more consistent with translocations within chromosomes (Table \ref{ConfirmedCoverage}).  Unannotated duplications in the reference are likely to be biased towards specific sizes, GO classes, or genomic locations and are likely to artificially influence statistical tests.  We removed duplications that were also identified in the reference, and did not include these in downstream analyses of duplication sizes or numbers, gene ontology, or  site frequency spectra.   
  
\subsection*{Polarizing ancestral state}
All tandem duplications identified are polymorphic in populations and are therefore expected to be extremely young. However, tandem duplications may be identified in sample strains relative to the references if they are new mutations in the sample or if they represent ancestral sequence that has returned to single copy in the reference through deletion.  In order to identify duplications that represent the ancestral state, we pulled reference sequence corresponding to the maximum duplicate span and ran a BLASTn comparison of the sequence against the \Dmel, \Dere, and \Dyak {} or \Dsim {} reference genomes at an $E$-value cutoff of $10^{-5}$.  

Ancestral tandem duplications were defined as any segment that has two hits on the same chromosome of the given reference that lie within 200 kb of one another, excluding unlocalized sequence and heterochromatin annotations where assembly and annotations are uncertain.  Ancestral duplications that are shared across species should be separated by moderate numbers of nucleotide differences, and therefore are expected to be correctly assembled across outgroups.  Hits must have at least 85\% nucleotide identity and must span at least 80\% of the contig spanned by divergently-oriented reads in the sample.  Based on these requirements, we removed 8.1\% of duplications in \Dsim {} and 3.3\% in \Dyak, suggesting that the vast majority of duplicates identified are recently derived, as expected.

\subsection*{Transposable Element Annotation and Repetitive Content}

	Transposable elements were initially identified in the \Dsim {} and \Dyak {} reference sequences by identifying all read pairs where one member of the pair aligned uniquely to the reference and the other member of the pair aligned multiply.  This has been previously demonstrated to be an indication of a TE breakpoint \citep{CridlandTEs, Cridland2013}  All locations to which these multiply aligning reads align were recorded and a fasta file of these putative TE locations was generated.  Putative TE sequences were then aligned to the set of annotated TEs in version 5 of the \Dmel {} reference downloaded from flybase (www.flybase.org) using tblastx \citep{BLAST} with the following parameters (-f 999 -F "" -e $10^{-4}$ -m 8).  Regions of the reference sequence that aligned to a \Dmel {} TE with an E-value of $\leq 10^{-9}$ were kept and annotated as TEs.  We then extracted 500 base pairs to either side of the annotated regions of the reference and aligned these regions to the set of \Dmel {} TEs, as above.  This procedure was performed twice to capture the full length of TE sequence in the reference genomes.  

	Once transposable elements were annotated in the reference genomes, TEs were detected in the 20 sample genomes \citep{CridlandTEs} which includes both TE presence and TE absence calls. Briefly, initial TE detection was done by identifying all read pairs where one member of the pair aligned uniquely to the reference and the other member aligned to any known TE sequence were identified.  Unique reads were clustered if they aligned to the same strand of the same chromosome, within a given threshold distance.  This threshold was defined as the 99th quantile of mapping distances observed for uniquely mapping read pairsÊthat are properly paired and lie in the expected orientation on opposite strands.  At least three read pairs from each of the left and right estimates, in the correct orientation and correct strand, were required to indicate a TE.  Phrap (version 1.090518) \citep{phrap} with the following parameters (-forcelevel 10 -minscore 10 -minmatch 10) was used to reassemble the local area around the TE insertion breakpoint. Contigs were classified according to TE family based on alignments in a BLASTn search \citep{BLAST}.

Following the initial TE detection phase we examined each position where a TE was identified, including TEs identified in the appropriate reference, in each other line in that species.  At this stage we were able to both identify TEs which had previously been missed by our pipeline as well as to make absence calls by reconstructing a contig that spans the TE insertion location.  Repetitive content independent of transposable elements was defined using an all-by-all BLASTn comparison \citep{BLAST} of sequence 500 bp upstream and downstream of duplication coordinates in the \Dyak {} and \Dsim {} references at an E-value $\leq 10^{-5}$ with no filter for low complexity or repetitive sequences.  We used a resampling approach to identify overrepresentation of duplications associated with direct repeats.   We  performed 10,000,000 replicates choosing the same number of tandem duplications at random and determining whether an equal or greater number were identified on the X chromosome.  

Tandem repeats which might putatively facilitate ectopic recombination may hinder identification of tandem duplicates if repeat sequences are identical.  We performed an all-by-all blastn of all chromosomes for each of the reference genomes at an E-value of $10^{-5}$ with low complexity filters turned off (-F F).  Ignoring identical self hits, we identified all directly repeated sequences with greater than 99.5\% nucleotide identity which are greater than 300 bp in length and which lie within 25 kb of one another, in accordance with the criteria for identifying duplicates.  

\subsection*{Identifying duplicated coding sequence} 
 Gene duplications were defined as any divergent read calls whose maximum span across all lines overlaps with the annotated CDS coordinates. \Dyak {} CDS annotations were based on flybase release \Dyak {} r.1.3.  Gene annotations for the recent reassembly of the \Dsim {} reference were produced by aligning all \Dmel {} CDS sequences to the \Dsim {} reference in a tblastx.  Percent coverage of the CDS was defined based on the portion of the corresponding genomic sequence from start to stop that was covered by the maximum span of divergent read calls across all strains.     
\subsection*{Gene Ontology}
We used DAVID gene ontology analysis software (http://david.abcc.ncifcrf.gov/, accessed Mar 2013) to determine whether any functional categories were overrepresented at duplicated genes.  Functional data for \Dyak {} and \Dsim {} are not readily available in many cases, and thus we identified functional classes in the \Dmel {} orthologs as classified in Flybase. Gene ontology clustering threshold was set to Low and significance was defined using a cutoff of EASE $\geq1.0.$  The DAVID clustering software uses Fuzzy Heuristic Partitioning to identify genes with related functional terms at all levels of Gene Ontology from cellular processes to known phenotypes.

\subsection*{Differences among chromosomes}
We calculated the size of duplications that span less than 25 kb in each sample strain, excluding duplications identified in the reference, as incorrectly assembled duplicates are likely be biased toward repetitive and low complexity sequence.  Significant differences in duplication sizes were identified using ANOVA and Tukey's  Honestly Significant Difference test on the log normalized distribution of duplication sizes by chromosome.

We calculated the number of duplications that span less than 25 kb on each major chromosomal arm for each sample strain, excluding duplications identified in the reference.  The number of duplications was then normalized by the number of mapped bases in the reference to adjust for different chromosome sizes and coverage.  Differences in the number of duplications per base pair were identified using ANOVA and Tukey's Honestly Significant Difference test.  There was no significant difference in the number of duplications present across lines.

\subsection*{Supporting Files}
Extended methods as well as detailed description of sequence data and confirmation is provided in Text S1.  All data files are available at molpopogen.org/Data.  Aligned bam files were deposited in the National Institutes of Health Short Read Archive under accession numbers SRP040290 and SRP029453.  Sequenced stocks were deposited in the University of California, San Diego (UCSD) stock center with stock numbers 14021-0261.38- 14021-0261.51 and 14021-0251.293- 14021-0251.311.

\section*{Acknowledgements}
We would like to thank Elizabeth G. King, Anthony D. Long, and Alexis S. Harrison, and Trevor Bedford  for helpful discussions, as well as Portola Coffee Labs for a supportive writing environment.  J. J. Emerson gave valuable advice on the use of lastz and the UCSC toolkit.  B. Ballard shared  \Dsim {}  fly stocks collected from Madagascar.  We would also like to thank several anonymous reviewers whose comments substantially improved the manuscript.  This work is supported by a National Institute of General Medical Sciences at the National Institute of Health Ruth Kirschtein National Research Service Award F32-GM099377 to RLR.  Research funds were provided by National Institute of General Medical Sciences at the National Institute of Health grant R01-GM085183 to KRT and R01-GM083228 to PA.  All sequencing and PacBio library preparations were performed at the UC Irvine High Throughput Genomics facility, which is supported by the National Cancer Institute of the National Institutes of Health under Award Number P30CA062203.  The content is solely the responsibility of the authors and does not necessarily represent the official views of the National Institutes of Health.  The funders had no role in study design, data collection and analysis, decision to publish, or preparation of the manuscript.\\


\bibliographystyle{MBE}
\bibliography{TandemDups}

\newpage
\captionsetup[table]{list=no}
\begin{table}
\caption{Duplicated Regions in \Dyak {} and \Dsim}
\begin{center}
\begin{tabular}{lrr}
& \Dyak & \Dsim \\
\hline
Whole gene& 248 & 296 \\
Partial gene & 745  & 462 \\
Intergenic & 745 & 577  \\
\hline
\end{tabular}
\end{center}
\label{GeneDups}
\end{table} 

\newpage

\newpage

\clearpage


\clearpage
\begin{figure}[h]
\begin{center}

\includegraphics[scale=0.70]{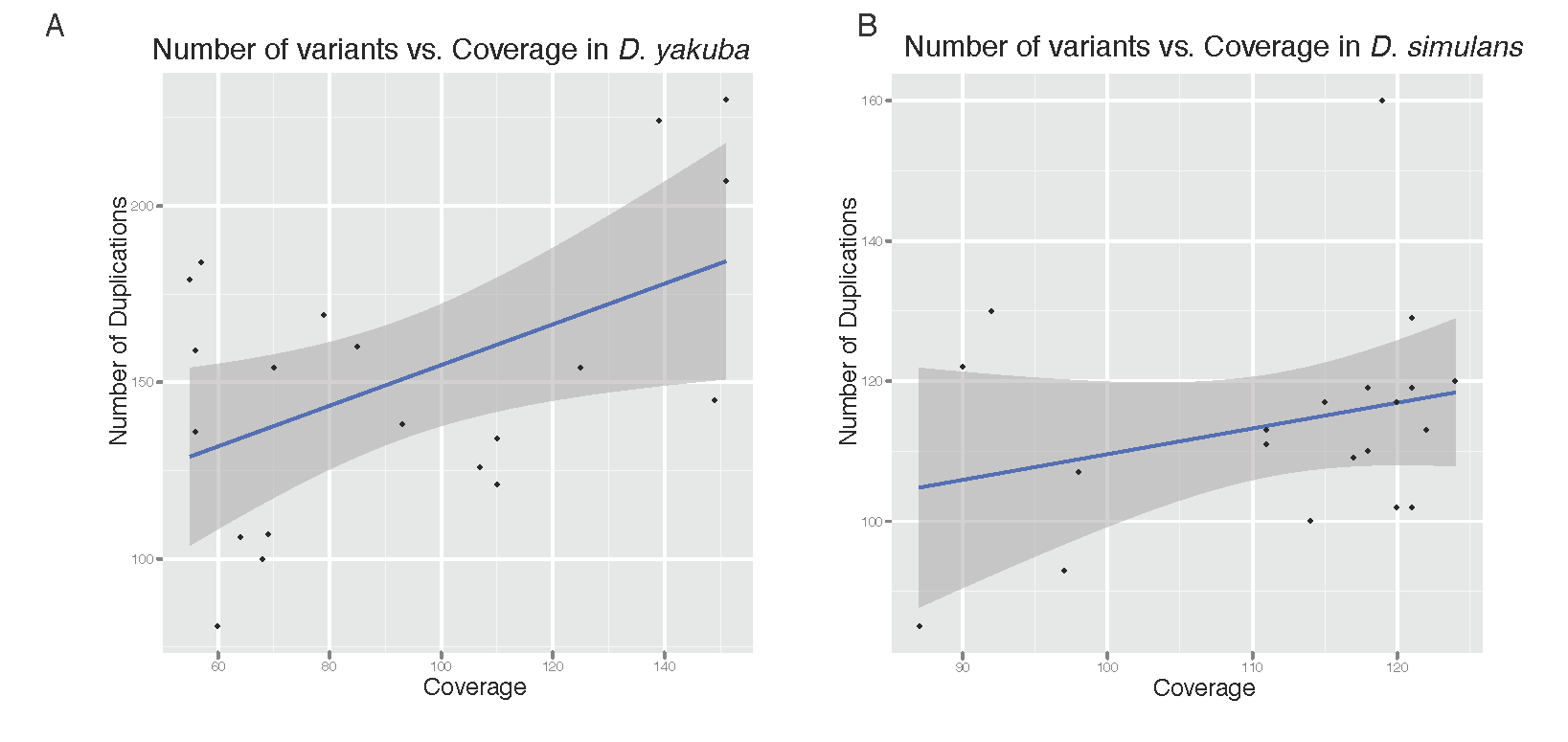}

\caption{\label{CovCorr}Number of variants vs. coverage by line in \Dyak {} (A) and \Dsim {} (B).  Regression line (blue) and 95\% CI (grey) are shown. Correlation between coverage and number of duplications is low (\Dyak {} adjusted $R^2=0.21$, \Dsim {} adjusted $R^2=0.03$). }
\end{center}

\end{figure}

\clearpage

\clearpage

\clearpage

\clearpage

\clearpage
\clearpage

\newpage

\clearpage
\clearpage
\renewcommand{\thefigure}{S\arabic{figure}}
\renewcommand{\thetable}{S\arabic{table}}
\setcounter{figure}{0}
\newpage


\clearpage
 
\chapter*{Supporting Information}
\renewcommand{\thefigure}{S\arabic{figure}}
\renewcommand{\thetable}{S\arabic{table}}
\setcounter{figure}{0}
\setcounter{table}{0}
\setcounter{page}{1}

%
\subsection*{Illumina sequencing and mapping}
	
The genomic DNA (gDNAs) used for the library preparations were extracted with the Gentra Puregene Cell Kit including the RNase  A Solution (Qiagen) or the reagents from the Puregene Accessories (Qiagen, as indicated below). The extraction and purification procedures were modified from the protocol for \Dmel (Qiagen Supplementary Protocol) and combined with column purification (Zymo) as described below, to give better and more constant yields with higher throughput. 

About 15 frozen (-80\textcelsius) female flies of each fly line were homogenized into small pieces in cold 300$\mu$L Cell Lysis Solution (158906, Qiagen), and then incubated for 15min at room temperature. The RNAs were digested by 4.5ul RNase A Solution (158922, Qiagen) and incubated for 15min at 37\textcelsius {}. The contaminated proteins were precipitated by mixing with 100$\mu$L Protein Precipitation Solution (158910, Qiagen) and centrifuged 4min at 16,000g. The extracted gDNAs in the supernatant were column purified with the DNA Clean and Concentrator Ð25 Kit (D4013, Zymo) followed its protocol, using 5 volumes of Binding Buffer. The quality of these purified gDNAs were checked with 1\% agarose gel, showed intact size of about 10 kb without degraded lower bands. 

About 3$\mu$g purified gDNAs in 100$\mu$L EB (Elution Buffer, 10mM Tris, pH8.5, Invitrogen) were sheared with Covaris at 300 bp setting and purified with Purelink PCR Purification (HC) Kit (K3100-01, Invitrogen), followed their protocols respectively. The distributions of these fragments of about 300-1000 bp were confirmed with 2\% agarose gel.

The purified fragments were end-repaired for 30min at 25\textcelsius {} with Quick Blunting Kit (E0542L, NEB), A-tailed for 30min at 37\textcelsius {} with Klenow Fragments (M02112L, NEB) and dATP (R0141, Fermentas), ligated to adaptors (see below) for 10min at 25\textcelsius {} with Quick Ligation Kit (E0542L, NEB), size selected between 350 bp to 450 bp on 2\% agarose gel Seakem LE (5000, Lonza) and extracted with ZymoClean Gel DNA Recover Kit (D4001, Zymo), and PCR amplified 15 cycles by Phusion High-Fidelity DNA Polymerase (M0530S, NEB) with dNTP (N0447L, NEB) and Primers (see below), followed these reagentÕs protocols respectively. The DNA Clean and Concentrator-5 Kits (D4013, Zymo) were used to purify the product after each reaction.

The libraries showed a band between 400 bp to 500 bp on 2\% agarose gel. All concentrations through the preparation were measured with Qubit Fluorometer (Invitrogen). Final libraries were diluted to 10nM before sequencing, based on the reads from Qubit and the sizes from agarose gel.

Adapters and PCR Primers for Paired-End Library:
All adapters and primers were HPLC purified and ordered from Integrated DNA Technologies (IDT):\\

\footnotesize \noindent PE-Ad1, 5'-/5Phos/GATCGGAAGAGCGGTTCAGCAGGAATGCCGAG \\
PE-Ad2, 5'-ACACTCTTTCCCTACACGACGCTCTTCCGATC*T \\
PE-PCR, (1)  5'-AATGATACGGCGACCACCGAGATCTACACTCTTTCCCTACACGACGCTCTTCCGATC*T \\
PE-PCR, (2)  5'-CAAGCAGAAGACGGCATACGAGATCGGTCTCGGCATTCCTGCTGAACCGCTCTTCCGATC*T \\

\normalsize
\subsection*{HMM Parameters}
Emission probabilities were defined using the probability density function of each respective normal distribution as below:
\\

\noindent $E_{deletion}=p(SampleCoverage- ReferenceCoverage)  \sim N(- \mu_{Ref}, 2 \sigma_{Ref}^2 )$ \\
$E_{singleton} =p(SampleCoverage- ReferenceCoverage )  \sim N(0, 2 \sigma_{Ref}^2) $ \\
$E_{duplicate}=p(SampleCoverage- ReferenceCoverage  ) \sim  N(\mu_{Ref}, 2 \sigma_{Ref}^2) $  \\
$E_{triplicate}=p(SampleCoverage- ReferenceCoverage  ) \sim N(2 \mu_{Ref}, 2 \sigma_{Ref}^2) $ \\ 

Where $\mu_{Ref}$ and $\sigma_{Ref}$ are the mean and standard deviation, respectively, for quantile normalized coverage in the reference strain for the window being evaluated.  In each iteration, the most likely path was predicted using the Viterbi algorithm \citep{Viterbi}, which identifies the most probable path for an HMM.

Initial state probabilities were set according to $\pi_0$ and initial transition probabilities were set according to $T_0$, where row and column indices ranging from 0 to 3 are indicative of copy number.  Initial probabilities are set such that the singleton state is initially most likely and states are initially most likely to remain constant during transitions.

$\pi_0 = \begin{bmatrix}
0.07 & 0.79 & 0.07 & 0.07 \\
\end{bmatrix}$
\\

$T_0 = \begin{bmatrix}
0.79 & 0.07 &  0.07 & 0.07 \\
0.07 & 0.79 & 0.07 & 0.07 \\
0.07 & 0.07 & 0.79 &  0.07 \\
0.07 & 0.07 &  0.07 & 0.79 \\
\end{bmatrix}$

\vspace{0.25in}
HMM parameters of transition probabilities (T) and state likelhoods ($\pi$) were optimized using Expectation Maximization based on the Viterbi predictions for each individual window until a steady state was reached.  Emission probability distributions were not updated as data in a given window will often be insufficient for one or more underlying states.   Coverage can change substantially from one region of the genome to another, and so state calls were estimated for 4 kb windows.  For larger windows, the variance in reference coverage across the window can render the emission probabilities less sensitive, and smaller windows can result in insufficient data to obtain adequate likelihoods. 

Transition probabilities serve to ground HMM output such that low transition probabilities lower the likelihood of state changes, resulting in smoother calls.  However in extreme cases low estimated transition probabilities can have an extreme chilling effect such that they will overwhelm the emission probabilities.  The minimum transition probability for any given state change was therefore set at $p=0.07$.  Minimum transition probabilities below a threshold of 0.05 can have a massive effect on the number of duplicated sites, but varying minimum transition probabilities between 0.05 and 0.15 have only a minor effect.  Similarly the minimum state probability was set to 0.00025 for all states and minimum emission probability was set to $10^{-10}$ to avoid zero probabilities.  

Coverage can be highly stochastic and will depend on the amount of divergence between sample and reference, GC content, sequence complexity, and uniformity of error rates across sites.  Requiring a more stringent mapping quality results in substantial coverage changes for many sites (Table \ref{tab:rawCovSumm}).  For even moderately divergent paralogs, higher mapping quality thresholds can greatly diminish our ability to observe two-fold increases in coverage.   

\begin{figure}[!h]
  \begin{center}
    \includegraphics{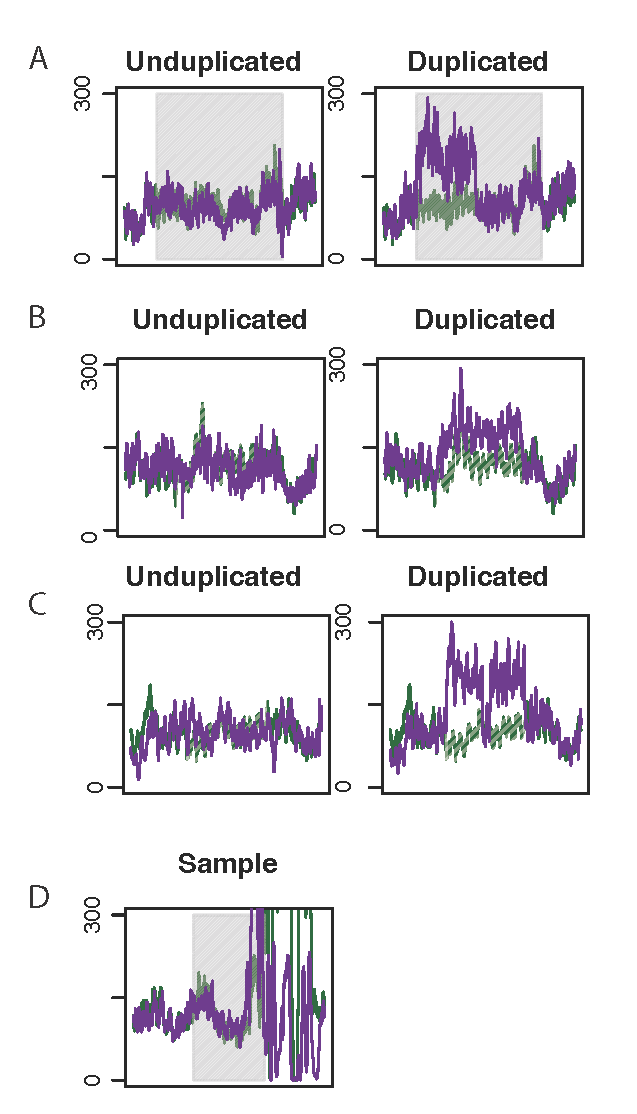}
  \caption{\label{CoverageFails} Cases where automated detection of duplicates using coverage becomes problematic.  A. A duplication has occurred but coverage does not increase for the full variant length.  Possible explanations include subsequent deletion, duplication combined with gene conversion in an adjacent region, or TE movement combined with duplication.  B.  Coverage increases, but there remains substantial overlap between sample and reference coverage.  Modest coverage increases such as these could putatively be hemizygous duplicates, high variance in coverage, or other factors.  C.  A region is duplicated with subsequent deletion in the interior of one copy.   D.  Reference and sample variance are typical in one region but variance is magnified in an adjacent region.  Automated detection for the typical region may be problematic due to an inability to properly estimate HMM parameters due to the abnormally high variance in the same window. }
\end{center}
\end{figure}

In some cases, highly divergent paralogs may not display increases in coverage at all, and in other cases high variance in reference strain coverage may make automated detection of elevated coverage in samples extremely difficult.  Many duplicated variants also show clear elevated coverage for a large portion of the region spanned by divergently-oriented reads, with drops in other regions.  Whether these represent cases where reads fail to map due to divergence or whether they are subsequent deletion at duplicated sites through replication slippage or through the large-loop mismatch repair system remains ambiguous.  Examples of these problematic regions are included in Figure \ref{CoverageFails}.  Additionally, duplicated regions that are substantially smaller than the 325 bp Illumina library insert size may not be identified easily using divergent read calls, but should be readily apparent in coverage changes.  Hence, there may be some disparity between the HMM output and divergent read calls, especially for very small or highly divergent duplications.

Some previous CNV calling schemes have not taken advantage of HMM models, but rather have simply relied on regions of high coverage relative to the genome average \citep{Alkan2009}.  These schemes cannot correct for local variation in coverage due to base composition or potential for mismapping, nor do they correct for natural variation in coverage across genomic regions.  Moreover, such models risk a high false positive rate among genomes with low levels of genomic duplication and high false negative rates for genomes that display rampant duplication.  We observe substantial variation in coverage across sites within the reference, which is naturally accounted for in the HMM emission probabilities without the need for additional models or coverage corrections. 

Previous work, especially analysis of microarrays, has clustered probes with two-fold coverage increases as indicative of a single duplication provided that they fell within a specific threshold distance \citep{Emerson2008, Dopman, Conrad2010}. Given the sparseness of microarray probes, such methods are prone to misidentify multiple duplications in a larger window as single duplications that span large sections of the genome.  However, given the precision provided by Illumina resequencing, we are able to identify increased coverage at individual sites, which in combination with divergently-oriented reads should allow us to identify duplications with greater precision and increased ability to identify multiple small duplications that lie adjacent to one another.

 \subsection*{Sample coverage} 
Illumina reads from the \Dsim {} reference stock consist of a single lane of paired-end data with a 104 bp read length, resulting in a median coverage of 55X (considering only the major chromosome arms X, 2L, 2R, 3L, 3R, and 4) when aligned to the assembly described in \citep{SimRef}.  The remaining 41 samples were sequenced with two to three lanes of paired-end sequencing, with read lengths being a mix of primarily 54 bp and 76 bp (Tables \ref{sfig:laneSummSim} and \ref{sfig:laneSummYak}).  Illumina reads cover 116 Mbp of the \Dsim {} reference and 99 Mbp of the \Dyak {} reference (Table \ref{BasesCovered})  The total sequencing thus consists of 123 lanes of Illumina data. Table \ref{tab:rawCovSumm} shows summary statistics of the sequencing at various mapping quality \citep{BWA} cutoffs.  With no quality filters, median coverage of the major arms ranges from 55x to 151x, covering $> 98\%$ of the genome with 95\% of sites with non-zero coverage having a coverage $\geq 7$ (Table \ref{tab:rawCovSumm}).  The variation in coverage from sample to sample, and the overall higher coverage in \Dsim {} compared to \Dyak {} reflects the order in which the samples were sequenced.   As the sequencing effort progressed, the throughput for Illumina runs increased, while the number of lanes and read length per lane were held constant, resulting in higher coverage.
 
\begin{figure}[!h]
\begin{center}
  \includegraphics[scale=0.5]{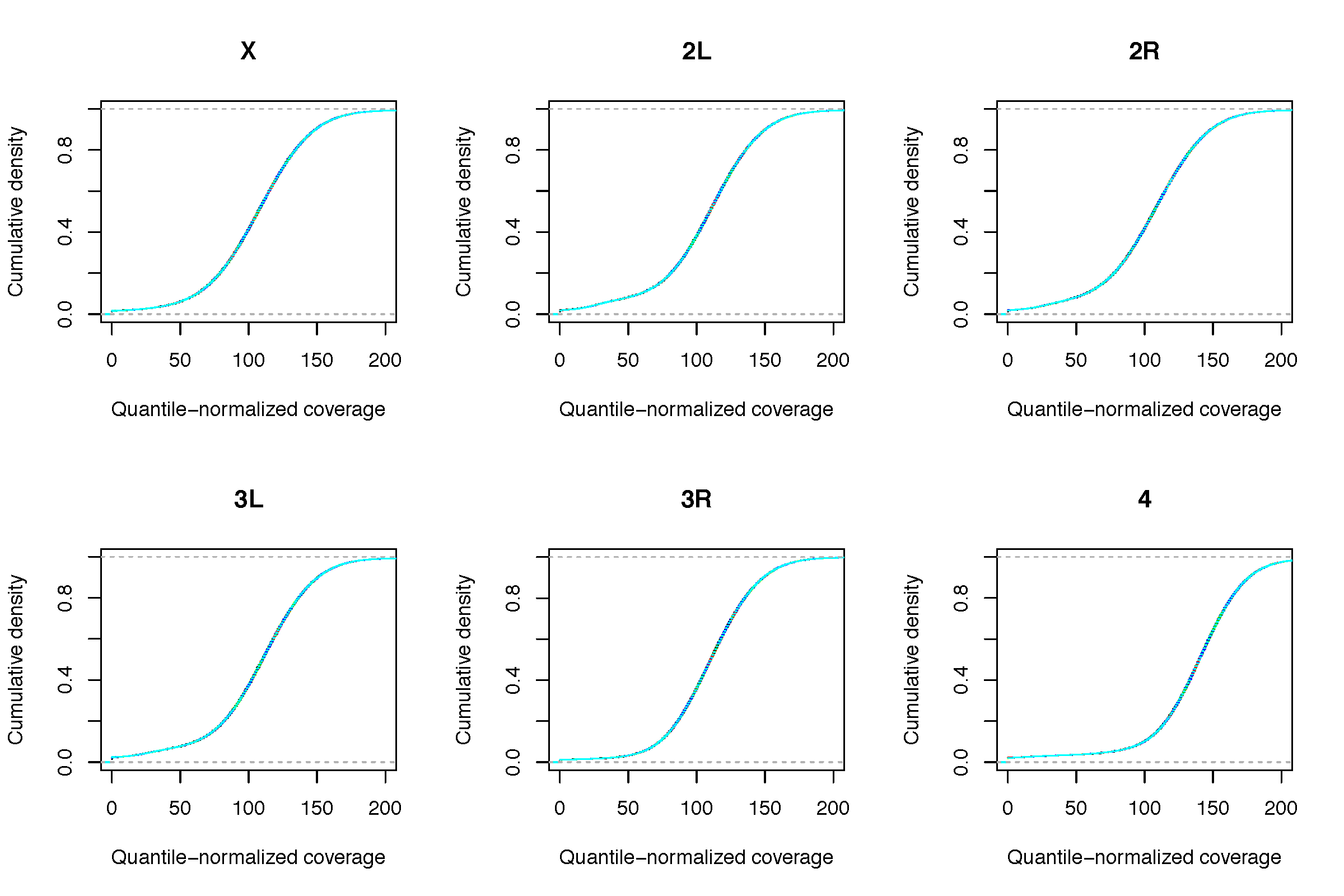}
  \end{center}
  \caption{\label{sfig:DsimNcov} Cumulative distribution plots of the result of quantile-normalization of raw sequence coverage for the \textit{D. simulans} samples.  All 21 samples are plotted, each with an arbitrary color. Compared to the variable median raw coverage values shown in Table \ref{tab:rawCovSumm}, the quantile normalization procedure results in the same distribution of coverage for all samples.}
\end{figure}

\begin{figure}[!h]
\begin{center}
  \includegraphics[scale=0.5]{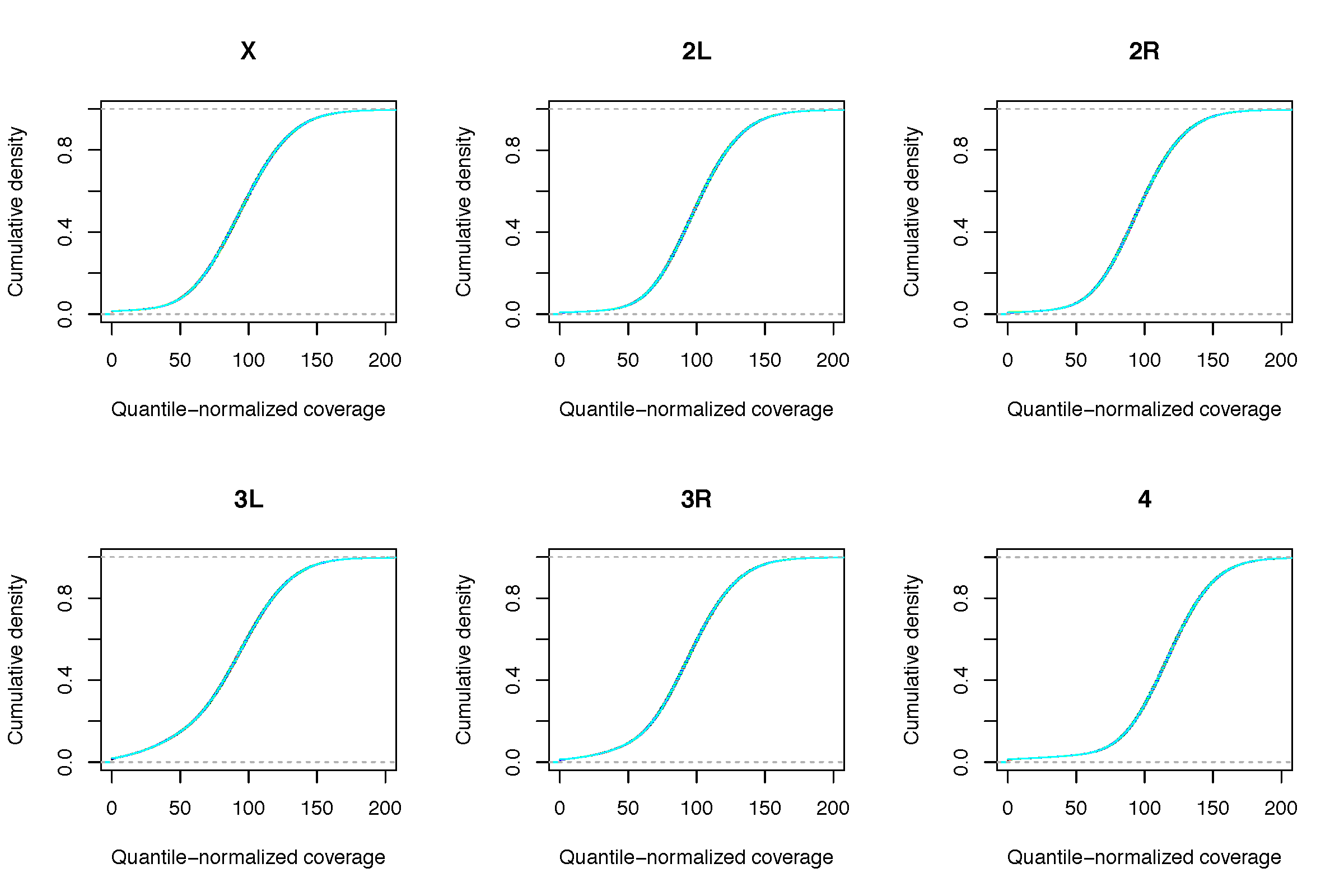}
  \end{center}
  \caption{\label{sfig:DyakNcov} Cumulative distribution plots of the result of quantile-normalization of raw sequence coverage for the \textit{D. yakuba} samples.  All 21 samples are plotted, each with an arbitrary color. Compared to the variable median raw coverage values shown in Table \ref{tab:rawCovSumm}, the quantile normalization procedure results in the same distribution of coverage for all samples.}
\end{figure}

\begin{figure}[h]
\begin{center}

\includegraphics[scale=0.8]{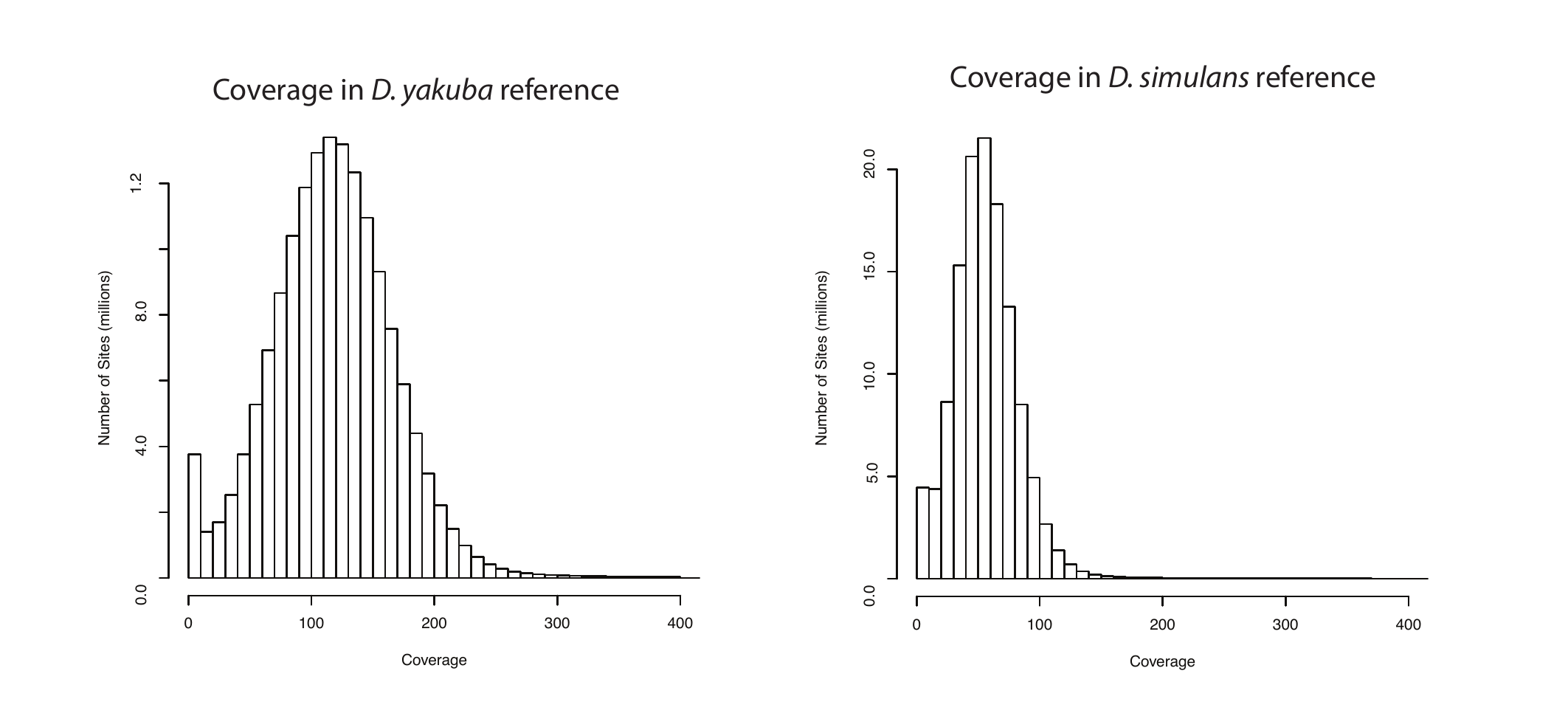}
\end{center}
\caption{Histogram of raw coverage in the \Dyak {} and \Dsim {} reference genomes.  Sites with coverage higher than 400X not shown.  Median coverage in the \Dyak {} reference was 115X whereas the \Dsim {} reference was sequenced to 55X. }

\label{RefCovHist}
\end{figure}

Median raw coverage for the \Dyak {} reference genome was 115X with a standard deviation of 404.25 and a range from 0 to 281347, whereas median raw coverage in the \Dsim {} reference was 55x with a standard deviation of 109.44 and a range from 0 to 8073 (Figure \ref{RefCovHist}).  While some regions have abnormally high coverage that inflates the standard deviation, the majority of the genome is sequenced to between 0 and 400X in both species (Figure \ref{RefCovHist}).  Excluding sites with raw coverage greater than 400X yields a median of 114X and a standard deviation of 54.99 in the \Dyak {} references and a median of 55 and a standard deviation of 28.80 in the \Dsim {} reference.  Distributions of raw coverage were highly similar for all strains (Figure \ref{sfig:DsimNcov}, \ref{sfig:DyakNcov}).
 
Increasing quality filters results in modest changes in median coverage, but has a noticeable effect on the number of sites with a coverage of zero ($f_{0,X}$) and on the first quantile of coverage for sites with coverage greater than zero ($Q_{1,X}$, Table \ref{tab:rawCovSumm}).  These data would suggest that a large number of sites with single digit raw coverage in the reference may be the product of sequencing errors and mismapping. The affect of imposing additional map quality filters is stronger in sample strains than in the reference (Table  \ref{tab:rawCovSumm}) and can impede mappings in the face of even a small number of mismatches.  Hence, for all analyses of coverage changes in sample strains relative to reference, we used coverage with no additional quality filters, a factor that is essential in detecting increased coverage for even modestly diverged regions.  Targeted insert size for Illumina libraries was 325 bp, median fragment size by line ranges from 270 bp-532 bp (Table \ref{InsertSize}).  
\section*{\textit{In silico} confirmation of rearrangements}
\subsection*{\textit{De novo} assembly of breakpoints}
A putative tandem duplication results in the formation of a single novel sequence junction \cite[Figure \ref{BreakReads}]{JCridland}.  Reads mapping to this novel junction will fail to map to an existing reference genome \cite{CridlandTEs}.  We mined the both the divergently-oriented reads and mapped/unmapped read pairs from the putative breakpoint region (Figure \ref{readsused}A) using \texttt{samtools} version 0.1.18 \cite{samtools} and fed them into \texttt{phrap} version 1.090518 (\url{http://www.phrap.org}) for assembly.  We used the following parameters for phrap: \texttt{-vector\textunderscore{}bound 0 -forcelevel 10 -minscore 10 -minmatch 10 -new\textunderscore{}ace -bypasslevel 1 -maxgap 45 -revise\textunderscore{}greedy -force\textunderscore{}high}.

\subsection*{Homology searches using \texttt{lastz}}
The contigs obtained from \texttt{phrap} were then subjected to a homology search against the reference genome using \texttt{lastz} version 1.02.00 (\url{http://www.bx.psu.edu/~rsharris/lastz/})  \cite{lastz}, and the resulting alignments chained together using \texttt{axtChain} and parsed with \texttt{chainNet} \cite{KentTools}.  These latter two tools combine the alignments from \texttt{lastz} to find the highest-scoring colinear alignment of a query to a target, and then separate alignments into separate files by target, respectively.  A single contig chaining within Xbp of the regions identified by divergently-oriented reads as flanking the breakpoints was considered to have confirmed the existence of the event.  In practice, we let X equal 0, 50, or 100 base pairs.

\newpage

\subsection*{CNV calling}

\begin{figure}[!h]
  \begin{center}
 \includegraphics[scale=0.75]{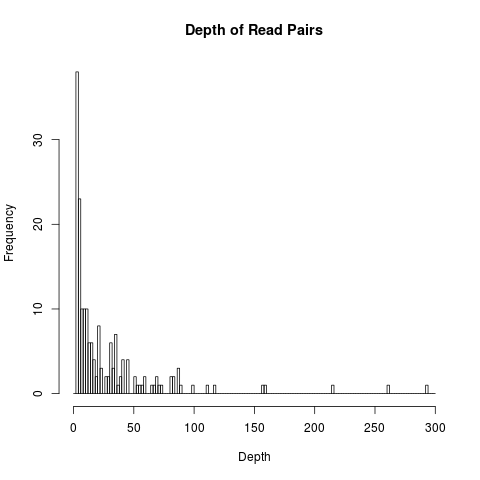}
  \caption{\label{DIVCov} Histogram of read pair depth indicating a tandem duplication for sample strain CY20A.  Number of read pairs is highly skewed and ranges from 3-300 read pairs supporting each event. Mean depth is 22.3, and median of 11 read pairs.}
\end{center}
\end{figure}

We identify 38 divergent read calls consistent with tandem duplications in the \Dyak {} reference strain, 20 of which are specific to the reference strain. Meanwhile the \Dsim {} reference strain has 92 divergent calls in the reference strain, 29 of which are specific to the reference.  These unannotated duplicates in the reference sequence were excluded from downstream analyses on the grounds that they are likely to be biased with respect to gene content, genomic location, base composition, and population frequency. Excluding unannotated duplicates in the reference and putative ancestral duplications, \Dyak {} has an average of 148.6 duplications per strain with a standard deviation of 41.7, while in \Dsim {} we find 113 duplications per strain with a standard deviation of only 15.6 (Table \ref{ByStrain}) and a weak correlation between coverage and number of variants per strain (Figure \ref{CovCorr}).  Hence the variance in number of duplications per strain is 7.5x greater in \Dyak, in addition to strains harboring on average greater numbers of tandem duplications.  Corresponding numbers with reference duplicates and ancestral duplications included are found in Table \ref{ByStrainWithRefs}. The distribution of read pair depth indicating events for a representative sample strain, CY20A, is in Figure \ref{DIVCov}.

We used a Hidden Markov Model that compares coverage in genomic resequencing of the reference to observed coverage in reference strains to identify regions with elevated coverage consistent with duplication of sequences spanned by divergently-oriented reads.  Coverage was quantile normalized prior to analysis so that each strain displays equal mean and variance, rendering tests of differential coverage robust in the face of differing sequence depth across samples or across sites  \citep{Bolstad2003}.  

In some cases, a duplication may in fact exist, but due to lower sequencing coverage in one strain, or due to stochastic effects of sampling, a duplication may not be identified via paired-end reads.  Such false negatives may be more common for variants that are significantly smaller than the library insert size where the likelihood of finding divergently-oriented reads is far lower.  To correct the frequency distribution for such false negatives, we used increased coverage to identify duplications in additional strains.  For duplications defined using 3 or more pairs of divergently-oriented reads that also showed at least one divergently-oriented read pair as well as two-fold or greater coverage increases in additional strains, frequencies were corrected to include a duplication in that strain.    

For \Dyak, frequency correction based on increased coverage resulted in additional calls for 10 duplications across a total of 17 calls in individual strains for duplications larger than 325 bp, yielding a final total of 1033 duplications across  2323 calls in individual strains greater than the targeted library insert size.   These estimates suggest that among duplications that can be identified using paired-end reads, the false negative rate is 0.732\% ($\frac{17}{2323}$) for duplications larger than the library insert size.  For duplications smaller than the targeted library insert size we corrected the frequency estimates for 53 duplications resulting in a total of 91 duplication calls in additional strains.  Compared to the total of 691 calls in individual strains across 382 duplications, the false negative rate for duplications smaller than the targeted library insert size is 13.2\% ($\frac{91}{691}$) for an overall false negative rate of 3.6\% ($ \frac{108}{3014}$).  In \Dsim, corrections result in 14 additional duplication calls across 11 duplicated sites for duplicates smaller than 325 bp and 4 additional calls across 2 sites for duplciates larger than 325 bp.  The resulting false negative rates are 4.3\% ($\frac{14}{329}$) for small duplications and 0.20\% ($\frac{4}{1964}$) for duplications larger than the library insert size for an overall false negative rate of 0.78\%($\frac{18}{2293}$).  Duplicate size and coverage with divergently-oriented reads are not correlated in individual samples ($R=-0.048$, $P=0.55$) suggesting that methods are unbiased with respect to duplicate size.  

We have excluded duplications larger than 25 kb as these are substantially less likely to display increased coverage across the span of divergently-oriented reads.  While there may be some duplications larger than 25 kb, such divergently-oriented reads may be caused by within-chromosome translocations or TE movement as well as tandem duplications, and thus the genetic constructs which produce these abnormal mappings are uncertain.  We, however, observe no duplications greater than 25 kb which have continuously elevated coverage across the span of divergently-oriented reads, consistent with size limits observed in previous surveys in \Dmel {} \citep{Dopman}.

\subsection*{Confirmation with long sequencing reads}
We generated PacBio genomic DNA libraries for four sample strains to roughly 8X coverage.  A total of 10ug Qiagen column purified genomic DNA was sheared using a Covaris g-tube according to PacBio.  Protocol low-input 10 kb preparation and sequencing (MagBead Station).  The Covaris protocol for g-tube was followed (Beckman Allegra 25R centrifuge).  AMPure magnetic beads were purchased directly from Beckman and manually washed according to the protocol in Pacific Biosciences Template Preparation and Sequencing Guide.  DNA template prep kit 2.0 (3 kb-10 kb) was used for library construction.  After construction of the 10 kb SMRTbell template, the concentration was measured by Qubit and the library size analysis performed using an Agilent Bioanalyzer.  The Pacific Biosciences calculator (version 1.3.3) determined the amount of sequencing primer to anneal and the amount of polymerase to bind (DNAPolymerase binding kit 2.0). The calculator also recommended sample concentrations for binding the polymerase loaded SMRTbell templates to MagBeads.  

The PacBio RS remote version 1.3.3 set up the sequencing reaction by identifying sample wells, sequencing protocol, number of SMRT cells and length of movie.  The sequencing protocol was MagBead standard seq v1. Reagents from DNA sequencing kit 2.0 were used for the sequencing protocol.  120 minute movies were taken for each SMRTcell.  The SMRT cells were version 3.  DNA control complex 2.0 (3 kb-10 kb) was the internal control.  Five SMRT cells were analyzed for the 10 kb preparation of \Dyak{} reference genome.  The movie which records the light pulses during nucleotide incorporation is delivered in real time to the primary analysis pipeline which is housed completely in the Blade Center.  Proprietary algorithms translate each pulse into bases with a set of quality metrics.  The data is then available for secondary analysis.  SMRT Portal version v1.4.0 build 118282 with RS\_only filter protocol generated the FASTQ files for the sequences and the circular consensus reads (CCS).  The default filters were removal of reads $<$50 bases and less than 0.75 accuracy.

The single-molecule sequencing resulted in an average genomic coverage of 8X for samples CY17C, CY21B3 and NY66, NY73.  The entire genome is spanned by only on the order of $10^7$ reads (Table \ref{ReadLengths}), offering low clone-coverage and providing sparse opportunities for confirmation.  The detailed results from long-read alignments are shown in Tables \ref{tab:confirmations}.  The single unconfirmed event in line NY66 has nonzero coverage throughout the duplicated region, but coverage drops to zero immediately 3' of the 3' locus containing divergent short reads and there is evidence of large deletions with respect to the reference, but no single read suggesting a tandem duplication.   Thus, only three events were not confirmed (one event per sample), suggesting that our protocol for identifying structural variants using short read data has a 96.1\% confirmation rate, and a low false positive rate.  For two of these unconfirmed rearrangements we do not have sufficient data to confirm or refute the existence of a rearrangement. Thus, the true false positive rate is likely to be less than the 3.9\% implied by these numbers.  Thus, single-molecule sequencing suggests that our protocol for detecting tandem duplications via short reads is highly accurate.

\subsection*{Short split read mapping}
For comparison of performance (see Results), we ran Pindel version 0.2.5a1 \citep{Ye2009} with command line options \verb!--max_range_index 4 --RP --report_inversions --report_duplications! \\ \verb!--report_long_insertions --report_breakpoints --report_close_mapped_reads! \verb!--min_inversion_size 500 --min_num_matched_bases 20 --additional_mismatch 1! \verb!--min_perfect_match_around_BP 3 --sequencing_error_rate 0.05! \verb!--maximum_allowed_mismatch_rate 0.02 --anchor_quality 20! \verb!--balance_cutoff 100 --window_size 0.01 --minimum_support_for_event 3! \verb!--sensitivity 0.99!  to identify duplications on each major chromosome arm independently.   We required that duplication breakpoints be spanned by at least one read on each strand with a total coverage depth of at least three reads, keeping only calls which map to regions with coverage across all strains.  These requirements are somewhat more lenient than the paired read mapping above in that they do not remove putative PCR duplicates, due to a smaller span in which split reads may reasonably be expected to map.  Overlap between Pindel and paired-end reads is low, with 11.7\% of duplicates (179 of 1415) in \Dyak, as defined by paired-end reads matching duplicates whose breakpoints lie within 100 bp of breakpoints defined by split read mapping of Illumina sequence data (Table \ref{PindelConf}).   These 179 variants capture 102 genes or gene fragments, including 21 whole or nearly whole duplicates ($\leq$ 90\% of CDS span).  Only 5 out of 179 (2.7\%) are flanked by 30 bp or longer direct repeats in the reference, and none are flanked by 100 bp or longer direct repeats.  The site frequency spectra for these confirmed duplicates is also significantly different from that of all duplicates defined by paired-end read mapping ($W = 136270$, $P=0.0369$).   

Yet, we observe a high confirmation rate for duplicates defined by paired-end read orientation, meaning that the false negative rate for Pindel is extremely high.  Thus, it would seem that paired-end reads mapping grossly outperforms Pindel for duplications greater than 50 bp when coverage is high. After clustering across strains, requiring that breakpoints in different strains fall within 100 bp of one another, consistent with criteria used for paired-end read orientation, we find 1620 duplications that are not represented among paired-end reads (Table \ref{PindelConf}), with an average length of 155 bp, in stark contrast to duplicates defined using paired-end reads, with a maximum span identified solely by Pindel of 5766 bp.  Thus, Pindel is likely to outperform paired-end read mapping for extremely small duplications but will perform poorly for duplications larger than the library insert size.  

\begin{figure}[h]
\begin{center}

\includegraphics[scale=0.50]{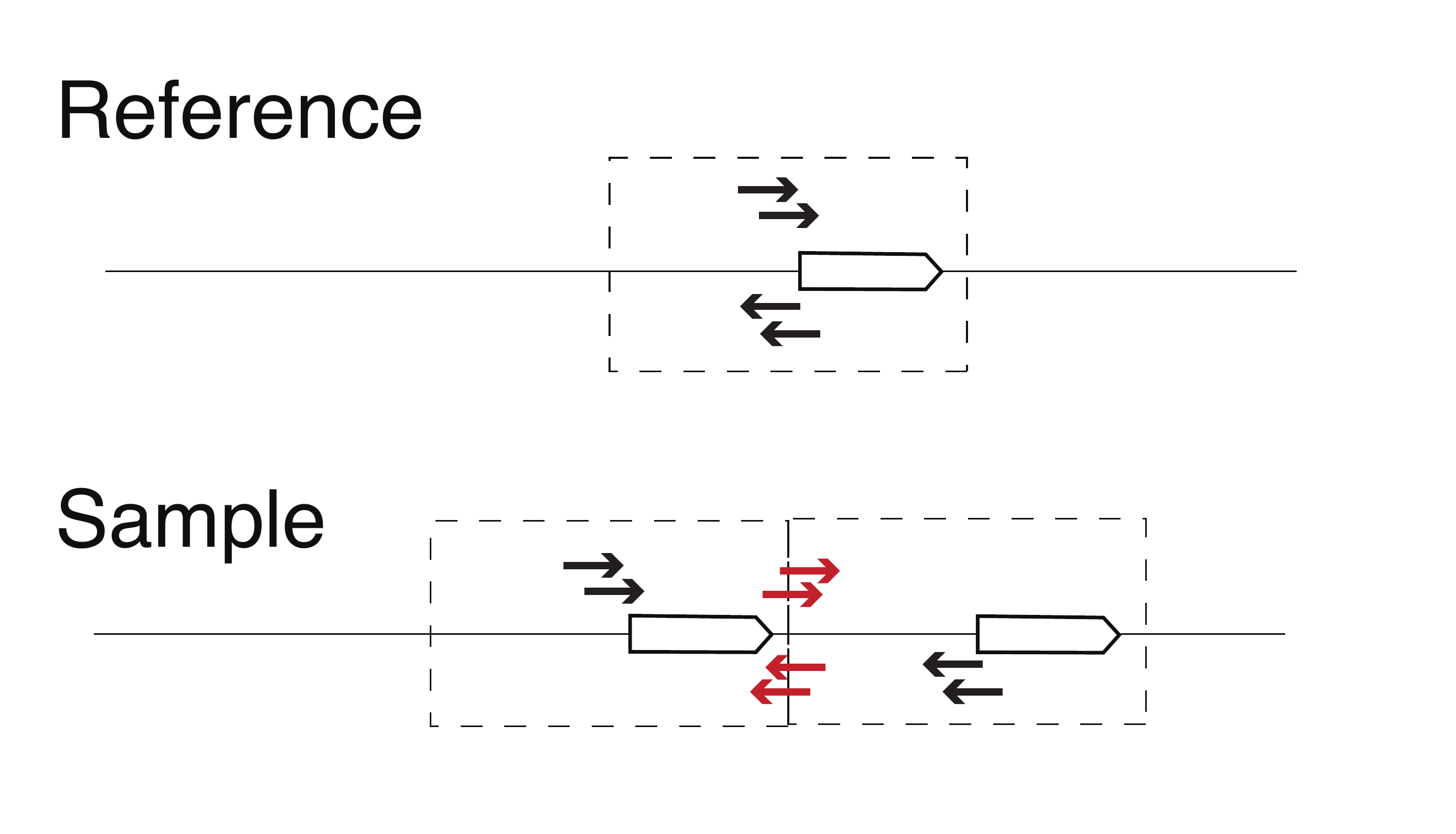}
\end{center}   
\caption{\label{BreakReads} Unique Reads at duplication breakpoints.  A duplication in the sample strain (below) is used to generate paired-end sequencing libraries.  Read pairs interior to duplication boundaries map uniquely to the genome. However read pairs which overlap with duplication boundaries (red) contain unique sequence which is not present in the reference and remain unmapped.  Breakpoint sequence can be assembled \emph{de novo} from unmapped partners of mapped reads.  }
\end{figure}

\begin{figure}[!h]
  \centering
  \includegraphics[scale=0.4]{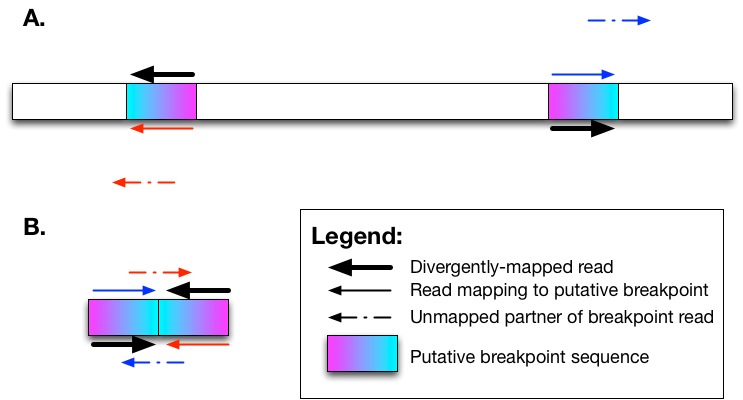}
\caption{\label{readsused}Cartoon of \textit{de novo} assembly procedure.  (a) We mined the alignment archives (bam files) for both the divergently-mapping reads (thick black arrows) and the mapped/unmapped pairs whose mapped reads align to the putative breakpoint regions.  (b) Once assembled using \texttt{phrap}, a single contig with reads spanning the novel junction of the breakpoint confirms the detection of a tandem duplication with a simple breakpoint sequence.}
\end{figure}

\begin{figure}[h]

\includegraphics[scale=0.80]{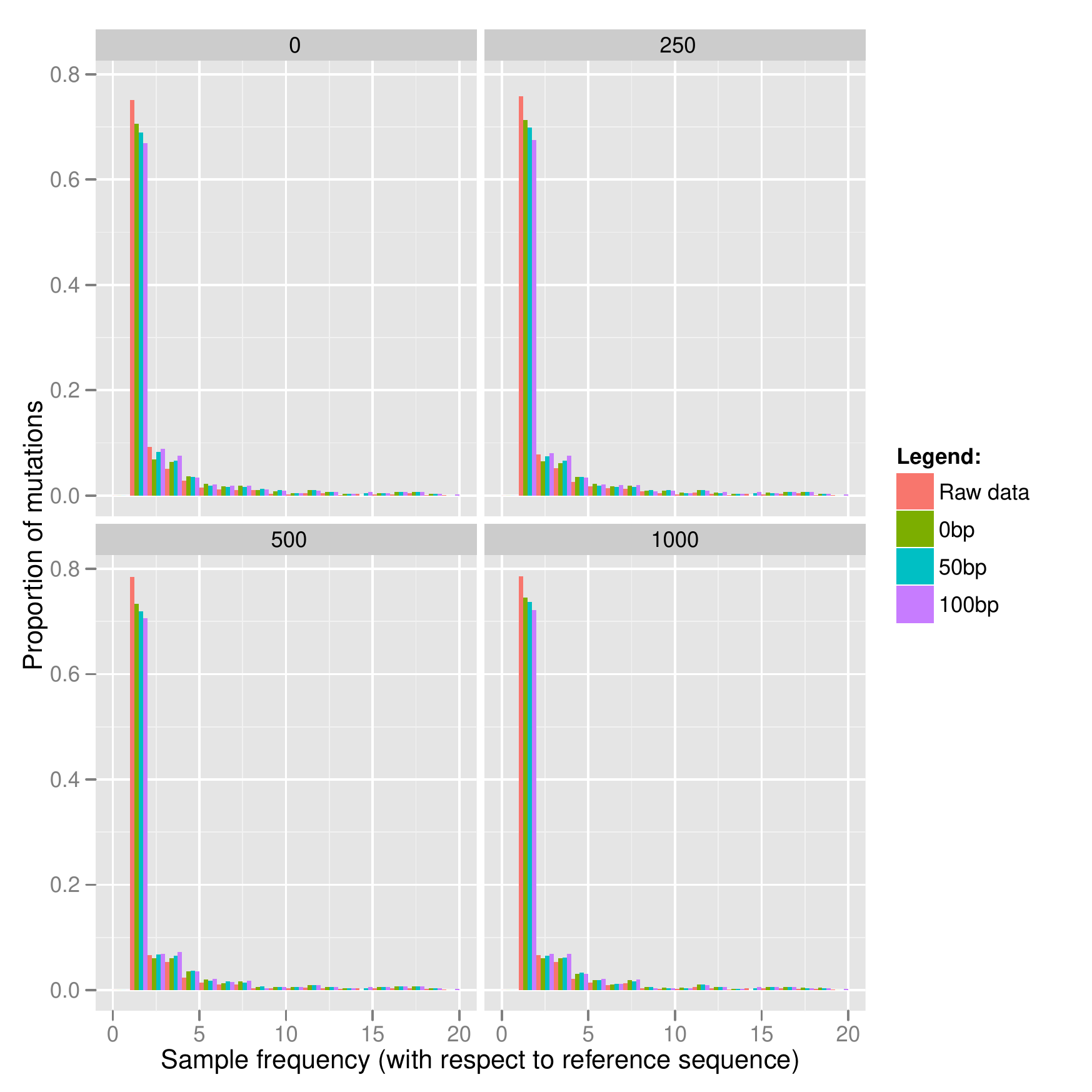}


\caption{\label{ConfirmationResults1}SFS for confirmed breakpoints in \Dyak}

\end{figure}

\begin{figure}[h]

\includegraphics[scale=0.80]{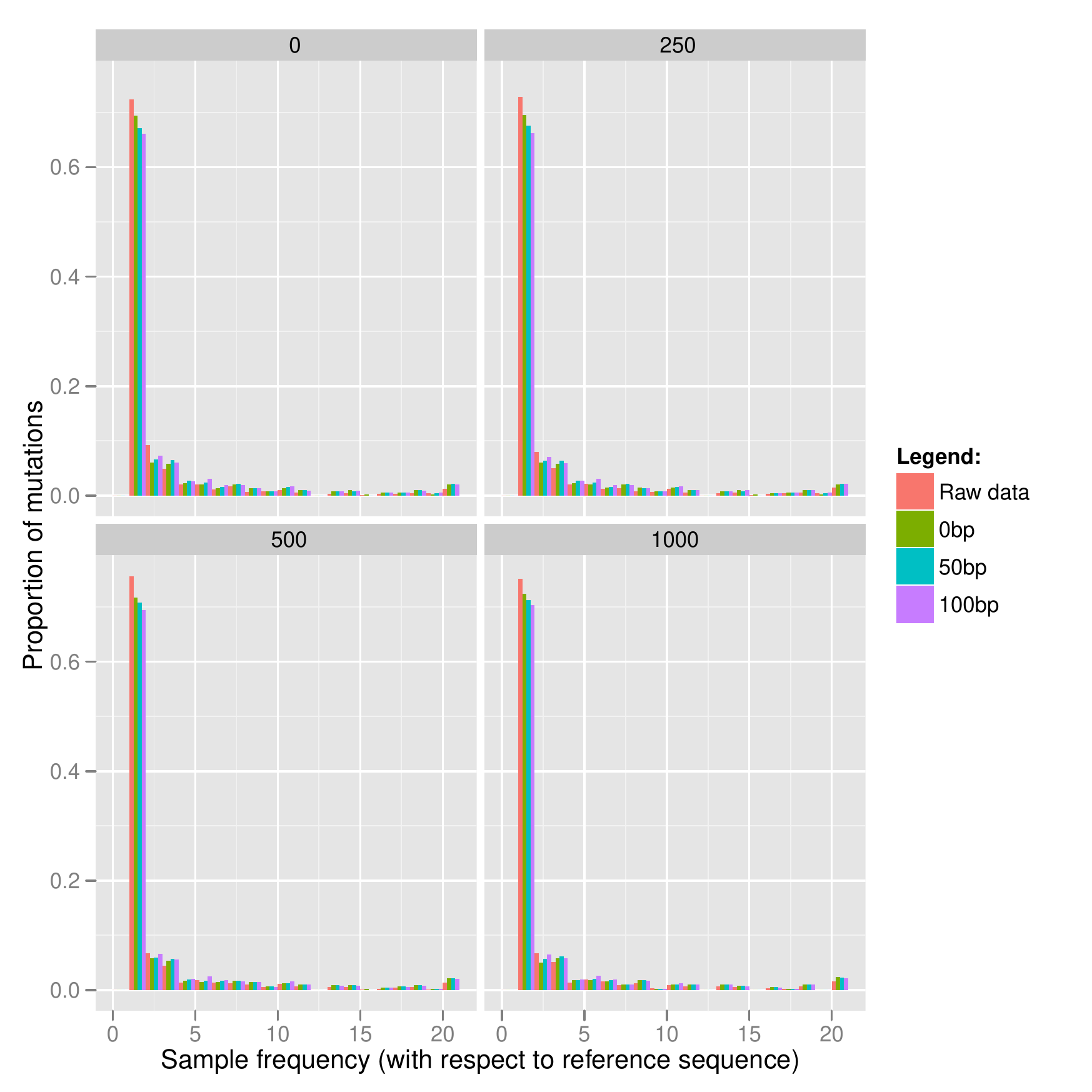}

\caption{\label{ConfirmationResults2}SFS for confirmed breakpoints in \Dsim}

\end{figure}

Only 11.7\% of duplicates identified with paired end reads  are identified through split read mapping (Table \ref{PindelConf}) \citep{Ye2009}, which is expected with short Illumina reads (Table \ref{sfig:laneSummYak}-\ref{sfig:laneSummSim}), especially in cases where variants are flanked by repetitive sequence (Text S1).  Yet, variants identified through paired end read mapping have a 96.1\% confirmation rate with long molecule sequencing.  Thus, split read mapping with short sequences has a high false negative rate, and was therefore not used for identification of tandem duplicates or following analyses.  In a further attempt to establish precise breakpoints from Illumina sequencing data, we attempted to assemble and confirm \emph{in silico} with short Illumina reads (Text S1, Figure \ref{BreakReads}-\ref{readsused}).  We have reconstructed 49.9\% of breakpoints in \Dyak {} and 58.1\% of breakpoints in \Dsim {} that are larger than the 325 bp targeted insert size.  Assembly rates increase for larger duplications (Table \ref{Reconstruct}) with 60.3\% of breakpoints in \Dyak {} and 71.5\% of breakpoints in \Dsim {} over 1 kb that can be assembled.  There are no apparent differences in breakpoint assembly between the X and the autosomes  (Table \ref{Reconstruct}) and no significant difference in the frequency spectrum of variants with assembled breakpoints (Figure \ref{ConfirmationResults1}-\ref{ConfirmationResults2}).  However, requiring breakpoint reconstruction would eliminate a large amount of the observed variation and likely be biased against repetitive sequences as well as against small variants.  

\subsection*{Genome wide surveys}
Some regions of the genome cannot be surveyed due to technical limitations.  We are unable to identify duplicates flanked by repeats in the reference genome which have zero divergence and are beyond the size limits of our Illumina sequencing library insert size.  We identify 121 direct repeats in the \Dyak {} reference with 99.5\% identity to one another which are 300 bp or larger and lie within 25 kb of one another, spanning a total of 678,707 bp (0.57\%) of the \Dyak {} reference (Table \ref{DirRep}).  In \Dsim {} we identify only 5 such direct repeats, covering 84,055 bp of sequence (0.09\%) of the reference sequence (Table \ref{DirRep}).  In principle, such divergence levels amount to one divergent site per 200 bp, and therefore are expected to be captured with paired end read data.  However, these criteria are extremely lenient and offer an upper bound of sequence where repeats might confound duplicate identification.  Additionally, duplications whose breakpoints lie within the span of repeats could potentially still be identified, and therefore we did not apply any filters to exclude these regions. These estimates therefore represent an upper bound of sequence in the reference that cannot be readily surveyed.  Assuming that the \Dyak {} reference genome is representative of strains in the population, the number of variants that are unidentified due to repetitive content is likely to be very low.  

The methods described here are similarly precluded from surveying sites which have zero coverage across strains and not associated with deletions but are simply a product of stochastic effects of library prep.  This filter removes 0.9-1.6\% of sequence per strain in \Dsim {} and 1.2-1.8\% of sequence in \Dyak {} and will therefore have limited effect.  We therefore suggest that the number of tandem duplicates identified via paired end read mapping in high coverage sequencing are therefore likely to be an accurate representation of genome wide variation in the population.  

\clearpage
\bibliographystyle{MBE}
\bibliography{TandemDups}

\begin{table}

 \caption{\label{sfig:laneSummYak}Sequencing statistics for \textit{Drosophila yakuba} libraries }

\centering
\begin{threeparttable}
\scriptsize
\begin{center}
\begin{tabular}{c|c|c|c}
Species & Sample & Read length & Number of read pairs \\
\hline
\textit{Drosophila yakuba} & reference & 54 & 27,771,582\\
 &  & 76 & 29,065,488\\
 &  & 105 & 68,660,482\\
  & CY20A & 76 & 86,628,715\\
 &  & 76 & 35,886,204\\
 &  & 76 & 39,262,696\\
  & CY28 & 54 & 31,908,008\\
 &  & 76 & 40,631,850\\
 &  & 76 & 33,242,246\\
  & CY01A & 48 & 25,239,444\\
 &  & 54 & 25,206,710\\
 &  & 76 & 25,393,295\\
 &  & 76 & 111,535,323\\
  & CY02B5 & 48 & 20,266,663\\
 &  & 54 & 23,669,407\\
 &  & 76 & 30,960,740\\
  & CY04B & 48 & 20,525,873\\
 &  & 54 & 19,048,177\\
 &  & 76 & 29,088,850\\
 &  & 76 & 81,634,715\\
  & CY08A & 48 & 22,518,030\\
 &  & 76 & 28,321,290\\
 &  & 54 & 31,630,793\\
  & CY13A & 48 & 21,632,572\\
 &  & 76 & 26,933,913\\
 &  & 54 & 31,252,102\\
  & CY17C & 48 & 25,560,116\\
 &  & 76 & 30,498,100\\
 &  & 54 & 29,303,889\\
 &  & 76 & 116,502,064\\
  & CY21B3 & 48 & 21,084,747\\
 &  & 76 & 28,094,292\\
 &  & 54 & 24,831,622\\
 &  & 76 & 92,234,832\\
  & CY22B & 54 & 21,924,434\\
 &  & 76 & 25,916,918\\
 &  & 54 & 26,611,105\\
  & NY66-2 & 54 & 24,894,912\\
 &  & 76 & 26,776,198\\
 &  & 76 & 88,481,425\\
  & NY81 & 54 & 25,534,953\\
 &  & 76 & 27,838,117\\
 &  & 76 & 85,887,835\\
  & NY48 & 54 & 23,129,714\\
 &  & 76 & 28,500,212\\
 &  & 76 & 30,114,197\\
  & NY56 & 54 & 26,486,845\\
 &  & 76 & 31,506,258\\
 &  & 76 & 32,601,617\\
  & NY62 & 54 & 29,743,963\\
 &  & 76 & 30,946,111\\
 &  & 76 & 31,529,718\\
  & NY65 & 54 & 27,490,843\\
 &  & 76 & 32,600,073\\
 &  & 76 & 28,468,446\\
  & NY73 & 54 & 27,534,698\\
 &  & 76 & 31,988,717\\
 &  & 76 & 29,556,013\\
  & NY42 & 54 & 33,715,271\\
 &  & 76 & 42,770,174\\
 &  & 76 & 35,701,540\\
  & NY85 & 54 & 27,811,873\\
 &  & 76 & 32,528,787\\
 &  & 76 & 33,047,656\\
  & NY141 & 54 & 25,298,106\\
 &  & 76 & 80,761,937\\
 &  & 76 & 30,145,168\\
\hline
\end{tabular}
\begin{tablenotes}
\item[]CY= Cameroon \Dyak, NY = Nairobi \Dyak
\end{tablenotes}
\end{center}
\end{threeparttable}
 \end{table}
\clearpage

\begin{table}
\centering
\caption{\label{sfig:laneSummSim}Sequencing statistics for \Dsim {} libraries}
\begin{threeparttable}
\scriptsize
 \begin{center}

\begin{tabular}{c|c|c|c}
Species & Sample & Read length & Number of read pairs \\
\hline
\textit{Drosophila simulans} & reference (\textit{w501}) & 104 & 46,855,159\\
 & MD221 & 76 & 41,341,930\\
 & & 76 & 43,274,500\\
 & & 54 & 42,866,171\\
 & MD06 & 54 & 40,300,257\\
 & & 76 & 36,879,386\\
 & & 76 & 44,594,779\\
 & MD63 & 54 & 29,033,259\\
 & & 76 & 34,548,640\\
 & & 76 & 32,326,841\\
 & MD251 & 76 & 42,795,040\\
 & & 54 & 42,248,728\\
 & & 76 & 43,250,191\\
 & MD105 & 54 & 28,957,390\\
 & & 76 & 35,059,033\\
 & & 76 & 30,862,150\\
 & MD199 & 54 & 39,280,145\\
 & & 76 & 41,798,120\\
 & & 76 & 42,708,601\\
 & MD106 & 76 & 45,264,776\\
 & & 76 & 45,425,267\\
 & MD73 & 54 & 42,630,510\\
 & & 76 & 43,929,651\\
 & & 76 & 44,473,268\\
 & MD15 & 76 & 24,882,026\\
 & & 76 & 38,422,860\\
 & & 76 & 38,035,776\\
 & MD233 & 76 & 44,330,417\\
 & & 76 & 38,302,627\\
 & & 76 & 38,525,975\\
 & NS40 & 54 & 41,655,021\\
 & & 76 & 42,810,725\\
 & & 76 & 44,451,089\\
 & NS05 & 54 & 43,132,030\\
 & & 76 & 42,000,020\\
 & & 76 & 43,987,326\\
 & NS137 & 54 & 31,917,670\\
 & & 76 & 39,650,018\\
 & & 76 & 34,951,602\\
 & NS39 & 54 & 41,063,490\\
 & & 76 & 42,341,695\\
 & & 76 & 45,140,717\\
 & NS67 & 76 & 42,962,176\\
 & & 76 & 36,357,328\\
 & & 76 & 39,932,110\\
 & NS50 & 76 & 40,491,844\\
 & & 76 & 32,795,824\\
 & & 76 & 37,809,667\\
 & NS113 & 76 & 37,031,543\\
 & & 76 & 34,325,117\\
 & & 76 & 37,031,994\\
 & NS78 & 76 & 41,381,084\\
 & & 76 & 36,771,570\\
 & & 76 & 38,763,692\\
 & NS33 & 76 & 37,225,499\\
 & & 76 & 34,271,146\\
 & & 76 & 36,656,786\\
 & NS79 & 76 & 38,326,489\\
 & & 76 & 36,965,044\\
 & & 76 & 40,146,925\\
\hline
 \end{tabular}
\begin{tablenotes}
\item[]MD = Madagascar \Dsim, NS = Nairobi \Dsim.
\end{tablenotes}
\end{center}
 \end{threeparttable}
 \end{table}
\clearpage

\begin{table}

\begin{center}
\caption{\label{BasesCovered} Bases with coverage in reference }
\begin{tabular}{lrr}

\hline
 chrom & \Dsim &\Dyak \\
 \hline
2L & 23275275 & 28575774 \\
2R & 21329557 & 23894902\\
3L & 23878518  & 21033386\\
3R & 26966593 & 22193704\\
4 & 1007160 & 1342752\\
X & 20607623 & 21512027 \\
\hline
\end{tabular}
\end{center}
\end{table}

\clearpage

\begin{table}
\caption{\label{tab:rawCovSumm}Summaries of raw sequence coverage for inbred lines at mapping quality thresholds}
\centering
\begin{threeparttable}
\footnotesize
\begin{singlespace}
\begin{center}
\renewcommand{\thefootnote}{\textit{\alph{footnote}}}
\renewcommand\footnoterule{}
\begin{tabular}{c|c||c|c|c||c|c|c||c|c|c}
Species & Sample & $M_0$\tnote{a} & $f_{0,0}$\tnote{b} & $Q_{1,0}$\tnote{c} & $M_{20}$\tnote{a}  & $f_{0,20}$\tnote{b}  & $Q_{1,20}$\tnote{c} & $M_{30}$\tnote{a}  & $f_{0,30}$\tnote{b} & $Q_{1,30}$\tnote{c} \\
\hline
\textit{Drosophila simulans} & reference & 55 & 0.012 & 8 & 55 & 0.046 & 15 & 49 & 0.047 & 10\\
 & MD221 & 118 & 0.017 & 19 & 116 & 0.051 & 25 & 90 & 0.071 & 5\\
 & MD06 & 115 & 0.017 & 17 & 113 & 0.051 & 24 & 85 & 0.071 & 4\\
 & MD63 & 90 & 0.017 & 15 & 89 & 0.051 & 22 & 70 & 0.071 & 3\\
 & NS40 & 121 & 0.017 & 22 & 119 & 0.051 & 32 & 95 & 0.070 & 5\\
 & MD251 & 118 & 0.017 & 19 & 116 & 0.051 & 26 & 90 & 0.071 & 5\\
 & MD105 & 92 & 0.016 & 14 & 90 & 0.051 & 22 & 71 & 0.070 & 3\\
 & MD199 & 114 & 0.017 & 19 & 112 & 0.051 & 27 & 90 & 0.069 & 5\\
 & NS05 & 119 & 0.017 & 20 & 118 & 0.051 & 27 & 92 & 0.069 & 5\\
 & NS137 & 98 & 0.017 & 17 & 96 & 0.051 & 26 & 76 & 0.071 & 4\\
 & NS39 & 121 & 0.016 & 20 & 119 & 0.051 & 30 & 94 & 0.069 & 5\\
 & MD73 & 120 & 0.016 & 20 & 118 & 0.051 & 29 & 93 & 0.069 & 5\\
 & MD106 & 87 & 0.018 & 14 & 86 & 0.052 & 18 & 64 & 0.078 & 4\\
 & NS67 & 124 & 0.018 & 21 & 122 & 0.051 & 34 & 97 & 0.075 & 6\\
 & NS50 & 117 & 0.017 & 19 & 116 & 0.052 & 29 & 91 & 0.076 & 6\\
 & NS113 & 111 & 0.017 & 18 & 109 & 0.051 & 27 & 85 & 0.074 & 5\\
 & NS78 & 122 & 0.018 & 19 & 120 & 0.051 & 25 & 94 & 0.075 & 6\\
 & MD15 & 97 & 0.018 & 16 & 95 & 0.052 & 21 & 71 & 0.078 & 4\\
 & NS33 & 111 & 0.017 & 18 & 109 & 0.051 & 26 & 86 & 0.074 & 5\\
 & NS79 & 120 & 0.018 & 21 & 118 & 0.051 & 29 & 93 & 0.075 & 6\\
 & MD233 & 121 & 0.017 & 21 & 119 & 0.051 & 30 & 90 & 0.076 & 5\\
\hline
\textit{Drosophila yakuba} & reference & 124 & 0.009 & 36 & 122 & 0.015 & 11 & 111 & 0.023 & 4\\
 & CY20A & 149 & 0.013 & 20 & 146 & 0.032 & 4 & 121 & 0.060 & 3\\
 & CY28 & 85 & 0.013 & 11 & 83 & 0.036 & 3 & 66 & 0.067 & 2\\
 & CY01A & 151 & 0.012 & 21 & 148 & 0.032 & 4 & 123 & 0.056 & 2\\
 & CY02B5 & 55 & 0.013 & 8 & 54 & 0.039 & 2 & 44 & 0.067 & 2\\
 & CY04B & 125 & 0.013 & 17 & 123 & 0.036 & 4 & 102 & 0.063 & 2\\
 & CY08A & 57 & 0.013 & 9 & 55 & 0.036 & 2 & 45 & 0.066 & 1\\
 & CY13A & 56 & 0.014 & 9 & 54 & 0.036 & 2 & 45 & 0.068 & 1\\
 & CY17C & 151 & 0.012 & 21 & 148 & 0.031 & 4 & 118 & 0.058 & 2\\
 & CY21B3 & 139 & 0.012 & 21 & 137 & 0.029 & 3 & 113 & 0.053 & 2\\
 & CY22B & 56 & 0.014 & 8 & 55 & 0.038 & 2 & 45 & 0.071 & 1\\
 & NY66-2 & 110 & 0.014 & 15 & 107 & 0.038 & 4 & 85 & 0.070 & 2\\
 & NY81 & 107 & 0.014 & 14 & 104 & 0.038 & 5 & 82 & 0.070 & 2\\
 & NY48 & 64 & 0.015 & 9 & 62 & 0.040 & 3 & 49 & 0.072 & 2\\
 & NY56 & 60 & 0.016 & 7 & 58 & 0.044 & 4 & 45 & 0.081 & 2\\
 & NY62 & 70 & 0.013 & 10 & 68 & 0.036 & 3 & 54 & 0.066 & 2\\
 & NY65 & 69 & 0.014 & 10 & 68 & 0.039 & 4 & 54 & 0.072 & 2\\
 & NY73 & 68 & 0.015 & 9 & 66 & 0.041 & 4 & 53 & 0.073 & 2\\
 & NY42 & 93 & 0.014 & 12 & 91 & 0.037 & 3 & 74 & 0.067 & 2\\
 & NY85 & 79 & 0.013 & 12 & 78 & 0.033 & 3 & 64 & 0.062 & 2\\
 & NY141 & 110 & 0.015 & 14 & 108 & 0.041 & 6 & 87 & 0.073 & 3\\
\hline
\end{tabular}
\begin{tablenotes}
\item[a]\noindent $M_X$ refers to median coverage of major chromosome arms at mapping quality $\geq X.$
\item[b]$f_{0,X}$ refers to the fraction of the major arms with coverage of zero at mapping quality $\geq X.$
\item[c]$Q_{1,X}$ refers to the first quantile of sites with coverage > 0 and mapping quality $\geq X$.  In other words, 99\% of sites with coverage > 0 have coverage $\leq Q_{1,X}$
\end{tablenotes}
\end{center}
\end{singlespace}
\end{threeparttable}
\end{table}
\clearpage
\begin{table}
\centering
 \caption{\label{InsertSize}Summary statistics of libary insert sizes.}
\begin{threeparttable}
\footnotesize
\begin{center}

\begin{tabular}{c|c|c|c}
Species & Sample\tnote{a}& Median fragment size & $99.9^{th}$ quantile of fragment sizes \\
\hline
\textit{Drosophila simulans}  &reference  (\textit{w501}) & 532 & 593\tnote{b}\\
 & MD221 & 318 & 527\\
 &MD06 & 319 & 551\\
 &MD63 & 302 & 636\\
 &MD251 & 325 & 580\\
 &MD105 & 316 & 702\\
 &MD199 & 344 & 584\\
 &MD73 & 317 & 537\\
 &MD106 & 319 & 529\\
 &MD233 & 301 & 501\\
 &MD15 & 313 & 511\\
 &NS40 & 316 & 602\\
 &NS05 & 324 & 558\\
 &NS137 & 317 & 647\\
 &NS39 & 318 & 549\\
 &NS67 & 314 & 517\\
 &NS50 & 330 & 571\\
 &NS113 & 334 & 581\\
 &NS78 & 326 & 550\\
 &NS33 & 327 & 543\\
 &NS79 & 332 & 562\\
\hline
\textit{Drosophila yakuba} & reference & 343 & 565\\
 &CY20A & 336 & 998\\
 &CY28 & 308 & 1,018\\
 &CY01A & 322 & 1,124\\
 &CY02B5 & 313 & 1,183\\
 &CY04B & 327 & 1,123\\
 &CY08A & 322 & 1,338\\
 &CY13A & 316 & 1,070\\
 &CY17C & 338 & 1,102\\
 &CY21B3 & 372 & 1,226\\
 &CY22B & 328 & 1,156\\
 &NY66-2 & 313 & 989\\
 &NY81 & 326 & 1,063\\
 &NY48 & 322 & 1,030\\
 &NY56 & 332 & 1,069\\
 &NY62 & 319 & 1,037\\
 &NY65 & 341 & 1,089\\
 &NY73 & 325 & 1,078\\
 &NY42 & 270 & 725\\
 &NY85 & 326 & 1,036\\
 &NY141 & 336 & 1,061\\
\hline
\end{tabular}
\begin{tablenotes}
\item[a]MD = Madagascar \Dsim, NS = Nairobi \Dsim, CY = Cameroon \Dyak, and NY = Nairobi \Dyak.
\item[b]For this sample, the actual value calculated from alignments was $> 10^6$.  The value we used, which is shown in the table, is the mean value for the non-reference samples.
\end{tablenotes}
\end{center}
\end{threeparttable}
\end{table}
\clearpage
\begin{table}[ht]
\caption{\label{ByStrain}Number of tandem duplications in sample strains. }

\begin{singlespace}
\footnotesize
\begin{center}
\begin{tabular}{c|c|c}

Species & Strain & Duplications \\
 \hline

\hline
& CY20A & 145 \\
& CY28 & 160 \\
& CY01A & 207 \\
& CY02B5 & 179 \\
& CY04B & 154 \\
& CY08A & 184 \\
& CY13A & 159 \\
& CY17C & 230 \\
& CY21B3 & 224 \\
& CY22B & 136 \\
& NY66-2 & 134 \\
& NY81 & 126 \\
& NY48 & 106 \\
& NY56 & 81 \\
& NY62 & 154 \\
& NY65 & 107 \\
& NY73 & 100 \\
& NY42 & 138 \\
& NY85 & 169 \\
& NY141 & 121 \\
\hline
&std &40.22\\
&mean& 151.0\\
&median &154.0\\
\hline
\Dsim& MD221 & 119 \\
& MD06 & 117 \\
& MD63 & 122 \\
& NS40 & 119 \\
& MD251 & 110 \\
& MD105 & 130 \\
& MD199 & 100 \\
& NS05 & 160 \\
& NS137 & 107 \\
& NS39 & 129 \\
& MD73 & 102 \\
& MD106 & 85 \\
& NS67 & 120 \\
& NS50 & 109 \\
& NS113 & 113 \\
& NS78 & 113 \\
& MD15 & 93 \\
& NS33 & 111 \\
& NS79 & 117 \\
& MD233 & 102 \\
\hline
& std &15.56\\
& Mean &113.63\\
&Median&113.0\\
\hline
\end{tabular}
\end{center}
\end{singlespace}
\end{table}

\clearpage
\begin{table}[ht]
\caption{\label{ByStrainWithRefs}Number of tandem duplications in sample and reference strains. }

\begin{singlespace}
\footnotesize
\begin{center}
\begin{tabular}{c|c|c}
 \hline
Species & Strain & Duplications \\
\hline
\Dyak & reference & 38\\
&CY20A & 154\\
&CY28 & 170\\
&CY01A & 225\\
&CY025B  &187\\
&CY04B & 172\\
&CY08A  & 201\\
&CY13A & 168\\
&CY17C & 248\\
&CY21B3 & 247\\
&CY22B & 148\\
&NY66-2 & 151 \\
&NY81 & 135\\
&NY48  & 107\\
&NY56 & 81\\
&NY62 & 166\\
&NY65 & 114\\
&NY73  & 114\\
&NY42 & 152\\
&NY85 & 182\\
&NY141 & 128\\
\hline
& mean & 162.5 \\
& median & 160 \\
& stdev & 43.53\\
\hline
\Dsim & $w^{501}$ & 93 \\ 
&MD221 & 171 \\
&MD06 & 169\\
&MD63 & 180\\
& MD251& 174\\
&MD105& 195\\
&MD199 & 159\\
& MD73 & 165\\
& MD106 & 136\\
& MD15 & 141\\
&NS40 & 174\\
&NS05 & 219\\
&NS137 & 165\\
& NS39 & 187\\
& NS67 & 172\\
& NS50 & 164\\
& NS113 & 161\\
& NS78 & 166\\
& NS33 & 166\\
& NS79 &  170\\
& MD233 & 156\\
\hline
& mean & 165.86\\
& median & 166\\
& stdev & 23.44\\
\hline
\end{tabular}
\end{center}
\end{singlespace}
\end{table}

\clearpage

\begin{table}
\caption{PacBio Confirmation}
\begin{center}
\begin{tabular}{ccrr}
Line & Chromosome & Total & Confirmed \\
\hline
CY17C & 2L &   41 & 40 \\
& 2R &  70 & 69 \\
 & 3L &  38 & 36\\
&3R & 30 & 25\\
&X & 48 & 46\\
&4 & 3 & 3 \\
\hline
CY21B3 & 2L &  53 & 43 \\
& 2R &  57 & 55 \\
 & 3L & 34 & 34 \\
& 3R &  36 & 36\\
& X &  43 & 43\\
&4 & 1 & 1\\
\hline
NY66-2 & 2L &   28 & 26 \\
& 2R &   30 & 29\\
& 3L & 17 & 16 \\
& 3R &  24 &22 \\
& X &  33 & 33\\
& 4 &  2  & 2\\
\hline
NY73 & 2L &  19 & 16 \\
& 2R  &  24 & 23 \\
& 3L &  21 & 20 \\
& 3R & 15 & 14\\
& X &  21 & 18 \\
& 4 &  0  & 0 \\
\hline
\end{tabular}

 \end{center}
 \label{tab:confirmations}

\end{table}

\clearpage

\begin{table}[ht]
\caption{\label{Downsample}Fraction of genome covered in 3 or more reads in downsampled sequences of line CY17C }
\begin{singlespace}
\footnotesize
\begin{center}
\begin{tabular}{cc}
Median coverage & Fraction Covered\\
 \hline
15X & 0.866 \\
30X & 0.898 \\
45X & 0.909  \\
60X & 0.914 \\
75X & 0.92 \\
90X & 0.92 \\
105X & 0.923  \\
120X & 0.923 \\
135X & 0.926 \\
 150X & 0.928 \\
\hline
\end{tabular}
\end{center}
\end{singlespace}
\end{table}

\clearpage

\begin{table}
\begin{center}
\caption{\label{FET} $P$-value for Fisher's Exact Test of PCR confirmation rates }
\begin{tabular}{lcccc}
 \hline
Study   & Confirmed &  Total &             \Dsim &       \Dyak \\
\hline
Emerson et al. 2008 & 64&74&  0.7855 & $2.089\times 10^{-5}$ \\
Cridland  et al 2010   & 75&78&  0.0319 & $1.642\times10^{-9}$  \\
Cardoso et al. 2011  & 18&24& 0.5230 & 0.0725 \\
Cardoso et al.  2012  & 32&32& 0.0169 & $3.4194\times10^{-7}$ \\
Zichner et al.  2013  & 22&23&   0.2422 &  $1.1579\times10^{-4}$\\
Schrider  et al. 2013   & 7&19&    0.0007 &  0.4165 \\
\hline
\end{tabular}
\end{center}
\end{table}

\clearpage
\clearpage

\clearpage

\begin{table}
\begin{center}
\caption{\label{ReadLengths} PacBio Sequencing Reads}
\begin{tabular}{lccccc}
\hline
Line & Cell & Avg Length & Min Length & Max Length & Total Reads\\
\hline
CY17C & 1 & 2424 & 50 & 21696 & 137839 \\ 
CY17C & 2 & 2418 & 50 & 21680 &134568\\
CY17C & 3 & 2175 & 50 &19084 & 103354 \\
CY17C & 4 & 2161 & 50 & 20519 & 112597 \\ 
CY17C & 5 & 2148 & 50 & 18949 & 118144\\
\hline
CY21B3 & 1 & 2571 & 50 & 24042 & 123592\\
CY21B3 & 2 &  2549 & 50 & 22398 & 129855\\
CY21B3 & 3 & 2224 & 50 & 22206 & 161532 \\
CY21B3 & 4 & 2152 & 50 & 20779& 15733\\
CY21B3 & 5 & 2060 & 50 & 18204 & 149234 \\
\hline
NY66-2 & 1 & 2556 & 50 & 22178 & 150291 \\
NY66-2 & 2 & 2485 & 50 & 21204 & 112095 \\
NY66-2 & 3 & 2648 &50 & 21161& 139571 \\
NY66-2 & 4 & 2059 & 50 & 20709 &184485\\
NY66-2 & 5 & 2699 & 50 & 22203& 149048 \\ 
\hline
NY73 & 1 & 2269& 50 & 17924 & 95028 \\
NY73 & 2 & 2200&  50 & 17927 & 92587\\
NY73 & 3 & 2175 & 50 & 17779 & 97116  \\
NY73 & 4 & 2137 & 50 & 19110 &  83372 \\
NY73 & 5 & 2175 & 50 & 19084 &  87212 \\
\hline
\end{tabular}
\end{center}
\end{table}

\clearpage
\clearpage
\begin{table}
\label{ChiSq} 
\caption{\label{DirRep} Direct repeats 300bp or larger within 25kb span in reference}
\begin{center}
\begin{tabular}{llr}
\hline
Species & Chrom & Number \\
\hline
\Dyak & 2L& 13 \\
&2R & 10\\
& 3L &41\\
& 3R & 43\\
&X& 13\\
\hline
\Dsim & 2L & 2 \\
& 2R & 0 \\
& 3L & 3 \\
&  3R & 0 \\
& X & 0\\
\hline
\end{tabular}
\end{center}
\end{table}

\clearpage
\begin{table}
\caption{\label{GOBias}Overrepresented GO categories among duplicated genes at EASE $\geq$ 1.0}
\begin{center}
\begin{tabular}{llr}
Species & Functional Category & Group EASE score\\
\hline
\Dyak & Alternative splicing & 3.54 \\
&Immunoglobulin and Fibronectin & 2.70 \\
&Chitins and aminoglycans & 2.00 \\
&Signal peptide or glycoprotein & 1.54 \\
&Immune response to wound healing & 1.44\\
&Drug and hormone metabolism & 1.37\\
&Extracellular matrix & 1.34 \\
&Immune response to pathogens  & 1.17 \\
&Neurodevelopment and morphogenesis & 1.17\\
&Chemotaxis & 1.12 \\
&Chorion Development & 1.07 \\
\hline
\Dsim  &Oxidation-reduction and secondary metabolites & 2.63 \\
 &Cytochromes, oxidoreductases, and toxin metabolism &  2.32\\
& Lipases & 1.61 \\
&Immune response to bacteria & 1.59 \\
& phospholipid metabolism & 1.45 \\
&Chemosensory processing & 1.37\\
&Gultathion transferase and drug metabolism & 1.21\\
& Carboxylesterases &  1.2\\
&Sarcomeres & 1.0 \\
& Cuticle development & 0.97\\
& Endopeptidases & 0.87\\
\hline
\end{tabular}
\end{center}

\end{table} 

\clearpage

\begin{table}
\caption{\label{IndependentDuplication} Overrepresented GO categories in independently duplicated genes}
\begin{center}
\begin{tabular}{llr}
&Multiple Independent Duplications\\
\hline
\Dyak &Chorion Development and oogenesis & 1.79  \\
& Cell signaling & 1.34 \\
& Sensory processing & 1.23 \\
& Immune response & 1.11 \\
& Development & 1.05 \\
\hline
\Dsim &Immune Response to Bacteria & 3.35\\
&Chorion Development and oogenesis & 1.84 \\
&Organic Cation Membrane Transport &  1.48\\
&Chemosensory Perception & 1.41 \\

\hline 
\end{tabular}
\end{center}
\end{table}
\clearpage
\begin{table}
\caption{\label{HighFreq} Gene Duplications Identified at a Sample Frequency $\geq\frac{17}{20}$}
\begin{center}
\scriptsize
\begin{tabular}{llllll}
Species & Chrom & Start & Stop & Genes & Ontologies \\
\hline
\Dyak 
&2L & 20867709 & 20868709 &\emph{GE19441}&  cation transport \\
&2L & 4561989 & 4563943 &\emph{GE14706} & \tiny{Epidermal growth factor; Follistatin-like;  Zona pellucida }\\
&2R & 1192252 & 1198186 & \emph{GE12923}& Adenyl cyclase\\
&2R & 8627195 & 8631730 &\emph{GE13451},\emph{GE13452}& AMP dependent ligases or synthetases \\
&2R & 9718894 & 9722579 &\emph{GE12353},\emph{GE12354},\emph{GE13533} & serine-type endopeptidases \\
&2L & 22229672 & 22240590  & - & - \\
&2R & 2456185 & 2468412  & - & - \\
&2R & 550564 & 555698  & - & - \\
&X & 2263061 & 2263383 & - & - \\
&X & 6027171 & 6028326  & - & - \\
&3L & 6853202 & 6857398  & - & - \\
&3L & 1631349 & 1632283  & - & - \\
&3R & 28797150 & 28798631 & - & - \\
\hline 
\Dsim & 2L & 15442908 & 15460870 & \emph{CG7653} & aminopeptidase; male reproduction  \\
& 2R & 19220441 & 19221209 & \emph{CG3510} & mitotic spindle movement and cytokinesis\\ 
& 2R & 6705454 & 6706432 & \emph{CG18445} & cellular calcium homeostasis, adult lifespan\\
& 3L & 1138079 & 1141222  & \emph{CG1179} & antimicrobial response \\
& 3L & 1138061 & 1151515 & \emph{CG9116}, \emph{CG1165}, \emph{CG1180}, \emph{CG1179} & antimicrobial response\\
& 3R & 12694264 & 12697515 & \emph{CG11600}, \emph{CG11598}, \emph{CG11608} &  lipase;  accessory gland \\
& X & 10711505 & 10714527 & \emph{CG1725} & development and morphogenesis \\
& X & 815864 & 816572 & \emph{CG11638} &EF-hand-like domain\\
& X & 3859088 & 3860485 & \emph{CG12691} & no data \\ 
& X & 4216151 & 4218438 & \emph{CG12179}, \emph{CG12184} & no data \\
& 2L & 22881359 & 22884258 & - & - \\
& 2R & 1782843 & 1783435 & - & - \\
& 2R & 1849682 & 1858187 & - & - \\
& 2R & 368797 & 369174 & - & - \\
& 2R & 6769473 & 6770112 & - & - \\
& 3L & 16974345 & 16975592 & - & - \\
& 3L & 23132385 & 23145527 & - & - \\
& 3L & 23499705 & 23500091 & - & - \\
& X & 10810038 & 10810658 & - & - \\
& X & 14784709 & 14784963 & - & - \\
& X & 1942550 & 1943315 & - & - \\
& X & 20173558 & 20177752 & - & - \\
& X & 3953557 & 3954555 & - & - \\

& X & 4353206 & 4354968 & - & - \\
& X & 7149884 & 7152082 & - & - \\
\hline

\end{tabular}
\end{center}
\end{table}
\clearpage

\begin{table}

\begin{center}\footnotesize
\caption{\label{DyakRecruitedI}Recruited Non-Coding Sequence in \Dyak {} I}

\begin{tabular}{lcrrcrcc}

\hline
Type & Gene & Chrom & Start & Stop & Strand  & Strains  \\
\hline
Recruited Non-Coding& GE18269 & 2L & 4024479 & 4026346 & + & 1 \\
& GE11906 & 2R & 15447621 & 15450093 & - & 1 \\
& GE13954 & 2R & 15422288 & 15424931 & + & 1 \\
& GE10773 & 2L & 11412985 & 11416198 & - & 1 \\
& GE12793 & 2R & 3641829 & 3654689 & - & 1 \\
& GE20665 & 3L & 4498924 & 4499419 & - & 1 \\
& GE20642 & 3L & 4748069 & 4750627 & - & 1 \\
& GE12985 & 2R & 1220523 & 1223522 & + & 3 \\
& GE25403 & 3R & 456693 & 474667 & + & 3 \\
& GE14641 & 2L & 5056039 & 5058911 & - & 1 \\
& GE24887 & 3R & 7559447 & 7567609 & - & 1 \\
& GE18834 & 2L & 12012286 & 12013371 & + & 1 \\
& GE14103 & 2R & 17064455 & 17068625 & + & 1 \\
& GE14204 & 2R & 18340872 & 18341767 & + & 6 \\
& GE19947 & 3L & 15944998 & 15948170 & - & 2 \\
& GE13585 & 2R & 10199336 & 10208268 & + & 1 \\
& GE17610 & X & 15169824 & 15171000 & + & 1 \\
& GE13833 & 2R & 13593056 & 13597666 & + & 1 \\
& GE12921 & 2R & 1296122 & 1299376 & - & 4 \\
& GE22103 & 3L & 16172247 & 16172603 & + & 2 \\
& GE20019 & 3L & 14602982 & 14605035 & - & 1 \\
& GE13348 & 2R & 7486324 & 7489683 & + & 1 \\
& GE23504 & 3R & 26016459 & 26016872 & - & 2 \\
& GE26314 & 3R & 14152097 & 14152483 & + & 1 \\
& GE17862 & X & 19312815 & 19315826 & + & 1 \\
& GE12906 & 2R & 1569408 & 1572051 & - & 1 \\
& GE24569 & 3R & 12091442 & 12096476 & - & 1 \\
& GE23710 & 3R & 23607275 & 23608177 & - & 1 \\
& GE19620 & 3L & 20846937 & 20859977 & - & 1 \\
& GE14485 & 4 & 1300833 & 1304379 & - & 1 \\
& GE24349 & 3R & 14703209 & 14705506 & - & 1 \\
& GE23444 & 3R & 26616440 & 26618338 & - & 1 \\
& GE12929 & 2R & 921000 & 926017 & - & 1 \\
& GE14093 & 2R & 16915401 & 16922730 & + & 1 \\
& GE18272 & 2L & 4137997 & 4143127 & + & 1 \\
& GE16590 & X & 157495 & 158134 & - & 1 \\
& GE18174 & 2L & 2726172 & 2731718 & + & 1 \\
& GE13324 & 2R & 7085103 & 7088180 & + & 1 \\
& GE19465 & 2L & 22202637 & 22205208 & + & 1 \\
& GE23519 & 2L & 16475111 & 16476986 & - & 1 \\
& GE21452 & 3L & 5038513 & 5043130 & + & 1 \\
& GE13128 & 2R & 3600589 & 3603993 & + & 1 \\
& GE19410 & 2L & 20057880 & 20060145 & + & 1 \\
& GE11423 & 2R & 20547562 & 20548621 & - & 1 \\
& GE24770 & 3R & 9392640 & 9392863 & - & 4 \\
& GE20162 & 3L & 11976337 & 11979310 & - & 1 \\
& GE25302 & 3R & 1445612 & 1450376 & - & 1 \\
& GE21054 & 3L & 115270 & 118715 & + & 1 \\
\hline
\end{tabular}
\end{center}

\end{table}

\begin{table}

\begin{center}\footnotesize
\caption{\label{DyakRecruitedII}Recruited Non-Coding Sequence in \Dyak {} II}

\begin{tabular}{lcrrcrcc}

\hline
Type & Gene & Chrom & Start & Stop & Strand  & Strains  \\
\hline
Recruited Non-Coding& GE16460 & 2L & 108667 & 108970 & - & 1 \\
& GE13445 & 2R & 8579441 & 8581528 & + & 2 \\
& GE13294 & 2R & 6537075 & 6541877 & + & 1 \\
& GE18814 & 2L & 11619990 & 11635209 & + & 1 \\
& GE18468 & 2L & 6924773 & 6926435 & + & 1 \\
& GE18653 & 2L & 9824910 & 9828235 & + & 1 \\
& GE12963 & 2R & 363102 & 367650 & + & 2 \\
& GE21059 & 3L & 156476 & 159503 & + & 1 \\
& GE19269 & 2L & 17827409 & 17829872 & + & 1 \\
& GE26141 & 2L & 12979610 & 12983013 & - & 1 \\
& GE19225 & 2L & 17288040 & 17288648 & + & 1 \\
& GE21286 & 3L & 3185710 & 3196869 & + & 2 \\
& GE10233 & 3R & 18382396 & 18384823 & + & 6 \\
& GE12947 & 2R & 392470 & 401311 & - & 1 \\
& GE19172 & 2L & 16627714 & 16628834 & + & 1 \\
& GE13092 & 2R & 2741691 & 2741955 & + & 1 \\
& GE24207 & 3R & 16710156 & 16712312 & - & 1 \\
& GE14560 & 4 & 808540 & 816641 & + & 1 \\
& GE18453 & 2L & 6552459 & 6556843 & + & 1 \\
& GE11989 & 2L & 9417833 & 9419230 & - & 1 \\
& GE12986 & 2R & 1265746 & 1283900 & + & 1 \\
& GE14704 & 2L & 4595905 & 4603145 & - & 1 \\
& GE10115 & 2L & 12647621 & 12647846 & - & 1 \\
& GE21984 & 2L & 18369608 & 18370669 & - & 1 \\
& GE12947 & 2R & 385328 & 401500 & - & 1 \\
& GE13641 & 2R & 11060164 & 11061595 & + & 2 \\
& GE23918 & 3R & 20559041 & 20560656 & - & 2 \\
& GE24208 & 3R & 16678984 & 16680305 & - & 1 \\
& GE13389 & 2R & 7860267 & 7865453 & + & 1 \\
& GE15418 & X & 18564946 & 18568426 & - & 1 \\
& GE17176 & X & 9412662 & 9415570 & + & 1 \\
& GE21334 & 3L & 3723736 & 3727362 & + & 1 \\
& GE26071 & 3R & 11008671 & 11012107 & + & 2 \\
& GE16233 & X & 6345642 & 6350148 & - & 1 \\
& GE10260 & 3R & 18673427 & 18675156 & + & 1 \\
& GE15086 & 2L & 2556400 & 2557925 & - & 1 \\
& GE14314 & 2R & 19547242 & 19551254 & + & 1 \\
& GE14531 & 4 & 41861 & 42853 & - & 1 \\
& GE17162 & X & 9152733 & 9154332 & + & 1 \\
& GE19996 & 3L & 15023729 & 15026407 & - & 5 \\
& GE25401 & 3R & 444738 & 446198 & + & 1 \\
& GE10771 & 3R & 25846233 & 25849137 & + & 1 \\
& GE18000 & 3L & 18980225 & 18991194 & - & 3 \\
& GE12947 & 2R & 385387 & 401772 & - & 1 \\
& GE15364 & X & 19336095 & 19338151 & - & 1 \\
& GE13453 & 2R & 8628288 & 8637097 & + & 6 \\
\hline
\end{tabular}
\end{center}

\end{table}

\begin{table}

\begin{center}\footnotesize
\caption{\label{DsimRecruitedI}Recruited Non-Coding Sequence in \Dsim {} I}

\begin{tabular}{lcrrcrcc}

\hline
Type & Gene & Chrom & Start & Stop & Strand  & Strains  \\
\hline
Recruited Non-Coding& CG5939 & 3L & 8520531 & 8500705 & - & 1 \\
& CG3955 & 2R & 9687651 & 9686889 & - & 2 \\
& CG30030 & 2R & 7961858 & 7960689 & + & 1 \\
& CG5925 & 3R & 12851812 & 12842698 & - & 1 \\
& CG3823 & X & 5828413 & 5826778 & + & 1 \\
& CG15252 & X & 9410423 & 9407893 & - & 1 \\
& CG12487 & 3L & 14581107 & 14579140 & - & 8 \\
& CG4004 & X & 11999000 & 11996338 & + & 2 \\
& CG3919 & 3L & 14069074 & 14065733 & + & 1 \\
& CG9692 & 3L & 16291819 & 16289912 & + & 1 \\
& CG7058 & X & 17623953 & 17619037 & + & 1 \\
& CG8859 & 2R & 8781970 & 8776750 & - & 2 \\
& CG10251 & 3R & 19075153 & 19071626 & - & 1 \\
& CG5685 & 3R & 4469843 & 4462120 & - & 1 \\
& CG31536 & 3R & 697054 & 692538 & + & 1 \\
& CG12691 & X & 3860485 & 3859088 & + & 18 \\
& CG5659 & X & 17126212 & 17124294 & - & 1 \\
& CG7678 & 3R & 7067181 & 7063437 & + & 1 \\
& CG18063 & 2L & 15538184 & 15535240 & + & 2 \\
& CG9431 & 2L & 12835101 & 12832723 & - & 1 \\
& CG4572 & 3R & 5597362 & 5594122 & + & 1 \\
& CG31450 & 3R & 16589187 & 16588657 & + & 1 \\
& CG2174 & X & 10322987 & 10304124 & + & 1 \\
& CG6134 & 3R & 22342992 & 22341162 & - & 1 \\
& CG3210 & 2L & 2472529 & 2463036 & - & 1 \\
& CG6308 & X & 14530096 & 14528791 & + & 1 \\
& CG1851 & 2R & 3965243 & 3961574 & - & 1 \\
& CG3208 & X & 5145948 & 5145204 & - & 6 \\
& CG13350 & 2R & 10556652 & 10549764 & - & 4 \\
& CG3647 & 2L & 14671041 & 14666003 & + & 1 \\
& CG7914 & X & 17986800 & 17986419 & - & 1 \\
& CG6643 & 3R & 19913528 & 19912399 & + & 2 \\
& CG6416 & 3L & 8421994 & 8418765 & + & 1 \\
& CG17446 & X & 8714153 & 8711653 & - & 1 \\
& CG3558 & 2L & 2878467 & 2874676 & + & 1 \\
& CG5939 & 3L & 8521640 & 8500872 & - & 1 \\
& CG10240 & 2R & 11466660 & 11464170 & + & 1 \\
& CG17927 & 2L & 16357771 & 16354240 & + & 1 \\
& CG13658 & 3R & 20627528 & 20624799 & + & 1 \\
& CG14724 & 3R & 13454925 & 13451797 & - & 1 \\
& CG42575 & 3L & 10870586 & 10866452 & + & 3 \\
& CG9689 & X & 9313627 & 9312603 & + & 1 \\
& CG12184 & X & 4215528 & 4213588 & - & 1 \\
& CG1486 & X & 19933936 & 19931912 & + & 1 \\
& CG17510 & 2R & 2451922 & 2448224 & - & 2 \\
& CG12065 & X & 8054167 & 8053646 & + & 1 \\
& CG1179 & 3L & 1141222 & 1138079 & + & 20 \\
& CG5939 & 3L & 8520189 & 8500848 & - & 1 \\
& CG32703 & X & 8703815 & 8700230 & - & 2 \\
\hline
\end{tabular}
\end{center}

\end{table} 

\begin{table}

\begin{center}\footnotesize
\caption{\label{DsimRecruitedII}Recruited Non-Coding Sequence in \Dsim {} II }

\begin{tabular}{lcrrcrcc}

\hline
Type & Gene & Chrom & Start & Stop & Strand  & Strains  \\
\hline
Recruited Non-Coding & CG33223 & X & 7929354 & 7921199 & - & 6 \\
& CG31705 & 2L & 11219262 & 11213928 & - & 2 \\
& CG4335 & 3R & 5071920 & 5071599 & + & 1 \\
& CG5939 & 3L & 8522666 & 8500416 & - & 1 \\
& CG9914 & X & 15320652 & 15317592 & - & 1 \\
& CG40486 & X & 20790004 & 20788766 & - & 1 \\
& CG6667 & 2L & 16997487 & 16991856 & - & 1 \\
& CG5772 & 2L & 9867394 & 9858502 & - & 1 \\
& CG5939 & 3L & 8526072 & 8502756 & - & 1 \\
& CG6680 & 3L & 19904005 & 19895648 & + & 2 \\
& CG17921 & 2R & 18143092 & 18133788 & - & 1 \\
& CG5939 & 3L & 8521943 & 8500015 & - & 1 \\
& CG31118 & 3R & 20168765 & 20167250 & - & 15 \\
& CG32452 & 3L & 22218538 & 22215418 & + & 3 \\
& CG6203 & 3R & 15121661 & 15121183 & + & 1 \\
& CG7398 & 3L & 6104564 & 6098227 & + & 1 \\
& CG31164 & 3R & 17481879 & 17479404 & + & 1 \\
& CG3726 & X & 5475457 & 5471459 & + & 4 \\
& CG34454 & 3L & 202674 & 199394 & - & 1 \\
& CG9380 & 2R & 21517959 & 21517361 & - & 1 \\
& CG13225 & 2R & 7858337 & 7855361 & + & 8 \\
& CG6511 & 3L & 8513193 & 8508493 & + & 1 \\
& CG1165 & 3L & 1142790 & 1141414 & - & 14 \\
& CG5409 & 2R & 13335070 & 13334296 & + & 1 \\
& CG1925 & 2R & 4531779 & 4527304 & + & 1 \\
& CG11158 & X & 12915981 & 12913430 & + & 1 \\
& CG1179 & 3L & 1151515 & 1138061 & + & 20 \\
& CG9126 & X & 14999000 & 14996974 & - & 1 \\
& CG11030 & 2L & 5566837 & 5563794 & + & 1 \\
& CG13225 & 2R & 7859420 & 7856642 & + & 2 \\
& CG8887 & 3L & 19234359 & 19234240 & - & 1 \\
& CG11325 & 2L & 6485615 & 6483354 & - & 1 \\
& CG2174 & X & 10323004 & 10304482 & + & 4 \\
& CG4026 & 2L & 9474185 & 9468950 & + & 8 \\
& CG4937 & X & 15950724 & 15945545 & + & 1 \\
& CG5939 & 3L & 8521697 & 8497398 & - & 1 \\
& CG14803 & X & 1624945 & 1623645 & + & 1 \\
& CG6231 & 3R & 5832844 & 5822195 & + & 2 \\
& CG33223 & X & 7930084 & 7921981 & - & 1 \\
& CG2174 & X & 10323255 & 10302508 & + & 1 \\
& CG10146 & 2R & 11339645 & 11337864 & + & 2 \\
& CG9331 & 2L & 20280987 & 20279429 & + & 1 \\
& CG5939 & 3L & 8519186 & 8500624 & - & 1 \\
& CG9761 & 3R & 493871 & 486754 & - & 1 \\
& CG4086 & 3L & 16204452 & 16202997 & - & 1 \\
& CG5939 & 3L & 8519120 & 8497461 & - & 1 \\
& CG30497 & 2R & 4466546 & 4459504 & - & 1 \\
\hline
\end{tabular}
\end{center}

\end{table}

\begin{table}

\begin{center}\footnotesize
\caption{\label{DyakDualProm}Putative Dual Promoter Genes in \Dyak}

\begin{tabular}{lllcrrcrcc}

\hline
Type & \fiveP {} Gene & \thrP {} Gene & Chrom & Start & Stop & Strand  & Strains  \\
\hline
Dual Promoter& GE25749 & GE24961 & 3R & 6426573 & 6429128 & & 1 \\
& GE17248 & GE15994 & X & 10111434 & 10113143 & & 1 \\
& GE25858 & GE24866 & 3R & 8159079 & 8162914 & & 1 \\
& GE21080 & GE21031 & 3L & 282949 & 286511 & & 1 \\
& GE10392 & GE23928 & 3R & 20485903 & 20487331 & & 1 \\
& GE19241 & GE22696 & 2L & 17479513 & 17482694 & & 1 \\
& GE10859 & GE23415 & 3R & 26875286 & 26881333 & & 1 \\
& GE26132 & GE24588 & 3R & 11912024 & 11916314 & & 3 \\
& GE17889 & GE15249 & 2L & 2014960 & 2019346 & & 1 \\
& GE14337 & GE11513 & 2R & 19768417 & 19770224 & & 1 \\
& GE16626 & GE16372 & 2L & 146434 & 147628 & & 1 \\
& GE13245 & GE12657 & 2R & 5880254 & 5883193 & & 1 \\
& GE14570 & GE14484 & 4 & 1314161 & 1319175 & & 1 \\
& GE16771 & GE16447 & X & 2623733 & 2628942 & & 1 \\
& GE14002 & GE11855 & 2R & 15989525 & 15990929 & & 1 \\
& GE13989 & GE11866 & 2R & 15777997 & 15781002 & & 1 \\
& GE25878 & GE24850 & 3R & 8267252 & 8270103 & & 1 \\
& GE10233 & GE24091 & 3R & 18383837 & 18386578 & & 2 \\
& GE18995 & GE25461 & 2L & 13658567 & 13660529 & & 4 \\
& GE13812 & GE12098 & 2R & 13042582 & 13049000 & & 1 \\
& GE13138 & GE12791 & 2R & 3686554 & 3691429 & & 3 \\
& GE21885 & GE20148 & 3L & 12173462 & 12175684 & & 1 \\
& GE17429 & GE15699 & 2L & 1162125 & 1162631 & & 1 \\
& GE19355 & GE21644 & 2L & 18836291 & 18841517 & & 1 \\
& GE14442 & GE11414 & 2R & 20628266 & 20630184 & & 1 \\
& GE13815 & GE12096 & 2R & 13052601 & 13054893 & & 1 \\
& GE17290 & GE15947 & X & 10693772 & 10694086 & & 2 \\
& GE18487 & GE13339 & 2L & 7166837 & 7171789 & & 1 \\
& GE19240 & GE22696 & 2L & 17477743 & 17483758 & & 3 \\
& GE26291 & GE24418 & 3R & 13841865 & 13844988 & & 1 \\
& GE21597 & GE20441 & 3L & 7731981 & 7736528 & & 1 \\
& GE13211 & GE12702 & 2R & 5347302 & 5349788 & & 2 \\
& GE26230 & GE24475 & 3R & 13110215 & 13112850 & & 1 \\
& GE13929 & GE11929 & 2R & 15128544 & 15131155 & & 1 \\
& GE21085 & GE21022 & 3L & 343902 & 348093 & & 1 \\
& GE10951 & GE23313 & 3R & 28355223 & 28357215 & & 1 \\
& GE26060 & GE24657 & 3R & 10734970 & 10739334 & & 1 \\
& GE16619 & GE16580 & X & 349074 & 349922 & & 1 \\
& GE13338 & GE12567 & 2R & 7221214 & 7224978 & & 1 \\
& GE25687 & GE25035 & 3R & 5444807 & 5445189 & & 1 \\
\hline
\end{tabular}
\end{center}

\end{table}

 \begin{table}

\begin{center}\footnotesize
\caption{\label{DsimDualProm}Putative Dual Promoter Genes in \Dsim {} I}

\begin{tabular}{lcrrcrcc}

\hline
Type & \fiveP {} Gene & \thrP {} Gene& Chrom & Start & Stop & Strand  & Strains  \\
\hline

Dual Promoter &  CG4433 & CG4845 & 3R & 5438512 & 5436655 & & 1 \\
&  CG10086 & CG18748 & 3R & 3506551 & 3501272 & & 1 \\
&  CG12230 & CG3917 & X & 18350597 & 18348732 & & 1 \\
&  CG9506 & CG9500 & 2L & 6166401 & 6152814 & & 1 \\
&  CG7034 & CG5760 & 3R & 4371721 & 4366499 & & 1 \\
&  CG13472 & CG5295 & 3L & 14429033 & 14427393 & & 1 \\
&  CG3423 & CG9211 & 2L & 6722948 & 6714828 & & 1 \\
&  CG5588 & CG5611 & 3R & 22888188 & 22885586 & & 1 \\
\hline
\end{tabular}
\end{center}

\end{table}

\begin{table}

\begin{center}\footnotesize
\caption{\label{DyakChimI} Putative Chimeric Genes in \Dyak {} I}

\begin{tabular}{lllcrrcrcc}

\hline
Type & \fiveP {} Gene & \thrP {} Gene & Chrom & Start & Stop & Strand  & Strains \\
\hline
Chimeric   & GE18811 & GE18814 & 2L & 11624699 & 11635716 & + & 2 \\
 & GE12444 & GE12447 & 2R & 8682740 & 8684098 & - & 1 \\
 & GE12828 & GE12831 & 2R & 2621213 & 2625329 & - & 5 \\
 & GE21209 & GE21211 & 3L & 2196531 & 2199567 & + & 1 \\
 & GE16478 & GE16479 & X & 2094815 & 2098815 & - & 1 \\
 & GE12441 & GE12442 & 2R & 8693878 & 8697950 & - & 16 \\
 & GE25203 & GE25214 & 2L & 14060529 & 14065910 & - & 1 \\
 & GE19619 & GE19620 & 3L & 20848448 & 20861465 & - & 1 \\
 & GE10625 & GE10626 & 3R & 24047193 & 24049829 & + & 1 \\
 & GE19343 & GE19344 & 2L & 18662414 & 18663086 & + & 1 \\
 & GE13385 & GE13386 & 2R & 7858807 & 7861090 & + & 2 \\
 & GE17209 & GE17211 & X & 9829925 & 9831752 & + & 1 \\
 & GE19944 & GE19947 & 3L & 15944832 & 15946907 & - & 1 \\
 & GE12275 & GE12276 & 2R & 10705351 & 10711729 & - & 10 \\
 & GE16764 & GE16765 & X & 2538959 & 2540881 & + & 3 \\
 & GE23781 & GE23782 & 3R & 22385664 & 22386968 & - & 2 \\
 & GE14304 & GE14305 & 2R & 19416793 & 19423595 & + & 1 \\
 & GE26358 & GE26359 & 3R & 14778214 & 14779539 & + & 1 \\
 & GE10773 & GE10784 & 2L & 11411312 & 11413493 & - & 1 \\
 & GE21046 & GE21047 & 3L & 153714 & 155228 & - & 1 \\
 & GE18427 & GE18428 & 2L & 6328720 & 6335333 & + & 1 \\
 & GE15774 & GE15775 & X & 13146906 & 13148951 & - & 1 \\
 & GE13156 & GE13157 & 2R & 4244726 & 4248299 & + & 1 \\
 & GE13078 & GE13079 & 2R & 2529175 & 2532039 & + & 1 \\
 & GE18932 & GE18933 & 2L & 13058912 & 13060325 & + & 1 \\
 & GE18810 & GE18812 & 2L & 11616544 & 11627893 & + & 8 \\
 & GE15105 & GE15108 & 2L & 2406236 & 2408538 & - & 1 \\
 & GE23853 & GE23854 & 3R & 21163271 & 21166358 & - & 1 \\
 & GE26117 & GE26119 & 3R & 11563789 & 11567495 & + & 1 \\
 & GE12353 & GE12354 & 2R & 9718894 & 9722579 & - & 19 \\
 & GE19681 & GE19682 & 3L & 20114778 & 20124632 & - & 1 \\
 & GE21817 & GE21818 & 3L & 11535053 & 11537096 & + & 1 \\
 & GE25203 & GE25214 & 2L & 14060197 & 14065508 & - & 1 \\
 & GE21123 & GE21125 & 3L & 1172484 & 1179048 & + & 1 \\
 & GE10460 & GE10461 & 3R & 21424020 & 21428007 & + & 7 \\
 & GE12766 & GE12767 & 2R & 4305793 & 4310361 & - & 1 \\
 & GE10682 & GE10684 & 3R & 25036642 & 25044043 & + & 4 \\
 & GE14192 & GE14193 & 2R & 18131313 & 18134201 & + & 1 \\
 & GE20247 & GE20248 & 3L & 11082344 & 11085513 & - & 8 \\
 & GE14235 & GE14236 & 2R & 18687558 & 18689712 & + & 4 \\
 & GE24523 & GE24524 & 3R & 12546325 & 12548762 & - & 1 \\
 & GE12451 & GE12452 & 2R & 8660527 & 8664855 & - & 1 \\
 & GE14191 & GE14192 & 2R & 18127656 & 18131007 & + & 1 \\
 & GE26194 & GE26195 & 3R & 12581548 & 12584840 & + & 5 \\
 & GE13732 & GE13734 & 2R & 12089189 & 12095200 & + & 1 \\
 & GE15325 & GE15326 & X & 20043093 & 20045334 & - & 1 \\
 & GE13428 & GE13439 & 2L & 7100699 & 7103913 & - & 1 \\
\hline
\end{tabular}
\end{center}

\end{table} 

\begin{table}

\begin{center}\footnotesize
\caption{\label{DyakChimII}Putative Chimeric Genes in \Dyak {} II}

\begin{tabular}{lcrrcrcc}

\hline
Type & Gene & Chrom & Start & Stop & Strand  & Strains  \\
\hline
Chimeric & GE11692 & GE11693 & 2R & 17841057 & 17843909 & - & 4 \\
 & GE16094 & GE16096 & X & 8950320 & 8954906 & - & 12 \\
 & GE10460 & GE10461 & 3R & 21423681 & 21427312 & + & 2 \\
 & GE23849 & GE23860 & 2L & 16088829 & 16091980 & - & 1 \\
 & GE10382 & GE10383 & 3R & 20409220 & 20411134 & + & 1 \\
 & GE21851 & GE21854 & 3L & 11841757 & 11854748 & + & 13 \\
 & GE23589 & GE23592 & 3R & 25372683 & 25379092 & - & 5 \\
 & GE19428 & GE19429 & 2L & 20518336 & 20525187 & + & 1 \\
 & GE12481 & GE12482 & 2R & 8432422 & 8439919 & - & 4 \\
 & GE16583 & GE16584 & X & 303209 & 310866 & - & 1 \\
 & GE25989 & GE25990 & 3R & 9654460 & 9658865 & + & 1 \\
 & GE20212 & GE20213 & 3L & 11535513 & 11537526 & - & 1 \\
 & GE18810 & GE18811 & 2L & 11616597 & 11622870 & + & 1 \\
 & GE26258 & GE26259 & 3R & 13421611 & 13426161 & + & 1 \\
 & GE13653 & GE13654 & 2R & 11216791 & 11218480 & + & 1 \\
 & GE18474 & GE18475 & 2L & 7043543 & 7048586 & + & 5 \\
 & GE15105 & GE15108 & 2L & 2405167 & 2407381 & - & 3 \\
 & GE11452 & GE11454 & 2R & 20345528 & 20347724 & - & 2 \\
 & GE13618 & GE13619 & 2R & 10699218 & 10701733 & + & 8 \\
 & GE10109 & GE10110 & 3R & 17036248 & 17039632 & + & 1 \\
 & GE11708 & GE11711 & 2R & 17603369 & 17607091 & - & 1 \\
 & GE11640 & GE11641 & 2R & 18527790 & 18532538 & - & 1 \\
 & GE13733 & GE13734 & 2R & 12091221 & 12095128 & + & 7 \\
 & GE16141 & GE16142 & X & 7932016 & 7934615 & - & 1 \\
 & GE22398 & GE22399 & 3L & 20111729 & 20121616 & + & 1 \\
 & GE21366 & GE21367 & 3L & 4011112 & 4013997 & + & 1 \\
 & GE18810 & GE18814 & 2L & 11616623 & 11633710 & + & 1 \\
 & GE18648 & GE18649 & 2L & 9807924 & 9808680 & + & 1 \\
 & GE18812 & GE18814 & 2L & 11628837 & 11634750 & + & 1 \\
 & GE18811 & GE18812 & 2L & 11623999 & 11627896 & + & 1 \\
 & GE25583 & GE25584 & 3R & 3641910 & 3644927 & + & 1 \\
\hline
\end{tabular}
\end{center}

\end{table}

\begin{table}

\begin{center}\footnotesize
\caption{\label{DsimChim}Putative Chimeric Genes in \Dsim}

\begin{tabular}{lllcrrcrcc}

\hline
Type & \fiveP {} Gene & \thrP {} Gene & Chrom & Start & Stop & Strand  & Strains \\
\hline
 Chimeric  & CG11608 & CG11598 & 3R & 12697515 & 12694264 & - & 20 \\
 & CG10243 & CG10240 & 2R & 11467239 & 11463409 & + & 1 \\
 & CG3523 & CG3524 & 2L & 2944012 & 2931797 & + & 6 \\
 & CG33162 & CG5939 & 3L & 8516515 & 8502674 & - & 2 \\
 & CG10958 & CG2116 & X & 7600264 & 7599314 & + & 1 \\
 & CG6607 & CG13618 & 3R & 19885309 & 19882164 & + & 3 \\
 & CG17970 & CG14032 & 2L & 5075704 & 5069930 & + & 1 \\
 & CG32261 & CG1134 & 3L & 3915856 & 3910469 & + & 1 \\
 & CG15797 & CG17754 & X & 8697294 & 8694573 & + & 1 \\
 & CG18330 & CG1049 & 3L & 1464803 & 1462282 & + & 2 \\
 & CG10246 & CG10247 & 2R & 11475068 & 11469403 & - & 1 \\
 & CG6533 & CG6511 & 3L & 8514363 & 8506572 & + & 1 \\
 & CG2071 & CG1304 & X & 19554392 & 19552342 & - & 1 \\
 & CG4231 & CG12193 & 2L & 1461505 & 1459358 & + & 1 \\
 & CG33162 & CG5939 & 3L & 8517417 & 8502361 & - & 1 \\
 & CG33162 & CG5939 & 3L & 8519388 & 8502448 & - & 1 \\
 & CG18779 & CG10534 & 3L & 6035588 & 6032981 & - & 5 \\
 & CG13946 & CG12506 & 2L & 739495 & 736512 & + & 1 \\
 & CG13160 & CG33012 & 2R & 8966253 & 8961795 & + & 1 \\
 & CG4215 & CG7480 & 2L & 13945968 & 13940471 & + & 1 \\
 & CG32257 & CG32261 & 3L & 3917056 & 3916219 & + & 1 \\
 & CG2767 & CG11052 & 3R & 3721593 & 3720813 & - & 1 \\
 & CG33013 & CG30043 & 2R & 8957288 & 8953622 & + & 1 \\
 & CG32383 & CG32382 & 3L & 7299216 & 7297924 & - & 1 \\
 & CG32252 & CG15009 & 3L & 4185415 & 4182425 & - & 1 \\
 & CG17760 & CG17759 & 2R & 9222236 & 9218438 & + & 1 \\
 & CG4200 & CG12223 & X & 15354878 & 15351613 & + & 1 \\
 & CG33162 & CG5939 & 3L & 8525442 & 8502118 & - & 1 \\
 & CG5196 & CG34402 & 3R & 12884563 & 12880752 & - & 1 \\
 & CG33115 & CG8942 & 2L & 13552279 & 13546280 & - & 7 \\
 & CG14935 & CG14934 & 2L & 11526363 & 11522417 & + & 1 \\
 & CG14987 & CG1134 & 3L & 3920390 & 3910549 & + & 3 \\
 & CG32751 & CG32754 & X & 5729860 & 5725985 & - & 1 \\
 & CG33162 & CG5939 & 3L & 8526534 & 8501627 & - & 1 \\
 & CG32280 & CG11495 & 3L & 3024302 & 3020851 & + & 1 \\
 & CG4231 & CG12193 & 2L & 1460323 & 1458239 & + & 1 \\
 & CG6533 & CG6511 & 3L & 8514355 & 8507883 & + & 1 \\
 & CG4381 & CG4181 & 3R & 12913682 & 12912587 & - & 1 \\
 \hline
\end{tabular}
\end{center}

\end{table}

\clearpage

\clearpage
\begin{table}
\caption{Number of duplications and association with repetitive sequence by chromosome}

\begin{tabular}{lcrrrrr}
\hline

Species & Chrom & Total & 100bp Repeat & 30bp Repeat  & TE within 1kb &TE within 100bp\\
 \hline
\Dyak & 2L & 334 & 16 & 45 & 50& 14 \\
&2R & 294 & 34  & 52 & 63 & 17\\
&3L & 225 & 22 & 35 & 13 & 3\\
&3R & 256 & 32 & 52 & 21 & 6\\
&X & 279 & 18 & 50 & 17 & 7\\
&4 & 27 & 3 & 3 & 15 & 4\\
\hline
&Total & 1415 & 125  & 237  & 179 & 52 \\
\hline
\Dsim & 2L & 154 & 9 & 23  & 18 & 8 \\
&2R & 200 & 8 & 23  & 41 & 17\\
&3L & 198 & 5 & 19 & 31 & 16\\
&3R & 178 & 8 & 18  & 3 & 1\\
&X & 231 & 26 & 67  & 25 & 9\\
&4 & 14 & 0 & 0 & 3 &1 \\
\hline
& Total& 975 & 56 & 150  & 121 & 52 \\
\hline
\end{tabular}
\label{RepContent}
\end{table}

\clearpage
\begin{table}
\caption{\label{SmallRep} Direct repeats 30bp or larger within 25kb span in reference}
\begin{center}
\begin{tabular}{llr}
\hline
Species & Chrom & Number \\
\hline
\Dyak & 2L & 117 \\
&2R & 94 \\
& 3L & 80 \\
& 3R & 154\\
&X& 78\\
\hline
\Dsim & 2L & 62 \\
& 2R & 48 \\
& 3L & 25 \\
&  3R & 46\\
& X & 24 \\
\hline
\end{tabular}
\end{center}
\end{table}
\clearpage

\begin{threeparttable}
\caption{\label{SizeDistStats}Tukey's HSD test for log normalized duplication size by chromosome}
\begin{center}
\footnotesize
\begin{tabular}{lcc|ccc|l}
\hline
Species & Chromosome & vs. Chromosome &        Difference     &     Lower   &     Upper  &   Adjusted $P$-value \\
\hline
\Dyak& 2R & 2L &0.399 &0.092 &0.706 & $2.95\times10^{-3}$\superscript{**} \\
& 3L & 2L &-0.128 &-0.459 &0.204 & 0.881 \\
& 3L & 2R &-0.527 &-0.867 &-0.187 & $1.5\times10^{-4}$\superscript{**} \\
& 3R & 2R &-0.444 &-0.772 &-0.115 & $1.67\times10^{-3}$\superscript{**} \\
& 3R & 2L &-0.045 &-0.363 &0.274 & 0.999 \\
& 3R & 3L &0.083 &-0.268 &0.434 & 0.985 \\
& X & 2R &-1.081 &-1.402 &-0.760 &$<10^{-16}$\superscript{**}  \\
& X & 2L &-0.682 &-0.994 &-0.371 & $8.14\times10^{-9}$\superscript{**} \\
& X & 3L &-0.554 &-0.898 &-0.210 & $6.80\times10^{-5}$\superscript{**} \\
& X & 3R &-0.638 &-0.970 &-0.305 & $7.74\times10^{-7}$\superscript{**} \\
& X & 4 &-0.574 &-1.348 &0.200 & 0.279 \\
& 4 & 2L &-0.108 &-0.876 &0.660 & 0.999 \\
& 4 & 2R &-0.507 &-1.279 &0.265 & 0.419 \\
& 4 & 3R &-0.063 &-0.840 &0.714 & 1.000 \\
& 4 & 3L &0.020 &-0.762 &0.802 & 1.000 \\
\hline
\Dsim& 2R & 2L &0.064 &-0.353 &0.480 & 0.998 \\
& 3L & 2L &0.444 &0.027 &0.862 & 0.029\superscript{*}  \\
& 3L & 2R &0.380 &-0.009 &0.770 & 0.060 \\
& 3R & 2R &-0.170 &-0.570 &0.231 & 0.832 \\
& 3R & 2L &-0.106 &-0.533 &0.322 & 0.981 \\
& X & 2L &-0.080 &-0.484 &0.324 & 0.993 \\
& X & 2R &-0.144 &-0.519 &0.231 & 0.884 \\
& X & 3L &-0.524 &-0.900 &-0.148 & $1.06\times10^{-3}$\superscript{**}  \\
& X & 3R &0.026 &-0.362 &0.413 & 1.000 \\
& X & 4 &1.023 &-0.047 &2.092 & 0.070 \\
& 3R & 3L &-0.550 &-0.951 &-0.149 & $1.36\times10^{-3}$\superscript{**}  \\
& 4 & 2L &-1.102 &-2.187 &-0.018 & 0.044\superscript{*}  \\
& 4 & 2R &-1.166 &-2.241 &-0.092 & 0.024\superscript{*}  \\
& 4 & 3L &-1.547 &-2.621 &-0.472 & $6.11\times10^{-4}$\superscript{**}  \\
& 4 & 3R &-0.997 &-2.075 &0.082 & 0.089 \\
\hline
\end{tabular}
\begin{tablenotes}
\item[]* $P< 0.05$, ** $P<0.01$
\end{tablenotes}
\end{center}
\end{threeparttable}

\clearpage

\begin{threeparttable}
\caption{\label{TukeyHSDPerChrom}Tukey's HSD test for number of duplications per mapped bp by chromosome}
\begin{center}
\footnotesize
\begin{tabular}{l|llccc|c}
\hline
Species &   Chromosome & vs. Chromosome &      difference      &   lower  &       upper  &   Adjusted $P$-value \\
\hline
\Dyak& 2R & 2L & $ 2.955 \times 10^{-7}$ & $ -2.506 \times 10^{-7}$ & $ 8.415 \times 10^{-7}$ & 0.620 \\
& 3L & 2L & $ -2.049 \times 10^{-7}$ & $ -7.510 \times 10^{-7}$ & $ 3.411 \times 10^{-7}$ & 0.885 \\
& 3R & 2L & $ -1.104 \times 10^{-7}$ & $ -6.565 \times 10^{-7}$ & $ 4.356 \times 10^{-7}$ & 0.992 \\
& 3L & 2R & $ -5.004 \times 10^{-7}$ & $ -1.046 \times 10^{-6}$ & $ 4.566 \times 10^{-8}$ & 0.092 \\
& 3R & 2R & $ -4.059 \times 10^{-7}$ & $ -9.519 \times 10^{-7}$ & $ 1.401 \times 10^{-7}$ & 0.267 \\
& 3R & 3L & $ 9.449 \times 10^{-8}$ & $ -4.516 \times 10^{-7}$ & $ 6.405 \times 10^{-7}$ & 0.996 \\
& X & 2L & $ 3.587 \times 10^{-7}$ & $ -1.873 \times 10^{-7}$ & $ 9.047 \times 10^{-7}$ & 0.404 \\
& X & 2R & $ 6.323 \times 10^{-8}$ & $ -4.828 \times 10^{-7}$ & $ 6.093 \times 10^{-7}$ & 0.999 \\
& X & 3L & $ 5.636 \times 10^{-7}$ & $ 1.756 \times 10^{-8}$ & $ 1.110 \times 10^{-6}$ & 0.039\superscript{*} \\
& X & 3R & $ 4.691 \times 10^{-7}$ & $ -7.692 \times 10^{-8}$ & $ 1.015 \times 10^{-6}$ & 0.135 \\
& X & 4 & $ -6.273 \times 10^{-7}$ & $ -1.217 \times 10^{-6}$ & $ -3.750 \times 10^{-8}$ & 0.030\superscript{*} \\
& 4 & 2L & $ 9.860 \times 10^{-7}$ & $ 3.962 \times 10^{-7}$ & $ 1.576 \times 10^{-6}$ &$4.969\times10^{-05}$\superscript{**} \\
& 4 & 2R & $ 6.905 \times 10^{-7}$ & $ 1.007 \times 10^{-7}$ & $ 1.280 \times 10^{-6}$ & 0.012\superscript{*} \\
& 4 & 3L & $ 1.191 \times 10^{-6}$ & $ 6.011 \times 10^{-7}$ & $ 1.781 \times 10^{-6}$ & $1.766\times10^{-6}$\superscript{**} \\
& 4 & 3R & $ 1.096 \times 10^{-6}$ & $ 5.066 \times 10^{-7}$ & $ 1.686 \times 10^{-6}$ &$7.319\times10^{-6}$\superscript{**} \\

\hline
\Dsim& 2R & 2L & $ 3.962 \times 10^{-7}$ & $ -4.214 \times 10^{-8}$ & $ 8.344 \times 10^{-7}$ & 0.101 \\
& 3L & 2L & $ 3.565 \times 10^{-7}$ & $ -8.180 \times 10^{-8}$ & $ 7.948 \times 10^{-7}$ & 0.180 \\
& 3R & 2L & $ -6.753 \times 10^{-8}$ & $ -5.058 \times 10^{-7}$ & $ 3.708 \times 10^{-7}$ & 0.998 \\
& 3L & 2R & $ -3.966 \times 10^{-8}$ & $ -4.780 \times 10^{-7}$ & $ 3.986 \times 10^{-7}$ & 1.000 \\
& 3R & 2R & $ -4.637 \times 10^{-7}$ & $ -9.020 \times 10^{-7}$ & $ -2.538 \times 10^{-8}$ & 0.032\superscript{*} \\
& 3R & 3L & $ -4.240 \times 10^{-7}$ & $ -8.623 \times 10^{-7}$ & $ 1.428 \times 10^{-8}$ & 0.064 \\
& X & 2L & $ 1.107 \times 10^{-6}$ & $ 6.687 \times 10^{-7}$ & $ 1.545 \times 10^{-6}$ & $5.742\times10^{-10}$\superscript{**} \\
& X & 2R & $ 7.109 \times 10^{-7}$ & $ 2.726 \times 10^{-7}$ & $ 1.149 \times 10^{-6}$ &  $1.0537\times10^{-4}$\superscript{**}\\
& X & 3L & $ 7.505 \times 10^{-7}$ & $ 3.123 \times 10^{-7}$ & $ 1.189 \times 10^{-6}$ & $3.574\times10^{-5}$\superscript{**}\\
& X & 3R & $ 1.175 \times 10^{-6}$ & $ 7.363 \times 10^{-7}$ & $ 1.613 \times 10^{-6}$ & $5.891\times10^{-11}$\superscript{**} \\
& X & 4 & $ -4.977 \times 10^{-7}$ & $ -9.417 \times 10^{-7}$ & $ -5.369 \times 10^{-8}$ & 0.019\superscript{*} \\
& 4 & 2L & $ 1.605 \times 10^{-6}$ & $ 1.161 \times 10^{-6}$ & $ 2.049 \times 10^{-6}$ &  $3.675\times10^{-14}$\superscript{**}  \\
& 4 & 2R & $ 1.209 \times 10^{-6}$ & $ 7.646 \times 10^{-7}$ & $ 1.653 \times 10^{-6}$ & $3.144\times10^{-11}$\superscript{**}\\
& 4 & 3L & $ 1.248 \times 10^{-6}$ & $ 8.042 \times 10^{-7}$ & $ 1.692 \times 10^{-6}$ & $8.245\times10^{-12}$\superscript{**} \\
& 4 & 3R & $ 1.672 \times 10^{-6}$ & $ 1.228 \times 10^{-6}$ & $ 2.116 \times 10^{-6}$ & $2.420\times10^{-14}$\superscript{**} \\

\hline 
\end{tabular}
\begin{tablenotes}
\item[]* $P< 0.05$, ** $P<0.01$
\end{tablenotes}
\end{center}
\end{threeparttable}

\clearpage

\begin{table}
\begin{center}
\caption{\label{PindelConf}Duplications confirmed via split read mapping with Pindel }
\begin{tabular}{lcrrr}
 \hline
Species&Chromosome &Confirmed & Total & Percent Confirmed \\
\hline
\Dyak & 2L & 42 & 334 & 12.5\%  \\
&2R & 26 & 294 & 8.8\%  \\
&3L & 31 &   225 &13.7\%  \\
&3R & 45 & 256 & 17.6\% \\
&X & 32 & 279 & 11.5\%  \\
&4 & 3 & 27 & 11.1\% \\
\hline
\Dsim& 2L & 30 & 154 & 19.5\% \\
&2R & 39 & 200 & 19.5\% \\
& 3L & 29 & 198 & 14.6\% \\
&3R & 29 & 178 & 16.3\% \\
& X & 31 & 231 & 13.4\% \\
&4 & 4 & 14 & 28.6\% \\
\hline
\end{tabular}
\end{center}
\end{table}

\clearpage

\clearpage

\clearpage
\begin{table}
\label{ChiSq} 
\caption{\label{Reconstruct} Breakpoints Reconstructed}
\begin{center}
\footnotesize
\begin{tabular}{llcrrc}
\hline
Species &  Min Size (bp) & Chromosome & Reconstructed & Total & Percent Reconstructed \\
\hline
\Dyak & 0 & 2L & 145 & 338 & 42.9 \\
&  & 2R & 133 & 296 & 44.9 \\
&  & 3L & 78 & 237 & 32.9 \\
&  & 3R & 97 & 260 & 37.3 \\
&  & X & 81 & 284 & 28.5 \\
&  & 4 & 8 & 27 & 29.6 \\
& 325 & 2L & 142 & 256 & 55.5 \\
&  & 2R & 133 & 253 & 52.6 \\
&  & 3L & 75 & 170 & 44.1 \\
&  & 3R & 95 & 195 & 48.7 \\
&  & X & 74 & 162 & 45.7 \\
&  & 4 & 8 & 21 & 38.1 \\
& 500 & 2L & 138 & 229 & 60.3 \\
&  & 2R & 131 & 231 & 56.7 \\
&  & 3L & 72 & 150 & 48.0 \\
&  & 3R & 93 & 174 & 53.4 \\
&  & X & 66 & 130 & 50.8 \\
&  & 4 & 8 & 20 & 40.0 \\
& 1000 & 2L & 127 & 194 & 65.5 \\
&  & 2R & 124 & 209 & 59.3 \\
&  & 3L & 62 & 111 & 55.9 \\
&  & 3R & 88 & 150 & 58.7 \\
&  & X & 58 & 95 & 61.1 \\
&  & 4 & 7 & 14 & 50.0 \\
\hline
\Dsim &0  & 2L & 87 & 157 & 55.4 \\
&  & 2R & 102 & 200 & 51.0 \\
&  & 3L & 85 & 198 & 42.9 \\
&  & 3R & 77 & 178 & 43.3 \\
&  & 4 & 3 & 14 & 21.4 \\
&  & X & 106 & 231 & 45.9 \\
& 325  & 2L & 85 & 118 & 72.0 \\
&  & 2R & 100 & 171 & 58.5 \\
&  & 3L & 85 & 162 & 52.5 \\
&  & 3R & 76 & 140 & 54.3 \\
&  & 4 & 3 & 8 & 37.5 \\
&  & X & 102 & 177 & 57.6 \\ 
& 500  & 2L & 84 & 109 & 77.1 \\
&  & 2R & 98 & 146 & 67.1 \\
&  & 3L & 82 & 147 & 55.8 \\
&  & 3R & 74 & 115 & 64.3 \\
&  & 4 & 3 & 6 & 50.0 \\
&  & X & 98 & 154 & 63.6 \\
&1000  & 2L & 71 & 87 & 81.6 \\
&  & 2R & 85 & 114 & 74.6 \\
&  & 3L & 76 & 128 & 59.4 \\
&  & 3R & 67 & 88 & 76.1 \\
&  & 4 & 2 & 2 & 100.0 \\
&  & X & 77 & 110 & 70.0 \\
\hline
\end{tabular}
\end{center}
\end{table}

\clearpage

\begin{table}
\begin{center}
\caption{\label{ConfirmedCoverage}Duplications with increased coverage compared to reference }
\begin{tabular}{lr|rrrr}
 \hline
 Species & total  & $>$25\% & $ > $50\% & $ >$75\% \\ 
 \hline
\Dyak & 1442&   833 & 646 & 430  \\
\Dsim & 978 &   641  & 507 & 321  \\
\hline
\end{tabular}
\end{center}
\end{table}

\clearpage
\section*{Supplementary Figures}

\clearpage

\clearpage

\clearpage

\begin{figure}[!h]
  \begin{center}
\includegraphics[scale=0.25]{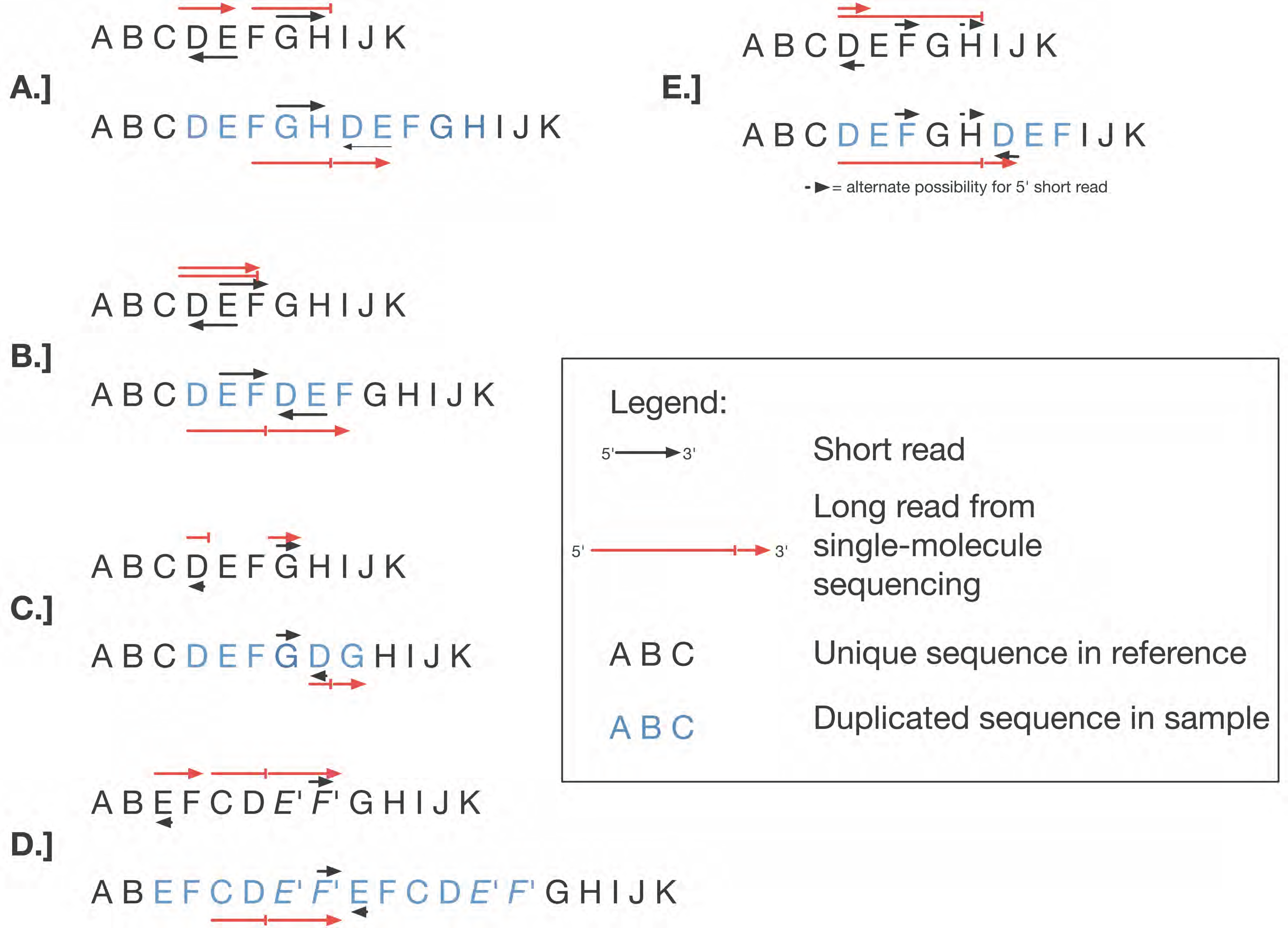}
  \caption{\label{fig:SMclasses}Confirmation of tandem duplications using single-molecule sequencing.  We considered five simple breakpoint structures resulting from simple tandem duplication events.  A.]  A simple tandem duplication.  Short reads (black arrows) from this type of breakpoint map in a divergent orientation, with reads aligning to different strand and pointing (5' to 3') away from each other \citep{JCridland} when aligned to the reference.  A long read (red arrow) spanning such a breakpoint will align discontinuously to the reference, with the fragment from the 3' section of the breakpoint aligning to the reference upstream of where the fragment from the 5' section of the breakpoint aligns.  These fragments are expected to align to the loci where the divergently-oriented reads aligned.  B.]  When the long read is long relative to the length of the duplication, the entire duplication event may be captured in a single read. In this case, the long read will contain two alignments to the region flanked by divergently-oriented short reads.  C.] The case of an incomplete duplication, where markers E and F are deleted from the tandem duplicate.  In this case, 5' fragment of a long read will align to the 5' segment of the reference containing divergently-oriented short reads (\textit{e.g.} short reads aligned to the minus strand), and the 3' fragment will align to the 3' cluster of divergently-oriented reads (which are aligned to the plus strand).  The two fragments of the long read will align to the same strand in the same orientation. D.]  The case of a tandem duplication involving semi-repetitive sequence.  Here, the regions EF and E'F' in the reference are assumed to be similar, but not identical, in sequence, such that short reads may align uniquely to positions differentiating the two loci.  A long read spanning the repetitive segment will result in a fragment aligning twice to the reference.  This multiply-mapping segment will cover both regions containing divergently-oriented short reads.  E.] The case of a tandem duplication separated by spacer sequence.  For this case, there are two possible configurations for divergently-oriented short reads, and a long read spanning the novel breakpoint will contain two different regions mapping to the same part of the reference sequence.}
    \end{center}
\end{figure}

\clearpage

\clearpage

\clearpage

\end{document}